\documentclass[twoside,11pt]{article}

\usepackage{amsmath}
\usepackage{amssymb}
\usepackage{amsthm}
\usepackage{indentfirst}
\usepackage[hidelinks]{hyperref}
\usepackage{paralist}
\usepackage{geometry}
\usepackage{setspace,lipsum}
\usepackage[title]{appendix}
\usepackage{color}
\usepackage{multimedia}
\usepackage{multirow}
\usepackage{bbm}
\usepackage{graphics}
\usepackage{graphicx}
\usepackage{adjustbox}
\newcommand{\fitdisplay}[1]{\adjustbox{max width=.91\linewidth}{$\displaystyle #1$}}
\usepackage{booktabs}
\usepackage{grffile}
\usepackage{float}
\usepackage{url}
\usepackage{mathabx}
\usepackage{supertabular}
\usepackage{rotating}
\usepackage{pdflscape}
\usepackage{verbatim}
\usepackage{listings}
\usepackage{xcolor}
\usepackage{multicol}
\usepackage[shortlabels]{enumitem}
\usepackage{caption,subcaption}
\usepackage{epstopdf}
\usepackage{algorithm}
\usepackage[round]{natbib}
\usepackage{etoolbox}
\usepackage{makecell}

\graphicspath{{Figures/}}
\newcommand{\bm}[1]{\boldsymbol{#1}}
\newcommand{\dd}{\mathrm{d}}

\newcommand{\E}{\mathbb{E}}
\newcommand{\p}{\mathbb{P}}

\newcommand{\g}{\mathcal{G}}
\newcommand{\f}{\mathcal{F}}

\newtheorem{theorem}{Theorem}
\newtheorem{lemma}{Lemma}

\newtheorem{remark}{Remark}
\usepackage{tikz}
\usepackage{multirow}
\usepackage{algpseudocode}

\algblock{Input}{EndInput}
\algnotext{EndInput}
\algblock{Output}{EndOutput}
\algnotext{EndOutput}

\usetikzlibrary{fit,positioning}

\begin{document}
\raggedbottom
\title{Mean--Variance Portfolio Selection by Continuous-Time Reinforcement Learning: Algorithms, Regret Analysis, and Empirical Study}
\author{Yilie Huang\thanks{Department of Applied Mathematics, The Hong Kong Polytechnic University, Hung Hom, Kowloon, Hong Kong. \texttt{yilie.huang@polyu.edu.hk}.}\qquad Yanwei Jia\thanks{Department of Systems Engineering and Engineering Management, The Chinese University of Hong Kong, Shatin, New Territories, Hong Kong. \texttt{yanweijia@cuhk.edu.hk}.}\qquad Xun Yu Zhou\thanks{Department of Industrial Engineering and Operations Research \& Data Science Institute, Columbia University, New York, NY, 10027, USA. \texttt{xz2574@columbia.edu}.}}
\date{}
\maketitle
\begingroup\renewcommand{\thefootnote}{}\footnotetext{Accepted for publication in \textit{Management Science}.}\endgroup
\begin{abstract}\singlespacing
We study continuous-time mean--variance portfolio selection in markets where stock prices are diffusion processes driven by observable factors that are also diffusion processes, yet the coefficients of these processes are unknown. Based on the recently developed reinforcement learning (RL) theory for diffusion processes, we present a general data-driven RL approach that learns the pre-committed investment strategy directly without attempting to learn or estimate the market coefficients. For multi-stock Black--Scholes markets without factors, we further devise an algorithm and prove its performance guarantee by deriving a sublinear regret bound in terms of the Sharpe ratio. We then carry out an extensive empirical study implementing this algorithm to compare its performance and trading characteristics, evaluated under a host of common metrics, with a large number of widely employed portfolio allocation strategies on S\&P 500 constituents. The results demonstrate that the proposed continuous-time RL strategy is consistently among the best, especially in a volatile bear market, and decisively outperforms the model-based continuous-time counterparts by significant margins.
\end{abstract}

\section{Introduction}
\label{sec:intro}
In this paper, we study portfolio selection (or asset allocation) in dynamically traded markets for an investor who aims to achieve mean--variance efficiency in a finite investment horizon using reinforcement learning (RL). Since \cite{markowitz1952portfolio} introduced the mean--variance (MV) framework for static (single-period) portfolio choice, it has become one of the central topics in both modern portfolio theory and quantitative investment practice. However, despite its profound theoretical appeal and implications,  practically implementing MV efficient strategies is challenging. First, most applications of the MV analysis are still restricted to the static setting to this day in practice \citep{kim2021mean}, whereas applying static strategies myopically is surely inefficient from the dynamic perspective \citep{kim1996dynamic}.
Second, accurately estimating the moments of asset returns is notoriously difficult, especially for the expected returns \citep{merton1980estimating,luenberger1998investment}.
Portfolios derived from analytical/numerical solutions of the MV problems in the static setting are known to be extremely sensitive to such estimation errors \citep{best1991sensitivity,best1991sensitivityms,britten1999sampling}, and become even worse for the dynamic one. Mitigating such errors and sensitivity and achieving MV efficiency in the dynamic environment remains largely an important open question.

Recent developments in machine learning have slowly but surely changed the thinking and the practice of decision-making under uncertainty in a fundamental way, and RL-based approaches have become more popular and better accepted in many application domains. One important feature of RL is to learn optimal actions (portfolio strategies in our case) {\it directly} via dynamic interactions with the environment (market) in a data-driven and model-free fashion, {\it without} estimating any parameter of a statistical/probabilistic model. In the setting of this paper, ``data" are both  exogenous (including  asset price data and other possibly time-varying but {\it observable/computable}, aggregate or individual covariates that affect the means and covariances of asset returns) and  {\it endogenous} generated by an agent's strategic interactions with the unknown market. The dynamic nature of RL aligns with the setting of dynamically traded markets and farsighted investors. More importantly, learning portfolio choices directly while bypassing model estimation provides a powerful remedy to the aforementioned drawbacks of estimation errors and sensitivity inherent in the classical MV approach.

This paper studies portfolio selection in a continuously traded market for an agent with an MV preference. The agent observes stock prices and market factors but has minimum knowledge about the market and is unable to form a precise statistical model about the law of motions as assumed in the conventional financial economics literature.  The only assumption about the market environment is that the stock prices are diffusion processes driven by observable factors that are also diffusion processes. The agent does not know the coefficients of these diffusion processes and aims to solve the continuous-time MV problem based on the observable data (such as the factors, stock prices, and wealth processes under different investment strategies) {\it only}.

The assumption of a diffusion-based framework is due to the following considerations: 1) asset prices governed by diffusion processes are commonly used and extensively studied in the finance literature; 2) the main idea and approach of this paper are adapted from the general continuous-time RL theory in \cite{wang2020reinforcement,jia2021policy,jia2021policypg}, which are developed for controlled diffusion processes. The diffusion model serves as a canonical benchmark in continuous-time portfolio
theory, enabling us to study regret -- one of the main objectives of this paper -- rigorously;\footnote{The general RL theory has recently been extended to jump-diffusions \citep{gao2024reinforcement} that can be employed to model asset prices more realistically. However, in this paper, we restrict ourselves  to It\^o's diffusions for simplicity and for staying focused  on the core objective, namely to study convergence and regret, which is already amply technical in the diffusion setting.} and 3) importantly, being ``model-free" does not mean there is no underlying probability distribution assumption; rather, it means that the algorithm's design and execution do not depend on estimating a model statistically, and the algorithm is applicable to a
broad class of models.\footnote{\citet{gao2024reinforcement} prove that one can apply the {\it same} algorithm designed for diffusions to jump-diffusions without having to verify in advance which one underlines the actual data generating process.}  Indeed, RL depends crucially on the Markov property to apply dynamic programming principle, which is why classical discrete-time RL always works with Markov chains. A diffusion process can be considered as  the continuous-time/space counterpart of a Markov chain. On the other hand, our theoretical regret analysis does depend on a specific diffusion dynamics.

The main contributions of this paper are three-fold. First, we propose procedures in developing RL algorithms for this problem by applying and adapting the general theory established by \cite{wang2020reinforcement} and \cite{jia2021policy,jia2021policypg} to the MV setting. The foundation of the algorithms is to solve moment conditions arising from certain martingale conditions. Yet, these moment conditions are profoundly different from those employed in conventional econometrics in terms of actively generating {\it new} data for learning.
Second, when the stock prices follow a multi-dimensional Black--Scholes environment without factors, we devise a specific RL algorithm and prove its convergence. Moreover, we show that the algorithm achieves a {\it sublinear} regret in terms of the Sharpe ratio. Here, ``regret" is the cumulative error over a number of learning episodes between the algorithm and the ``oracle" one (i.e., the theoretically optimal one under the complete knowledge of the market environment). The sublinearity ensures that the RL algorithm will achieve nearly optimal results after a sufficiently long training period, even with an unknown market. This is, to our knowledge, the first sublinear regret analysis of a model-free learning algorithm in continuous-time MV portfolio choice. While the learning algorithm itself operates without estimating market parameters, we establish this theoretical guarantee specifically within a Black-Scholes environment, utilizing specific function approximators whose proof is premised upon a highly delicate analysis of diffusion processes and stochastic approximation techniques.
Then, we carry out a comprehensive empirical study to compare the resulting RL strategy with 13 alternative popular, mostly econometric methods, using multiple performance metrics on S\&P 500 constituents for the period 2000--2020, with  1990-2000 as the burn-in period for pre-training. These alternatives include the market portfolio, equally weighted portfolio, sample-based estimations, factor models, Bayesian estimation, model-based continuous-time MV, a linear predictive model, and two general-purpose RL algorithms. The evaluation criteria cover both performance and trading metrics. The performance measures include annualized return, Sharpe ratio and its variants, maximum drawdown, and recovery time, while the trading metrics include gross exposure to risky assets, turnover, concentration, and bankruptcy probability. An unequivocal conclusion from the extensive empirical study is that our RL strategy significantly outperforms the classical model-based, plug-in continuous-time counterpart across all evaluation criteria regardless of market conditions. Our strategy also consistently dominates the others in most metrics, especially in  volatile and downturn markets. The superiority of our approach does not stem from the use of predictive factors or complex neural networks but rather from our fundamentally distinct decision-making paradigm: learning the optimal strategy without learning the model.

\subsection*{Related Literature}
\paragraph{\textbf{Methodologies to improve static MV}} There are two main directions in the literature for mitigating  sample-based estimation issues for (static) MV problems.  The first is to develop more efficient estimators, including  Bayesian inference and shrinkage estimators \citep{james1992estimation} to reduce estimation errors. The latter has been particularly popular for portfolio selection, e.g., shrinkage estimators for mean \citep{jorion1986bayes,black1990asset}, covariance matrix \citep{ledoit2003improved,ledoit2017nonlinear}, and covariance matrix of idiosyncratic error in factor models \citep{fan2008high,fan2012vast}. The second direction takes the robust optimization approach. The idea is, instead of pinpointing a fixed model for optimization, to consider a {\it family} of models (also known as the ambiguity set) that contain the true but unknown model and optimize the objective in the worst scenario among these many models. Applying to MV portfolio selection, this approach modifies the original MV preference to a max-min MV objective  \citep{garlappi2007portfolio,goldfarb2003robust}.
Other works along this line include portfolio weight norm regularization  \citep{demiguel2009generalized}, performance-based regularization  \citep{ban2018machine}, and many others. Most of the related formulations, however, need to set the width/radius of the ambiguity set as an exogenous hyper-parameter. \citet{blanchet2022distributionally} employ a distributional robust approach and propose a statistical inference way of determining this uncertain set endogenously with a performance guarantee. However, robust approaches have been developed predominantly for static optimization, which become amply complex and intractable when dealing with a dynamic environment. On the other hand,  \citet{demiguel2009optimal}, in a thorough empirical study, show that most of these approaches do not consistently outperform the na\"ive equally-weighted portfolio. \citet{blanchet2022distributionally} corroborate the competitive performance of the equally-weighted portfolio but find their distributional robust portfolios achieve a higher Sharpe ratio on average. However, they stop short of experimenting with other popular metrics such as maximum drawdown and recovery time. Above all, all these studies are on static MV problems. By contrast, we investigate forward-looking and dynamically planning investment policies, while providing a more comprehensive empirical study.

\paragraph{\textbf{Econometric methods for estimating diffusion models}} With high-frequency observations, various econometric methods have been developed to estimate diffusion processes, such as the generalized method of moments - GMMs \citep{hansen1995back,kessler1999estimating}, approximate maximum likelihood estimation \citep{lo1988maximum,ait2002maximum,ait2008closed,ait2007maximum}, non-parametric regression with approximate moment \citep{stanton1997nonparametric}, and Monte Carlo Markov Chain (MCMC) based simulation \citep{eraker2001mcmc}. However, even for the simplest model, an accurate estimation of drift coefficients requires unrealistically large datasets due to the so-called ``mean-blur" problem (see \citealt{luenberger1998investment} for estimating stock returns, and \citealt{baek2021limits} for epidemic and marketing models). Estimation errors, in turn, profoundly affect MV portfolio choices; see \cite{best1991sensitivity, britten1999sampling, chan1999portfolio} and \cite{chopra2013effect} for the static setting and  \citet{blanchet2022distributionally} for the dynamic one.

\paragraph{\textbf{Financial economics literature on dynamic portfolio choice}} Dynamic portfolio choice has been studied extensively in the conventional financial economics and financial engineering literature. However, the research typically focuses on  specific models, such as \citet{zhou2000continuous,lim2002mean,basak2010dynamic,wachter2002portfolio,liu2007portfolio,gennotte1986optimal,cvitanic2006dynamic}, among many others, by assuming the agent has complete or at least partial knowledge about the underlying market environments. In the case when, for instance, the agent knows that the stock prices follow geometric Brownian motions but the drift and/or volatility coefficients are unknown, she employs Bayesian  learning to estimate the unknown coefficients. By contrast, the RL framework distinguishes itself  by considering a ``model-free'' paradigm; that is, the agent only has the minimum knowledge about the market (such as that stock prices are diffusion processes) and learns optimal/efficient portfolio strategies directly which is not guided by statistical principles (such as Bayesian learning).

\paragraph{\textbf{Machine learning in portfolio related problems}} Despite the long history of machine learning research, applications to finance only started recently in the wake of AI and FinTech boom. For example, deep neural networks have been employed to study empirical asset pricing \citep{lettau2020factors,gu2020empirical,gu2021autoencoder,bianchi2021bond,guijarro2021deep,leippold2022machine,chen2024deep}. These works focus largely on building nonlinear predictive models for asset returns or constructing trading signals to learn complex patterns in historical data.
However, RL has been hitherto barely used by the asset management industry \citep{snow2020machine}, largely due to its lack of interpretability/explainability and lack of theoretical guarantee even under the simplest Black--Scholes environment. Most existing literature (e.g., \citealt{gao2000algorithm,jin2016portfolio,ritter2017machine} among others) on RL for portfolio optimization are based on ad-hoc adoptions of existing general-purpose RL algorithms without theoretical formulation nor analysis, and the empirical investigations of their performances are not comprehensive. This paper is a part of the on-going effort that aims to provide rigorous underpinnings for RL with diffusion processes since \citet{wang2020reinforcement,jia2021policy,jia2021policypg}. The algorithm proposed in this paper is an adaptation of the policy gradient-based actor--critic algorithm in \citet{jia2021policypg} by incorporating an expectation constraint and a covariance matrix update formula. By virtue of delicately tailoring to the specific MV problem, this paper is the first to obtain a regret upper bound to ensure the performance of a model-free algorithm in the diffusion setting.\footnote{\cite{wang2020continuous} is the first to propose an RL algorithm for continuous-time MV portfolio selection built on the rigorous mathematical foundation established by \cite{wang2020reinforcement} with an entropy-regularized relaxed control formulation for general continuous-time RL. However, \cite{wang2020continuous} employ the commonly used mean--square temporal--difference error (MSTDE) as the objective to perform policy evaluation (PE). Later, \cite{jia2021policy} point out that minimizing MSTDE for diffusion processes is equivalent to minimizing the expected quadratic variation of a martingale, which is not a proper objective. Instead, \cite{jia2021policy} prove some martingale conditions that theoretically support their proposed offline and online PE algorithms. The present paper is based on \cite{jia2021policy} for PE and the subsequent \cite{jia2021policypg} for policy gradient, leading to an entirely different algorithmic paradigm than \cite{wang2020continuous}. }

\paragraph{\textbf{Theoretical results on regret in RL}} For episodic RL in discrete-time with finite state--action spaces, it has been established in the literature that the typical ``optimal" regret order is $\sqrt{N}$; see e.g. \citet{dann2017unifying,jin2018q,li2021breaking,agrawal2024optimistic}. \citet{gao2025square}  and \citet{jin2023provably} obtain the same order for continuous-time finite-state Markov chain and a class of general-state, discrete-time Markov decision processes respectively. For diffusion processes, the only work we can find is \citet{szpruch2024optimal} that derives a regret order of $\sqrt{N}$ for a {\it model-based} linear--quadratic RL algorithm. All these works study expected reward maximization problems. For a class of risk-aware problems, the same $\sqrt{N}$ order has also been obtained in the discrete-time literature, such as in \citet{fei2020risk,fei2021exponential,liu2022distributionally,wang2023finite,xu2023regret,wenhaoregret}. By contrast, here we consider a mean--variance problem. To our best knowledge, the present paper is the first to study the regret of the Sharpe ratio (which is probably the most appropriate in the MV context) in a model-free, continuous-time setting.

\medskip

The rest of the paper is organized as follows. In Section \ref{sec:overview portfolio approach}, we present the MV formulation in a continuous-time multi-stock market environment with factors, and discuss the fundamental differences between the conventional model-based plug-in paradigm and that of RL. Section \ref{sec:rl foundation} explains the key steps in a general RL algorithm to solve the MV problem. Section \ref{sec:proveable} presents an algorithm and its theoretical guarantee on the convergence of the learned policies, along with a regret analysis in terms of the Sharpe ratio, for a  multi-stock Black–Scholes market. Section \ref{sec:empirical} reports and discusses the results of an extensive comparative empirical study. Finally, Section \ref{sec:conclusion} concludes. Proofs and additional numerical results are included in the Appendix.

\section{Dynamic Mean--Variance Portfolio Choice}
\label{sec:overview portfolio approach}
In this section, we describe the market environment and the objective of an MV agent in the continuous-time setting with minimum assumption, and review two paradigms -- those of the conventional plug-in and the RL, respectively.

\subsection{Market environment and mean--variance agents}
\label{sec:market}
We first describe a general continuously traded market. There are $d+1$ assets, of which the 0-th asset is risk free whose price is $S^0(t)$ with a short interest rate $r(t)$. The other $d$ assets are risky stocks whose prices at $t$ are denoted by $S^1(t),\cdots,S^d(t)$. In addition, there are $m$ additional {\it observable} covariates $F(t)\in \mathbb{R}^m$ that are associated with the interest rate, mean and covariance of the asset returns, referred to as the (market) factors.

In general, a \textit{model} for the financial market makes further structural assumptions about the dynamics of asset prices and factors. For example, the celebrated Black--Scholes model assumes stock prices follow geometric Brownian motions, and Heston's model stipulates stochastic volatility as factors. However, we do not assume agents know concrete forms of the market models, other than that $S(t) = (S^0(t),S^1(t),\cdots,S^d(t))^\top$ and $F(t)$ are It\^o's diffusions.\footnote{We assume these processes are all well-defined and adapted in a given filtered probability space $\left(\Omega, \mathcal{F}, \p ;\left(\mathcal{F}_t\right)_{t \geq 0}\right)$ satisfying the usual conditions. An It\^o's diffusion belongs to a wide class of Markov processes, which can be represented as the solution to a stochastic differential equation driven by a (multi-dimensional) Brownian motion. It satisfies the strong Markov property and admits an infinitesimal generator. We do not consider non-Markov processes here, which can be equivalently formulated as path-dependent Markov ones where certain factors can be summary statistics of the path history (e.g. the momentum).} As a consequence, it is extremely difficult if not impossible to apply conventional statistical methods including Bayesian learning to estimate/learn the models.

Consider such a ``model-free'' agent with initial wealth $x_0$ and a pre-specified investment horizon $T > 0$. We denote the agent's portfolio  choice at time $t$ by $u(t)=(u^1(t), u^2(t), \cdots ,u^d(t))^\top\in \mathbb{R}^d$, where $u^i(t)$ is the discounted dollar amount (equivalently, $u^i(t) S^0(t)$ is the nominal dollar amount) invested in the $i$-th risky asset at time $t$, $1 \leq i \leq d$. Denote by $x^{u}\equiv \{x^u(t): 0 \leq t \leq T \}$ the discounted self-financing wealth process of the agent under a portfolio process $u\equiv \{u(t): 0 \leq t \leq T \}$. Then the agent's discounted wealth process satisfies the wealth equation
\begin{equation}
	\dd x^{u}(t)= \sum_{i=1}^d u^i(t) \frac{\dd S^i(t)}{S^i(t)} - \bm{e}_d ^\top u(t) \frac{\dd S^0(t)}{S^0(t)},
	\label{classical_dynamics}
\end{equation}
where $\bm e_d = (1,\cdots,1)^\top\in\mathbb{R}^d$ is a $d$-dimensional unit vector, and $\frac{\dd S^i(t)}{S^i(t)}$ is the return of the $i$-th asset.\footnote{The nominal wealth process $x^{u}(t) S^0(t)$ satisfies $\dd ( x^{u}(t) S^0(t) ) = \sum_{i=1}^d S^0(t)u^i(t) \frac{\dd S^i(t)}{S^i(t)} + (x^u(t) S^0(t) -\bm{e}_d ^\top S^0(t)u(t)) \frac{\dd S^0(t)}{S^0(t)}$. Applying stochastic calculus leads to \eqref{classical_dynamics}.} Note that the wealth equation \eqref{classical_dynamics} follows from a simple fact that the change of wealth is caused by the changes in asset prices; hence it is very general, {\it independent} of any model about the stock prices or factors.

We take the continuous-time framework for several reasons. First of all, it captures the fact that investors can trade continuously nowadays, which is also reflected by the ample amount of study on continuous-time portfolio choice in the literature. Second, the importance and merits of developing a continuous-time RL framework, {\it despite} the existence of a rich and extensive literature on discrete-time RL, have been explained in great details in \cite{wang2020reinforcement,jia2021policy,jia2021policypg}. The general theory and algorithms from those papers in turn provide a foundation for the specific application in this paper.
Third, we carry out all the theoretical analysis in continuous time and perform time discretization only at the {\it final} stage for numerical implementation. This differs fundamentally from discretizing time at the outset, which has been known to suffer from  sensitivity with respect to the discretization step and even collapse when the step size is sufficiently small (see, e.g., \citealt{tallec2019making,park2021time}). Finally, the last-stage discretization scheme is detailed in Appendix \ref{appendix:algo}, and the impact of this discretization is rigorously analyzed in \citet{jia2025accuracy}.

For a small investor (a price taker), the asset prices and market factors are exogenous that are unaffected by her actions (portfolios). By contrast,  a large investor's portfolio choice can alter the price and factor processes, e.g., through temporary or permanent price impact, and other frictions from the market microstructures. Such trading frictions and microstructures are an important part of the market environment,
which is not assumed to be known by the investor. In our RL setting, the only assumption about the market is that $(S(t), F(t) )$ are It\^o's diffusions.

Given the investment horizon $T$, the agent aims to find MV efficient allocations in this dynamically traded market. As the continuous-time counterpart to the Markowitz problem, the classical model-based continuous-time MV problem is formulated as follows. {\it Assuming} a model for $S(t),F(t)$ and the wealth equation \eqref{classical_dynamics} are known and given, to minimize the variance of the portfolio while achieving a given expected target return:
\begin{equation}
	\fitdisplay{\begin{aligned}
		&\min_{u} \operatorname{Var}\left( x^{u}(T)\right) \\
		&\text{subject to } \mathbb{E}\left[x^{u}(T)\right]=z
		\label{classical}
	\end{aligned}}
\end{equation}
where $z>x_0$ is the target expected terminal wealth that is pre-specified at $t=0$ by the agent as a part of the agent's preference. A larger $z$ indicates that the agent pursues higher return and hence is less risk-averse.

As a remark, the above is not a standard stochastic control problem (in contrast to the expected utility maximization) due to the presence of the variance term in \eqref{classical}. This term causes time-inconsistency so the dynamic programming principle does not apply directly. \cite{zhou2000continuous} introduce a method of using a Lagrange multiplier $w$ to transform the problem into an unconstrained expected quadratic (dis)utility minimization problem:
\begin{equation}
	\min_{u} \mathbb{E} [(x^{u}(T) - w)^2] - (w - z) ^2\\
	\label{classical_new}
\end{equation}
and then finding a proper multiplier $w$ to enforce the mean constraint. This problem is a standard stochastic control problem and is time consistent, whose solution gives rise to a {\it pre-committed} investment strategy to the original problem \eqref{classical}.\footnote{See a recent survey \cite{he2022time} on  pre-committed strategies and other types of strategies under time-inconsistency.}

\subsection{The conventional plug-in paradigm}
\label{sec:plug-in}
The solution to the asset allocation problem such as \eqref{classical} can be computed when the market model is completely specified, thanks to the well-developed stochastic control methodologies. How to mathematically solve \eqref{classical} and what the economic implications are have been the focus of conventional research on quantitative finance and financial economics. Conceptually, these works take the rational expectation point of view so that agents can form their belief about the market environment correctly and, hence, behave optimally. As a result, the optimal policy (i.e. the plan of optimally reacting to the state) can be determined/prescribed even before the investment starts.

Practically, agents are always limited by their knowledge about the ``true'' market model (let there be one). The traditional approach to asset allocation (or indeed more general decision making problems) is to first propose and estimate a specific model (for the joint dynamics of stock prices $S$ and factors $F$) and then plug the estimated model parameters (such as the drifts and volatilities of the dynamics) into the optimal solution for the corresponding model. This is usually referred to as the \textit{model-based} or {\it plug-in} approach which combines two steps/techniques: certain algorithms to estimate the model parameters (e.g. maximum likelihood or Kalman filtering) and certain algorithms to solve the stochastic control problem (e.g. analytical or numerical solutions to the Hamilton--Jacobi--Bellman (HJB) equation). Such a paradigm, however, suffers from problems of model misspecification, estimation errors due to limited data and inadequate statistical methods, and sensitivity of optimal solutions to model parameters, as discussed in Section \ref{sec:intro}.

\subsection{The reinforcement learning paradigm}
\label{sec:rl paradigm}
The RL version of the problem \eqref{classical} is to solve it based only on the observable data including the price/factor processes and the agent's own wealth processes under various portfolios, without any knowledge about the market other than that the underlying processes are It\^o's diffusions. Hence, it addresses the task of a model-free agent rather than one having rational expectations. Recall that  the plug-in approach first estimates (explores) and then optimizes (exploits), carrying out these two steps separately and {\it sequentially}. By contrast, the RL approach does exploration and exploitation {\it simultaneously} all the time.
With RL, the agent interacts  with the unknown (market) environment directly by trial and error, and improves strategies by incorporating the responses of the environment to the exploration.

The classical stochastic control theory (\citealt{YZbook,fleming2006controlled}) stipulates that, (under mild conditions) for a Markov system, the optimal portfolio choice can be written as a {\it deterministic} function of the time and the wealth--factor variables, that is, $u(t) = \bm{u}^*(t, x(t), F(t))$ for some measurable, deterministic function $\bm{u}^*$, also known as a \textit{policy}. Ultimately, both plug-in and RL approaches attempt to find this function, but in profoundly different ways. The former first estimates a market model and then optimizes accordingly by solving HJB PDEs (both steps are numerically challenging). The latter skips estimating a model and solving a PDE; instead, it approximates the policy function by a suitable class of  functions with finite-dimensional parameters (e.g., using polynomials or neural networks), and learns/updates directly these parameters via exploration and exploitation.
This idea underpins the \textit{model-free} approach. We reiterate here that a model-free approach does not mean there are no models; rather, there is an underlying {\it structural} model (e.g., a Markov chain, an It\^o diffusion, or a jump-diffusion) for the data-generating process but we do not know the model parameters, {\it nor do we attempt to estimate them}. Any provable performance guarantee including regret bounds must be established upon this structural assumption {\it only}.

To sum up, the plug-in approach focuses on learning the environment to make optimal decisions, whereas RL learns to optimize policies directly based on  performance -- it ``learns by experience".

\section{Foundation of Reinforcement Learning Algorithms}
\label{sec:rl foundation}

A typical RL algorithm involves answering three questions: how to choose actions to strategically interact with the environment for the purpose of exploration, how to evaluate the performance of a given policy, and how to update the policy to improve its performance. We follow the framework of continuous-time RL in \cite{wang2020reinforcement} and \cite{jia2021policy,jia2021policypg} to address these three questions, respectively. While applying those general results, there is an additional Lagrange multiplier $w$ in \eqref{classical_new} that needs to be learned, which is specific to the MV setting.

\subsection{Deterministic versus stochastic policies}
\label{sec:stochastic policy}

In the classical model-based setting, an optimal policy is a deterministic mapping from the time--wealth--factor triplet to an action (a portfolio choice). However, when the market environment is unknown, the RL agent undergoes exploration by {\it randomizing}
the policies in order to broaden the search space. Mathematically, these exploratory policies are now mappings that map  time--wealth--factor triplets to  probability (density) distributions on the action space:
\[ \left\{ \bm\pi: (t,x,F)\mapsto \bm\pi(\cdot|t,x,F) \in \mathcal{P}(\mathbb{R}^d)  \right\},   \]
where $\mathcal{P}(\mathbb{R}^d)$ denotes the set of all probability density functions on $\mathbb{R}^d$.\footnote{In this paper, we restrict randomization to distributions having density functions because they are the most commonly used and compatible with the entropy regularizer to be introduced momentarily. With more involved notation, the density functions can be replaced by distribution functions. }
Note that the agent wealth process $x(t)$ and the factor process $F(t)$ are both observable (i.e. they are {\it data}) at any time $t$.
Given a mapping $\bm\pi$, at each time $t$, a portfolio $u(t)$ is independently sampled  from the distribution given by $\bm\pi(\cdot|t,x(t),F(t))$, denoted by $u^{\bm\pi}(t) \sim \bm\pi(\cdot|t,x(t),F(t))$. Such a rule $\bm\pi$ to generate portfolios is called a \textit{stochastic policy}. When $\bm\pi$ is a point mass (aka Dirac measure), it reduces to the conventional deterministic policy. Once $\bm{\pi}$ is specified, the portfolio processes to be {\it actually} executed could be sampled according to $u^{\bm\pi}(t) \sim \bm{\pi}(\cdot|t,x(t),F(t))$ and the resulting wealth trajectories, while following \eqref{classical_dynamics}, can be directly observed, both not requiring the knowledge of the market coefficients. We denote the wealth trajectories under a stochastic policy $\bm\pi$ by $x^{u^{\bm\pi}}$. The statistical properties of this process are described in Appendix \ref{appendix:formulation rigorous}.

Technically, a stochastic policy $\bm\pi$ is a $\mathcal P(\mathbb R^d)$-valued feedback control adapted to the state variables $(x,F)$ and used to generate conventional $\mathbb R^d$-valued portfolios $u^{\bm\pi}$ to manage  the wealth process. One can think of $\bm\pi$ as a random device (e.g. a dice with infinitely many sides). Each portfolio choice $u^{\bm\pi}$ is an independent random draw from $\bm\pi$ and, hence, is adapted to not only the original state variables but also the policy  randomization. Therefore, the executed portfolios are adapted to a larger information set than the conventional counterpart due to a new source of randomness. However, this additional randomness stems from the random device used to generate portfolios which is {\it independent} of the market. Thus, being adapted to a larger information set alone does not necessarily yield more information about the market nor better performance. For a rigorous account for the probabilistic setup used in continuous-time RL, see \citet[Section 3]{jia2025accuracy}.

As the essence of RL is to balance exploration and exploitation while the former is needed during the entire time horizon, we further add an entropy regularizer to the objective function to encourage and, indeed, enforce exploration. The entropy regularization is closely related to the soft-max approximation in the RL literature \citep{ziebart2008maximum,haarnoja2018soft,wang2020reinforcement}, as well as the choice model and the perturbed utility \citep{hotz1993conditional,fudenberg2015stochastic,feng2017relation} in microeconomic theory. This leads to the following entropy regularized objective function:
\begin{equation}
\mathbb{E}\left[\left({x}^{u^{\bm{\pi}}}(T)-w\right)^{2}+\gamma \int_{0}^{T} \log \bm{\pi}(u^{\bm\pi}(t) |t,{x}^{u^{\bm{\pi}}}(t),F(t))  \dd t\right]-(w-z)^{2}
\label{rl_formulation}
\end{equation}
where $\gamma \geq 0$ is an exogenous parameter, called a {\it temperature parameter}, that specifies the weight put on exploration. Clearly, a larger $\gamma$ favors more exploration and vice versa.

Let us conclude this subsection by noting some important points about stochastic policies.
In optimization broadly, randomization is taken for conceptual and/or technical reasons; see \cite{zhou2023} for an assay on this. In RL specifically, randomization is used primarily for exploration or information collection (e.g., $\epsilon$-greedy policy for the bandit problem; see \citealt{sutton2018reinforcement}), as randomization broadens search space and enables an agent to experience counterfactuals.\footnote{For example, in the bandit problem, randomization allows the agent to play a currently sub-optimal machine that otherwise would have never been played (and therefore whose information would have remained unknown).} Intuitively, by trying out different trading portfolios the agent gets to know more about the market including  market impact which in turn guides her to optimize gradually.

However, if the agent is a {\it small} investor, the current portfolio selection problem has a {\it distinctive} feature in this aspect. Recall that the (discounted) wealth equation is described by \eqref{classical_dynamics}, where $\frac{\dd S^i(t)}{S^i(t)}$ and $\frac{\dd S^0(t)}{S^0(t)}$ are the (instantaneous) returns of the risky and risk-free assets respectively that can be observed directly from the market {\it without} having to know the market coefficients. So wealth change is jointly caused by portfolio choice and price movement in a {\it known}, multiplicative way. However, with a small investor, the price movement is purely {\it exogenous} and observable regardless of what portfolios she applies. Therefore, \eqref{classical_dynamics} reveals all the counterfactuals under alternative portfolios without actually having to execute them.\footnote{This is a very specific feature of this specific setting (a small investor), which is not owned by most stochastic control problems including portfolio choice with large investors.}  To wit, {\it there is no informational motive for exploration/randomization for a price taker}. That said, there are important {\it technical} reasons to use stochastic policies for learning. Generally, randomization convexifies policy spaces and facilitates differentiation. Specifically, in this paper, we will apply the policy gradient algorithms developed in \cite{jia2021policypg} to update policies, whereas the key idea of \cite{jia2021policypg} is to turn the policy gradient into a policy evaluation problem which works {\it only} for stochastic policies. Therefore, we will train our algorithms using stochastic policies and implement portfolios with deterministic policies.\footnote{In RL terms, this is a type of {\it off}-policy learning (\citealt[Chapter 6]{sutton2018reinforcement}), i.e. we use stochastic policies -- called the behavior policies -- to improve deterministic policies which are the target policies. }

\subsection{Policy evaluation}
\label{sec:pe}

Policy evaluation stands for estimating/predicting the payoff function of a given policy, based on which the agent decides how to update and improve the policy. In our case, it is to estimate the expected payoff \eqref{rl_formulation} for a given stochastic policy $\bm{\pi}$, a given multiplier $w$ and a given temperature parameter $\gamma$, {\it based on data only}. Moreover, the policy evaluation requires learning the expected payoff starting from {\it any} initial time--wealth--factor triplet and hence calls for estimating the {\it entire} objective function (instead of function {\it values} at some given triplets). Precisely, based on the Markov property, we need to learn a function $J$ of $(t,x,F)$, known as the value function, where
\[\fitdisplay{
J(t,x,F;\bm{\pi};w) = \mathbb{E}\left[\left(x^{u^{\bm{\pi}}}(T)-w\right)^{2} + \gamma \int_t^{T}  \log\bm\pi(u^{\bm\pi}(s)|s,x^{u^{\bm{\pi}}}(s),F(s)) \dd s\Big| x^{u^{\bm{\pi}}}(t) = x, F(t) = F \right] -(w-z)^{2}.
}\]

\cite{jia2021policy} show that the value function is characterized by two conditions. First, it satisfies a known terminal condition: $J(T,x,F;\bm{\pi};w) = (x - w)^2 - (w - z)^2$. Second, it maintains the following process
\[J(t,x^{u^{\bm{\pi}}}(t),F(t);\bm{\pi};w) + \gamma \int_0^t  \log\bm\pi(u^{\bm\pi}(s)|s,x^{u^{\bm{\pi}}}(s),F(s)) \dd s,\]
where $u^{\bm{\pi}}(s)\sim \bm{\pi}(\cdot| s,x^{u^{\bm{\pi}}}(s),F(s) )$, to be  a martingale with respect to the filtration generated by $x^{u^{\bm{\pi}}}(s)$ and $F(s)$. For a more general and rigorous description about the various filtrations involved, see \citet[Section 3]{jia2025accuracy}.

As computers cannot process learning functions that are infinite-dimensional objects, in RL, one uses function approximation to approximate the value function by a class of parameterized functions $J(\cdot,\cdot,\cdot;w; \bm\theta)$, where $\bm\theta$ is a finite-dimensional parameter. The choice of approximators may depend on the special structure of each problem or be through neural networks. Note that, for a given policy $\bm{\pi}$ and a function approximator $J(\cdot,\cdot,\cdot;w; \bm\theta)$, both $J(t,x^{u^{\bm{\pi}}}(t),F(t);w;\bm\theta)$ and $\log\bm\pi(u^{\bm\pi}(s)|s,x^{u^{\bm{\pi}}}(s),F(s))$ can be computed by observable samples or data. \cite{jia2021policy} develop several data-driven ways to learn or update $\bm\theta$ based on the aforementioned martingality. In this paper, we will apply one of them that is consistent with the well-known temporal--difference (TD) learning: to force $\dd J(t,x^{u^{\bm{\pi}}}(t),F(t);w;\bm\theta) + \gamma \log\bm\pi(u^{\bm\pi}(t)|t,x^{u^{\bm{\pi}}}(t),F(t)) \dd t$ to be a ``martingale difference sequence'' so that it is orthogonal to any adapted process. More precisely, it means
\begin{equation}
\label{eq:pe moment test function}
\E\left[ \int_0^T \mathcal{I}(t) \left\{ \dd J(t,x^{u^{\bm{\pi}}}(t),F(t);w;\bm\theta) + \gamma \log\bm\pi(u^{\bm\pi}(t)|t,x^{u^{\bm{\pi}}}(t),F(t)) \dd t \right\} \right] = 0,
\end{equation}
for all adapted processes $\{\mathcal{I}(t):0\leq t\leq T\}$, called \textit{test functions} (or \textit{instrumental variables} in the econometrics literature). While theoretically one needs to choose infinitely many test functions, for implementation one can take $\mathcal{I}(t) = \frac{\partial }{\partial \bm\theta}J(t,x^{u^{\bm{\pi}}}(t),F(t);w;\bm\theta)$ which is a vector having the same dimension as $\bm\theta$. The task of estimating $\bm\theta$ from the system of equations \eqref{eq:pe moment test function} can thus be accomplished by the well-developed \textit{generalized methods of moments (GMMs)}.

\subsection{Policy gradient}
\label{sec:pg}
Now that we have learned the value function of a given stochastic policy, the next step is to improve the policy. For that, following the general gradient-based approach in optimization, we need to estimate the gradient of the value function with respect to the policy. However, the policy itself lies in an infinite dimensional space of probability distributions; so it is infeasible to compute the derivative directly. As before, we parameterize policies by a finite-dimensional vector $\bm\phi$:
\[ \bm{\pi}^{\bm\phi}\equiv \bm\pi(\cdot|t,x,F;w;\bm\phi).\]
Denote by $J(\cdot,\cdot,\cdot;\bm\pi^{\bm\phi};w)$ the value function under $\bm\pi^{\bm\phi}$. It now suffices to consider $\frac{\partial }{\partial \bm\phi}J(0,x_0,F_0;\bm\pi^{\bm\phi};w)$, the gradient of  $J(0,x_0,F_0;\bm\pi^{\bm\phi};w)$ in $\bm\phi$. \citet{jia2021policypg} derive the policy gradient representation as follows
\begin{equation}
\label{eq:pg representation with test function}
\fitdisplay{\begin{aligned}
	& \frac{\partial }{\partial \bm\phi}J(0,x_0,F_0;\bm\pi^{\bm\phi};w) \\
	= & \E\bigg[\int_0^T    \left[ \frac{\partial }{\partial \bm\phi} \log \bm\pi(u^{\bm\pi^{\bm\phi}}(t)|t,x^{u^{\bm\pi^{\bm\phi}}}(t),F(t);w;\bm\phi ) + \mathcal{H}(t) \right]  \Big[ \dd J(t,x^{u^{\bm\pi^{\bm\phi}}}(t),F(t);\bm\pi^{\bm\phi};w) \\
	& \quad \quad \quad \quad + \gamma \log\bm\pi(u^{\bm\pi^{\bm\phi}}(t)|t,x^{u^{\bm{\pi}^{\bm\phi}}}(t),F(t);w;\bm\phi) \dd t  \Big]  \bigg] ,
\end{aligned}}
\end{equation}
for all test functions $\mathcal{H}$. Compared with \cite{jia2021policypg}, here we have added the test functions $\mathcal{H}$ (by virtue of \eqref{eq:pe moment test function}) to make the approximation of policy gradient more flexible while the default value is $\mathcal H(t) = 0$.\footnote{See, e.g., the discussion of the notion of ``eligibility traces" in \citet[Chapter 12]{sutton2018reinforcement}.}  The right-hand side terms inside the expectation in \eqref{eq:pg representation with test function} can be computed using observable state samples under the policy $\bm\pi^{\bm\phi}$ and the known parametric form $\bm\pi^{\bm\phi}$, together with an estimated value function from the policy evaluation step discussed in Section \ref{sec:pe}, without knowing the market coefficients.

\subsection{Actor--critic learning by solving moment conditions}
\label{sec:general algorithm}
Alternating policy evaluation and policy gradient iteratively leads to what is called an actor--critic type of learning in RL. More precisely, there are three equations that need to be satisfied by the optimal value function, the optimal policy, and the Lagrange multiplier:
\begin{equation}
\label{eq:joint moment}
\left\{
\fitdisplay{\begin{aligned}
	&  \E\left[ \int_0^T \mathcal{I}(t) \left\{ \dd J(t,x^{u^{\bm{\pi}^{\bm\phi}}}(t),F(t);w;\bm\theta)+ \gamma\log\bm\pi(u^{\bm\pi^{\bm\phi}}(t)|t,x^{u^{\bm{\pi}^{\bm\phi}}}(t),F(t);w;\bm\phi)\dd t \right\} \right] = 0,  \\
	& \E\bigg[\int_0^T  \left[ \frac{\partial }{\partial \bm\phi} \log \bm\pi(u^{\bm\pi^{\bm\phi}}(t)|t,x^{u^{\bm\pi^{\bm\phi}}}(t),F(t);w;\bm\phi ) + \mathcal{H}(t) \right]  \Big[ \dd J(t,x^{u^{\bm\pi^{\bm\phi}}}(t),F(t);w;\bm\theta) \\
	& \quad \quad \quad \quad + \gamma \log\bm\pi(u^{\bm\pi^{\bm\phi}}(t)|t,x^{u^{\bm{\pi}^{\bm\phi}}}(t),F(t);w;\bm\phi) \dd t  \bigg] = 0, \\
	& \E\left[ x^{u^{\bm\pi^{\bm\phi}}}(T) - z \right] = 0.
\end{aligned}} \right.
\end{equation}

The first equation in \eqref{eq:joint moment} follows from the martingale condition \eqref{eq:pe moment test function} by substituting the policy by its approximation $\bm\pi^{\phi}$, with test function $\mathcal{I}$. The second equation follows from \eqref{eq:pg representation with test function}, implying that the gradient of the optimal value function with respect to the parameters $\bm\phi$ (with test function $\mathcal{H}$) is zero, which is the usual first-order condition for optimality. In applying  \eqref{eq:pg representation with test function} we replace  the true value function under $\bm\pi^{\bm\phi}$ with its approximator, i.e. (with a slight abuse of notation),
\[ J(t,x,F;w;\bm\theta) \approx  J(t,x,F;\bm\pi^{\bm\phi};w) .\]
The last equation in \eqref{eq:joint moment} arises from the expected return constraint in the original MV problem \eqref{classical}, which requires additional treatment beyond the general RL methods considered in \cite{jia2021policy,jia2021policypg}.

\section{A Provably Efficient Algorithm for the Black--Scholes Market}
\label{sec:proveable}

Establishing a model-free theoretical guarantee of the efficiency of an RL algorithm is generally extremely hard, due to complicated function approximations (e.g., with neural networks) and possible non-stationarity of state processes involved. In this section, we present an RL algorithm for a frictionless, multi-stock Black--Scholes market without any factor $F$, i.e., the stock prices follow multi-dimensional geometric Brownian motions.
We prove that a stochastic approximation type algorithm  with  specific actor and critic function approximations converges to a Sharpe ratio maximizing policy, and derive a sublinear regret bound in terms of the Sharpe ratio. Our algorithm is model-free in the sense that it is based on the model-free characterization of the optimal policy \eqref{eq:joint moment}; yet the proof depends on the specific Black--Scholes market structure. We leave the question of empirical performance to Section \ref{sec:empirical}, where the distributions of stock returns are unknown and unverifiable.

\subsection{The algorithm}
\label{subsec:baseline_algo}
To recap what was introduced in Section \ref{sec:general algorithm}, an RL algorithm consists of approximating the value function (critic) and policy (actor), sampling/generating data, and updating/improving the approximation. Approximation or parametrization can be, in general, constructed through neural networks or by exploiting the specific problem structure, such as with the present case. A theoretical analysis of the exploratory MV problem with the Black--Scholes environment is presented in Appendix \ref{appendix:solutions}. The {\it theoretical} optimal value function and optimal policy given by \eqref{eq:optimal_value_function} and \eqref{eq:optimal_policy} involve the unknown model parameters so they cannot be used as the final solutions. However, the specific {\it forms} of these functions suggest that we can
apply the following approximations for the value functions and stochastic policies:
\begin{equation}
\label{eq:value function parameterize}
J(t, x ; w;\bm\theta)=(x-w)^{2} e^{-\theta_{3}(T-t)}+\theta_{2}\left(t^{2}-T^{2}\right)+\theta_{1}(t-T)-(w-z)^{2},
\end{equation}
\begin{equation}
\bm\pi(\cdot \mid t, x ; w;\bm\phi)=\mathcal{N}\left(\cdot \mid-\phi_{1}(x-w), \phi_{2} e^{\phi_{3}(T-t)}\right),
\label{eq:policy parameterize}
\end{equation}
where $(\theta_1,\theta_2,\theta_3)^\top \in \mathbb{R}^3$ and $ (\phi_1,\phi_2,\phi_3)\in \mathbb{R}^d \times \mathbb{S}^d_{++} \times \mathbb R$ are two sets of parameters,  $w\in \mathbb R$ is the Lagrange multiplier, and $\mathcal{N}(\cdot \mid\mu,\Sigma)$ is the multivariate normal distribution with mean vector $\mu$ and covariance matrix $\Sigma$.
We reiterate  that while our learning procedure is model-free relying solely on observed wealth trajectories and policy gradients rather than estimating model primitives, this form of policy is inspired by  the theoretical Black--Scholes solutions (presented in Appendix \ref{appendix:solutions}). In particular, the mean part of the policy (dollar values invested in the stocks) is affine in wealth $x$, which is consistent with the model-based MV solutions \citep{zhou2000continuous} and  {\it different} from the constant risky proportion policy under CRRA utility maximization.  This particular structure, arising from the Black--Scholes environment,  is a necessary simplification that grants us the mathematical tractability required to conduct a rigorous convergence and regret analysis.

If we are to completely reconcile the approximated solutions \eqref{eq:value function parameterize} and \eqref{eq:policy parameterize} with the theoretical solutions \eqref{eq:optimal_value_function} and \eqref{eq:optimal_policy}, then
these parameters should not be entirely mutually independent. However, we do not enforce their relations based on the theoretical solutions and treat them largely independently  in our learning procedure for generality and flexibility. One exception is that we
let $\phi_3=\theta_3$, inspired by \eqref{eq:optimal_value_function} and \eqref{eq:optimal_policy}.
Moreover, in our algorithm, we set $\phi_3=\theta_3$ to be a sufficiently large constant (a hyperparameter) without  updating it, since this parameter plays no role in the convergence analysis (see Theorems \ref{thm:convergence_both}--\ref{thm:regret0}). Thus, we denote $\bm\theta=(\theta_1,\theta_2)^\top \in \mathbb{R}^2$ and $ \bm\phi= (\phi_1,\phi_2)^\top\in \mathbb{R}^d \times \mathbb{S}^d_{++}$ which, together with $w$, are to be updated and learned.

The algorithm we devise relies on the whole trajectory, meaning that in each iteration, parameters $(\bm\theta,\bm\phi,w)$ are updated after the data is generated during the entire episode $[0,T]$. It is a stochastic approximation algorithm based principally on the moment conditions \eqref{eq:joint moment}. Instead of directly applying the policy gradient methods used in \cite{jia2021policypg}, we take some modification for tractability of the subsequent   theoretical analysis. Specifically, in applying \eqref{eq:joint moment} we reparameterize $\bm\phi = (\phi_1,\phi_2)$ to  $\tilde{\bm\phi} = (\phi_1, \phi_2^{-1})$ and turn the second equation in \eqref{eq:joint moment} in terms of the gradient in $\phi_2$ to
\begin{equation}
\label{eq:gradient wrt inverse matrix}
\fitdisplay{\begin{aligned}
	0 =&	\E\bigg[\int_0^T   \left[ \frac{\partial }{\partial \phi_2^{-1}} \log \bm\pi(u^{\bm\pi^{\bm\phi}}(t)|t,x^{u^{\bm\pi^{\bm\phi}}}(t);w;\bm\phi ) + \mathcal{H}(t) \right]  \Big[ \dd J(t,x^{u^{\bm\pi^{\bm\phi}}}(t),w;\bm\theta) \\
	& \quad \quad  + \gamma \log\bm\pi(u^{\bm\pi^{\bm\phi}}(t)|t,x^{u^{\bm\pi^{\bm\phi}}}(t);w;\bm\phi )  \dd t  \Big] \bigg] .
\end{aligned}}
\end{equation}
The above follows from the chain rule and the fact that the extra term $\frac{\partial \phi_2^{-1}}{\partial \phi_2}$, resulting from the chain rule, is a deterministic, time-invariant constant, and hence can be removed. Thus, our stochastic approximation algorithm for the component $\phi_2$ will be based on \eqref{eq:gradient wrt inverse matrix}, the gradient in $\phi_2^{-1}$, instead of $\phi_2$ as in the original conditions \eqref{eq:joint moment}. This trick of using  the inverse covariance matrix will prove instrumental in the proof of our convergence  results.

We use subscript $n$ to represent the $n$-th iteration. For example, $\phi_{1,n}$ is the value of the parameter $\phi_1$ in its $n$-th iteration. At the first iteration $n=1$, we initialize  $\bm\theta_1 = (\theta_{1,1},\theta_{2,1})^\top$, $\bm\phi_1 = (\phi_{1,1},\phi_{2,1})^\top$ and $w_1$ to be some constants. At the $(n+1)$-th iteration, with the current parameters $(\bm\theta_n,\bm\phi_n,w_n)$,  we use the policy $\bm\pi(\cdot \mid t, x;w_n;\bm\phi_n)$ determined by \eqref{eq:policy parameterize} to generate the portfolio--wealth process $\{ (u_n(t) , x_n(t)): 0\leq t \leq T \}$, where $x_n$ satisfies \eqref{classical_dynamics} under $u=u_n$ with $u_n(t)
\sim \bm\pi(\cdot \mid t,x_n(t);w_n;\bm\phi_n)$.

By choosing two specific test functions $\mathcal{I}(t) = \frac{\partial }{\partial \bm\theta} J(t,x(t);w;\bm\theta)$ and $\mathcal H(t) = 0$, the learnable parameters are then updated by the following rules:
\begin{equation}
\label{eq:theta_update1}
\bm \theta_{n+1} \leftarrow \Pi_{K_{\theta, n}} \biggl(\bm \theta_n+a_{n} \int_{0}^{T} \frac{\partial J}{\partial \theta}\left(t, x_n(t) ; w_n ; \bm \theta_n\right)\left[\mathrm{d} J\left(t, x_n(t) ; w_n; \bm \theta_n\right)+\gamma \log\bm\pi(u_n(t)|t,x_n(t);w_n;\bm\phi_n ) \mathrm{d} t\right]\biggl),
\end{equation}
\begin{equation}
\label{eq:phi1_update1}
\phi_{1, n+1} \leftarrow \Pi_{K_{1,n}} \biggl( \phi_{1,n}-a_{n} Z_{1,n}(T) \biggl),
\end{equation}
\begin{equation}
\label{eq:phi2_update1}
\phi_{2,n+1} \leftarrow \Pi_{K_{2,n}} \biggl( \phi_{2,n}+a_{n} Z_{2,n}(T) \biggl),
\end{equation}
\begin{equation}
\label{eq:w_update1}
w_{n+1} \leftarrow \Pi_{K_{w,n}} \biggl( w_n - a_{w,n} (x_{n}(T)-z) \biggl),
\end{equation}
where
\begin{equation}
\label{eq:z1_def1}\fitdisplay{
Z_{1,n}(t) = \int_{0}^{t}\biggl\{\frac{\partial }{\partial \phi_1}\log \bm\pi\left(u_n(s) \mid s, x_n(s) ; w_n; \bm\phi_n\right)
\left[\mathrm{d} J\left(s, x_n(s) ; w_n ; \bm\theta_n \right)+\gamma \log\bm\pi(u_n(s)|s,x_n(s);w_n;\bm\phi_n ) \mathrm{d} s\right] \biggl\},
}\end{equation}

\begin{equation}
\label{eq:z2_def1}\fitdisplay{
Z_{2,n}(t) = \int_{0}^{t}\biggl\{\frac{\partial }{\partial \phi_2^{-1}}\log \bm\pi\left(u_n(s) \mid s, x_n(s) ; w_n, \bm \phi_n\right)
\left[\mathrm{d} J\left(s, x_n(s) ; w_n; \bm \theta_n\right)+\gamma \log\bm\pi(u_n(s)|s,x_n(s);w_n;\bm\phi_n ) \mathrm{d} s\right] \biggl\},
}\end{equation}
and $\Pi_K(z): = \arg\min_{y\in K}|y - z|^2$ is the projection of a point $z$ onto the subset $K$. The subsets involved in the above are:
\[\fitdisplay{\begin{aligned}
& K_{\theta,n}=\left\{ (\theta_{1}, \theta_{2})\in \mathbb{R}^2 \Big| |\theta_{1}| \leq c_{\theta_1}, |\theta_{2}| \leq c_{\theta_2} \right\}, \  K_{1, n} = \left\{ \phi_{1}\in \mathbb{R}^d \Big| |\phi_{1}| \leq c_{1, n} \right\}, \\
& K_{2, n} = \left\{ \phi_{2}\in \mathbb{S}^d_{++} \Big| |\phi_{2}| \leq c_{2, n}, \phi_{2} - \frac{1}{b_n}I \in \mathbb{S}^d_{+}  \right\}, \ K_{w, n} = \left\{ w\in \mathbb{R} \Big| |w| \leq c_{w, n}  \right\}.
\end{aligned}} \]
In this procedure, the constants $a_n$, $a_{w,n}$, $c_{\theta_1}$, $c_{\theta_2}$,  $c_{1,n}$, $c_{2,n}$, $c_{w,n}$ and $b_n$ are hyperparameters that can be set according to Theorem \ref{thm:convergence_both} below. Note that the second equation in \eqref{eq:joint moment} represents the gradient with respect to $\bm\phi$ to {\it minimize} the variance, and hence each iteration should move in the opposite direction of the gradient. This is why there is a negative sign in \eqref{eq:phi1_update1} that updates $\phi_1$. However, in \eqref{eq:phi2_update1}, the increment  $Z_{2,n}(T)$ is with respect to the gradient in $\phi_2^{-1}$, which is {\it decreasing} in $\phi_2$; so the sign  in \eqref{eq:phi2_update1}
is changed back from negative to positive.

We now present Algorithm~\ref{alg:offline}, with the derivation and theoretical analysis deferred to Appendix~\ref{appendix:baseline algo}. Although the algorithm is developed based on continuous-time analysis, reflecting the continuous-time nature of financial markets, it must ultimately be implemented in discrete time. The time discretization in Algorithm~\ref{alg:offline} serves a dual purpose: the discrete timestamps define the rebalancing schedule and portfolio updates, while also enabling numerical approximation of the integrals involved in the solution. In our analysis, the impact of discretization is ignored. For the general analysis of such discretization error, see \citet{jia2025accuracy}.

\begin{algorithm}[h]
\caption{CTRL Algorithm}\label{alg:offline}
\begin{algorithmic}
	
	
	\State Initialize $\bm\theta, \bm\phi$ and $w$.
	
	\For{\texttt{iter = 1 to N}}
	\State Initialize $k=0$, time $t=t_k = 0$, wealth $x(t_k) = x_0$.
	\While {$t < T$}
	\State Generate action $u(t_k) \sim \bm\pi\left(\cdot \mid t_{k}, x(t_{k}); w;\bm\phi\right)$ in \eqref{eq:policy parameterize}.
	\State Apply action $u(t_k)$ and get new wealth $x(t_{k+1})$ by dynamics \eqref{classical_dynamics}.
	\State Update time $t_{k+1} \leftarrow t_k + \Delta t$ and $t \leftarrow t_{k+1}$.
	\EndWhile
	\State Collect the whole trajectory $\{ (t_k, x(t_k), u(t_k)) \}_{k\geq 0}$.
	\State Update $\bm\theta$ using 
	\eqref{eq:theta_update_discrete}.
	\State Update $\bm\phi$ using 
	\eqref{eq:phi1_update_discrete} and \eqref{eq:phi2_update_discrete}.
	\State Update $w$ by \eqref{eq:w_update1}.
	
	\EndFor
	
\end{algorithmic}
\end{algorithm}

The remainder of this section is devoted to the statements of the main theoretical results of the paper: the convergence of Algorithm~\ref{alg:offline} and the resulting regret bound in terms of Sharpe ratio, whose proofs can be found in Appendix \ref{appendix:proof}.
Note that while Algorithm~\ref{alg:offline} is essentially a (nonlinear) stochastic approximation algorithm (\citealt{chau2014overview}), there are substantial difficulties in adapting the general stochastic approximation theory, which requires strong assumptions, to our specific setting.
First, recall \(Z_{1,n}(T)\) defined in \eqref{eq:z1_def1}, which represents the ``direction" of updating the learnable parameter \(\phi_{1,n}\) or the policy improvement signal.
In standard stochastic approximation theory this term is usually  assumed to be bounded or square-integrable, but in our problem \(Z_{1,n}(T)\) is inherently state-dependent, unbounded and increases fast in \(\phi_{1,n}\). To analyze the sampling error, we need to engage  very delicate stochastic analysis to estimate how fast this term grows (Theorem \ref{thm:tradeoff}).
Second, unlike in the general stochastic approximation theory,  our problem involves unbounded and continuous parameter spaces. To address this, we introduce certain iterative projections in the learning process, a technique pioneered by \citet{andradottir1995stochastic}. More importantly, to find the best possible regret rate, we need to carefully choose how fast the bounds of these projection sets grow {\it without} relying on any prior knowledge about the market environment (Theorem \ref{thm:convergence_both}). Finally, the Sharpe ratio as a function of the policy parameters is not globally concave. We need to carefully exploit some local concavity and analyze the high-probability concentration of the
policy around the optimal one to obtain a regret bound (Theorem \ref{thm:regret0}), which requires a very delicate and technical analysis.

\subsection{Convergence and convergence rate of the algorithm}
\label{sec:analysis of algorithm}
Theorem \ref{thm:tradeoff} below focuses on the mean and variance of \(Z_{1,n}(T)\)  and reveals the tradeoff between exploration and exploitation in terms of  \(\phi_{2,n}\) that controls the level of exploration through the variance of the stochastic policy.

\begin{theorem}
\label{thm:tradeoff}
The (conditional) mean of \(Z_{1,n}(T)\) is given by
\begin{equation}
	\label{eq:mean for update phi 1}
\mathbb{E}\left[Z_{1, n}(T) \mid \boldsymbol{\theta}_n, \boldsymbol{\phi}_n, w_n\right] = -R(\phi_{1,n}, \phi_{2,n}, w_n) (\mu-r-\Sigma \phi_{1,n}),
\end{equation}
where the expression of $R(\phi_1,\phi_2,w)$ is presented in Appendix \ref{appendix:tradeoff_thm}.

Moreover, the (conditional) variance of \(Z_{1,n}(T)\) is bounded by
\begin{equation}
\Big|\operatorname{Var}\left( Z_{1,n}(T) \Big| \bm\theta_n, \bm\phi_n, w_n \right)\Big| \leq C \bigg( 1 + |w_n|^{16} + |\phi_{1,n}|^8 + |\phi_{2,n}|^8 + |\phi_{2,n}^{-1}|^8 \bigg) e^{C|\phi_{1,n}|^8}
\end{equation}
where \(C\) is a constant independent of $n$.
\end{theorem}

A proof is given in Appendix~\ref{appendix:tradeoff_thm}. The key observation is that the upper bound of the variance of \(Z_{1,n}(T)\) exhibits a U-shaped dependence on the exploration level \(\phi_{2,n}\). Indeed, this U-shape is not only in the upper bound but also in the variance itself, as demonstrated numerically in Figure~\ref{fig:tradeoff} of  Appendix~\ref{appendix:tradeoff}. As a result, both very small and very large values of \(\phi_{2,n}\) lead to high variances of the iterates \(\phi_{1,n}\). When \(\phi_{2,n}\) is too small, the policy is closer to being deterministic rendering insufficient policy improvement. Conversely, excessive exploration with a large \(\phi_{2,n}\) ``injects" too much noise hindering learning efficiency. This underscores the exploration--exploitation tradeoff: optimal learning requires balancing the two to avoid both underfitting and overfitting.

The following theorem, whose proof (for a more general version of the theorem) is relegated to Appendix~\ref{appendix:proof covergence},  presents the convergence and convergence rates of the parameters updated according to Algorithm \ref{alg:offline}.
\begin{theorem}
\label{thm:convergence_both}
Assume that the stock prices follow a multi-dimensional geometric Brownian motion with constant return and volatility rates, and the risk-free rate is a constant. In Algorithm \ref{alg:offline}, let the parameters $c_{\theta_1}$, $c_{\theta_2}$ be some positive constants, and $a_n$, $a_{w,n}$,  $c_{1,n}$, $c_{2,n}$, $c_{w,n}$ and $b_n$ be  set as follows:
\[
\fitdisplay{\begin{aligned}
	(i)& \quad  a_n = a_{w, n}=\frac{\alpha}{n+\beta}, \text{ for some sufficiently large } \alpha>0 \text{ and } \beta>0;\\
	(ii) & \quad b_n = L_n, c_{1, n} =  c_{2, n} = L_n, c_{w, n} = \sqrt{L_n}, \text{ where for sufficiently large }n, L_n = 1\vee (\log\log n)^{1/8}. \\
\end{aligned}}
\]
Then

\begin{enumerate}[(a)]
	\item As $n\to \infty$, $\phi_{1,n}$ and $\phi_{2,n}$ almost surely converge to the true values $\phi_1^{*} = ({\sigma} {\sigma}^\top)^{-1} (\mu - r)$ and $\phi_2^{*} = \frac{\gamma}{2} ({\sigma} {\sigma}^\top)^{-1}$ respectively, and $w_n$  almost surely converges to the true value $w^* = \frac{z \exp\{(\mu-r)^\top (\sigma \sigma^\top)^{-1} (\mu-r) T\} - x_0}{\exp\{(\mu-r)^\top (\sigma \sigma^\top)^{-1} (\mu-r) T\} - 1}$.
	\item For sufficiently large $n$, $\E [|\phi_{1, n+1} - \phi_1^*|^2] \leq C \frac{(\log n)^p(\log \log n)}{n}$, where $C$ and $p$ are positive constants independent of $n$.
\end{enumerate}
\end{theorem}

Note  the convergence rate of $\phi_{1,n}$ is of the order $\frac{(\log n)^p(\log \log n)}{n}$, which nearly matches the typical optimal convergence rate of stochastic approximation algorithms (e.g. \citealt{broadie2011general}) and differs only by a factor $(\log n )^p(\log\log n)$ which is very small relative to $n$. 
Theorem \ref{thm:convergence_both} gives guidance about properly choosing the sequences
$\{a_n, a_{w,n},  c_{1,n}, c_{2,n}, c_{w,n}\}$ in order to achieve this convergence rate.
Moreover, Assumptions (i)-(ii) in Theorem \ref{thm:convergence_both} can be relaxed to prove the first statement about the almost sure convergence of $\phi_{1,n}$, $\phi_{2,n}$ and $w_n$; see Appendix \ref{appendix:proof covergence} for weaker conditions \eqref{eq:thm1_assumption} and \eqref{eq:thm3_assumption}.

\subsection{Regret bound in Sharpe ratio}

We first present the following result.
\begin{theorem}
\label{thm: more random higher variance}
Consider two policies $\bm\pi$ and $\hat{\bm\pi}$ in the same form as \eqref{eq:policy parameterize}, given by
$\bm\pi = \mathcal{N}\biggl(- \phi_1(x-w),  C(t)\biggl)$, $\hat{\bm\pi} = \mathcal{N}\biggl(-\phi_1(x-w), \hat C(t)\biggl)$,
where $C(\cdot), \hat C(\cdot)\in \mathbb{S}^d_{++}$ are two deterministic functions satisfying $C(t) - \hat C(t) \in \mathbb{S}^d_+$ for all $t\in [0,T]$, along with their respective wealth trajectories $\{x^{u^{\bm\pi}}(t):0\leq t\leq T\}$ and $\{x^{u^{\hat{\bm\pi}}}(t):0\leq t\leq T\}$. Then $\hat{\bm\pi}$ mean--variance dominates $\bm\pi$; i.e.
$\E[x^{u^{\bm\pi}}(T)] = \E[x^{u^{\hat{\bm\pi}}} (T)]$ and $\operatorname{Var}\left( x^{u^{\bm\pi}}(T) \right) \geq \operatorname{Var}\left( x^{u^{\hat{\bm\pi}}} (T) \right)$.
\end{theorem}

A proof of this theorem is delayed to Appendix \ref{appendix:proof variance}. The theorem indicates that, for the same entropy-regularized MV problem (with the same temperature parameter), even though the two policies with the same mean generate the same expected terminal wealth, the one with a lower level of exploration has a more stable result in terms of the variance of the terminal wealth. So more exploration is worse off from the MV perspective. In particular, we only need to use {\it deterministic} policies for {\it actual} execution of a portfolio (instead of using a random sampler from the learned optimal stochastic policy).
Note that this feature is specific to the current setting of the problem (i.e. a frictionless market with a small investor), and is not necessarily true in general where actual actions also need to be sampled from stochastic policies in order to broaden search space and observe the responses to actions from the environment. As discussed at the end of  Section \ref{sec:stochastic policy}, the dynamics \eqref{classical_dynamics} with a small investor dictate that the investor would know the consequence of executing any portfolio even if she does not actually execute it. This is in sharp contrast to, say,  a bandit problem, in which an agent has no knowledge about counterfactuals. Thus, intuitively, in the MV problem one does not need to do exploration {\it per se} for the purpose of trial and error; she could do it {\it on paper}. However, as explained earlier, stochastic policies are necessary for computing the policy gradient due to technical reasons; so we will use them for training (i.e. for updating the parameters). We do this by randomly generating portfolio processes from the current stochastic policy and simulating the corresponding (counterfactual) wealth trajectories based on \eqref{classical_dynamics}.

Denote a deterministic policy for the original (non-exploratory) wealth equation \eqref{classical_dynamics} by
\begin{equation}
\label{eq:deterministic_policy}
\bm u(t, x ; w;\bm\phi)=-\phi_{1}(x-w),
\end{equation}
which is a degenerate stochastic policy with a Dirac distribution and coincides with the mean of the policy $\bm\pi^{\bm\phi}$ defined in \eqref{eq:policy parameterize}. The Sharpe ratio of the terminal wealth of \eqref{classical_dynamics} under this policy is defined as
\begin{equation}
\label{eq:def for sharpe ratio}
\frac{\E\left[ x^{\bm u}(T)/x^{\bm u}(0)\right] - 1 }{\sqrt{\operatorname{Var}(x^{\bm u}(T)/x^{\bm u}(0)) }}	,
\end{equation}
which depends only on  $\phi_1$, and is denoted by  $\operatorname{\operatorname{SR}}(\phi_1)$.

\begin{theorem}
\label{thm:regret0}
Under the same setting of Theorem \ref{thm:convergence_both},  we have
$$\E \left[\sum_{n=1}^{N} (\operatorname{SR}(\phi_1^*) - \operatorname{SR}(\phi_{1,n}))\right] \leq C + C \sqrt{N (\log N)^p \log\log N},\;\;\forall N,$$
where $C>0$ is a constant independent of $N$, and $p$ is the same constants appearing in Theorem \ref{thm:convergence_both}.
\end{theorem}

A proof of Theorem \ref{thm:regret0} is given in Appendix \ref{appendix:proof regret}.
The result stipulates that, in terms of the Sharpe ratio, the cumulative gap between the iterates of our algorithm and the ``oracle" (i.e. the theoretically optimal portfolio should all the market parameters be known) up to the $N$th iteration is of the order of $\sqrt{N (\log N)^p \log\log N}$. The sublinearity of this gap implies   that in the long run, the algorithm performs almost optimally. This bound matches the typical regret results for {\it discrete-time} RL algorithms of \(O(\sqrt{N})\) (up to logarithm factors; see the related literature review in Section \ref{sec:intro}), and is the first {\it model-free} regret result in {\it continuous-time} MV portfolio choice to our best knowledge.
Theorem \ref{thm:regret0} also reveals the importance of the parameter $\phi_1$. Indeed, the theoretical value of the vector $\phi_1^*=(\sigma \sigma^\top)^{-1}(\mu-r)$ (see \eqref{eq:optimal_policy}) constitutes the amount allocated to the risky assets and hence the composition of an MV efficient mutual fund. This composition in turn determines the Sharpe ratio of the resulting portfolio, noting that any MV efficient portfolio has the same Sharpe ratio.

\section{Empirical Performance and Comparisons}\label{sec:empirical}

To assess the efficacy of our CTRL algorithm, we carry out an empirical study to compare its performance with other well-established asset allocation strategies.

\subsection{Experimental setup}\label{subsec:setup}
\paragraph{\textbf{Data and asset universe}}
The dataset used in this study is obtained from the Wharton Research Data Services and Yahoo Finance. Our asset universe consists of stocks that were constituents of the S\&P 500 index and remained continuously listed with available daily trading data from January 1, 1990 to January 1, 2020. From this universe, we construct a pool of the first 300 stocks sorted alphabetically by their tickers. Daily returns are computed based on dividend-adjusted closing prices.

\paragraph{\textbf{Alternative allocation strategies}}
We compare our CTRL strategy with 13 existing alternative portfolio allocation strategies or benchmarks that are widely studied and employed in both theory and practice, including buy-and-hold market index (S\&P500), equal weight (ew), sample-based single-period mean-variance (mv), sample-based minimum variance (min\_v), James--Stein shrinkage estimator (js), Ledoit--Wolf shrinkage estimator (lw), Black--Litterman model (bl), Fama--French three-factor model (ff), risk parity (rp), sample-based continuous-time mean--variance (ctmv), predictive mean-variance (pmv), deep deterministic policy gradient (ddpg), and proximal policy optimization (ppo). The details of these alternative strategies are described in Appendix~\ref{alternative}.

\paragraph{\textbf{Sampling and evaluation protocol}}
To avoid selection bias in the construction of our portfolios, for each experiment, we randomly sample 10 stocks from this pool and apply the strategies under comparison over the test period from 2000 to 2020. All the methods use 1990--2000 for pre-training or estimation as the burn-in period. Over the testing period, all the methods update their parameters or estimates using data available up to the current date, while the corresponding portfolios are rebalanced monthly. If a method is based on a static optimization problem, then it always uses the most recent 10-year data when estimating the relevant model parameters. For our RL strategy, we follow exactly Algorithm~\ref{alg:offline} whose performance guarantee is established in our theory. This procedure is repeated 100 times to obtain statistical summaries. For all the strategies, whenever a wealth trajectory hits zero, the remainder of the trajectory is set to zero, which is treated as a bankruptcy event.

\paragraph{\textbf{Target return and initialization}}
For algorithms that require a specified mean target return, including CTRL  and other MV-based strategies, we set the target annualized return to be $15\%$. This corresponds to a monthly target return $\mu^* = (1.15)^{1/12}-1 \approx 1.17\%$, and thus $z = 1.0117$ in our model. This choice is consistent with the approximate annualized return of the S\&P~500 over the pre-training period 1990--2000. For simplicity, we set the risk-free interest rate $r$ to be zero. The initial wealth is normalized to \$1 in all experiments.

\paragraph{\textbf{Portfolio specification and decision horizon}}

Except for four strategies that inherently involve no risk-free allocation (namely buy-and-hold S\&P 500 index, equally weighted portfolio, risk parity, and sample-based minimum variance), all the other methods, including CTRL, are implemented without any modifications (in particular, holding and borrowing a risk-free asset are permitted). We set the planning horizon to be one month (i.e., $T=1/12$ year), and implement all the methods on a rolling one-month horizon basis.\footnote{We choose one-month horizon so that the model-based mean--variance variants have reasonably sufficient data to estimate the model parameters, versus for instance the one-year horizon. We thus take the same one-month rolling horizon for all the methods to ensure a fair comparison.}

\paragraph{\textbf{Evaluation criteria}}
We use a comprehensive set of performance metrics including annualized return, volatility, Sharpe ratio, Sortino ratio, Calmar ratio, maximum drawdown (MDD), and recovery time (RT). In addition, we report trading and exposure metrics including bankruptcy rate (BKR), turnover (TO), maximum single-asset long and short positions (MaxL and MaxS), and gross exposure (GE), together with its maximum (GEMax). The exact definitions of these metrics are given in Appendix~\ref{appendix:performance metric} and Appendix~\ref{appendix:trading metric}.

We now present a detailed analysis of the empirical study.

\subsection{Overall performance}\label{subsec:full}

\begin{figure}[h!]
\centering
\includegraphics[width=1\textwidth]{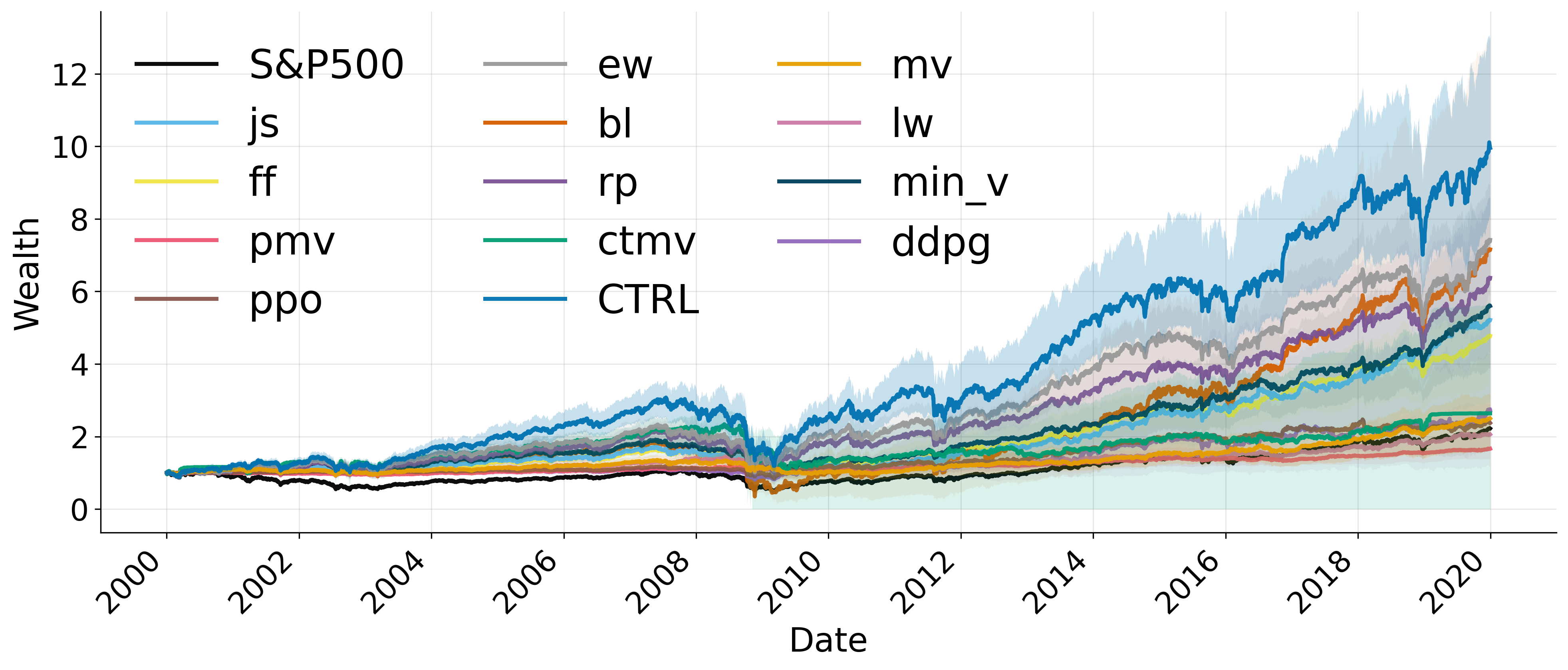}
\caption{Wealth trajectories under the proposed CTRL algorithm and 13 alternative methods over 100 independent experiments, each with 10 randomly selected stocks (except the S\&P 500 index), from 2000 to 2020. The solid line shows the median across experiments at each time, and the shaded region is the interquartile range.}
\label{fig:wealth}
\end{figure}

\paragraph{\textbf{Wealth trajectories}}
First, we compare the wealth trajectories under different strategies over 100 independent experiments, each (except the S\&P 500 index) with 10 randomly selected S\&P 500 constituents. They are depicted in Figure \ref{fig:wealth}. These trajectories summarize cross-instance performance and provide first impressions of the respective strategies. (For further visual clarity, the Appendix \ref{subsec:focused-wealth} presents wealth-trajectory plots for a focused comparison among CTRL, ew, mv, and ctmv.) CTRL clearly outperforms all the other methods, achieving the highest portfolio terminal value by the end of the testing period. Moreover, CTRL ranks among the best-performing methods {\it throughout} the entire period. Beyond this aggregate view, when comparing the final value on an instance-by-instance basis, CTRL outperforms the second-best method, ew, in 70 out of 100 instances.\footnote{The outperformance of the na\"ive ew strategy is known in the literature (e.g. \citealp{demiguel2009optimal}).} From Figure \ref{fig:wealth}, we also observe that CTRL falls sharply during the 2008 financial crisis, but it is also the quickest to recover. This visual inspection will be consistent with the recovery-time metric reported below.

\paragraph{\textbf{Performance summary}}\label{par:perf}

\begin{table}[!htbp]
\centering
\begin{tabular}{lrrrrrrr}
\toprule
 & Return & Volatility & Sharpe & Sortino & Calmar & MDD & RT \\
\midrule
S\&P500 & \makecell[r]{4.077\% \\ (0.000\%)} & \makecell[r]{0.189 \\ (0.000)} & \makecell[r]{0.216 \\ (0.000)} & \makecell[r]{0.341 \\ (0.000)} & \makecell[r]{0.072 \\ (0.000)} & \makecell[r]{0.568 \\ (0.000)} & \makecell[r]{1021 \\ (0)} \\
ew & \makecell[r]{10.343\% \\ (0.185\%)} & \makecell[r]{0.218 \\ (0.002)} & \makecell[r]{0.483 \\ (0.011)} & \makecell[r]{0.789 \\ (0.018)} & \makecell[r]{0.182 \\ (0.005)} & \makecell[r]{0.585 \\ (0.008)} & \makecell[r]{544 \\ (31)} \\
mv & \makecell[r]{4.226\% \\ (0.274\%)} & \makecell[r]{0.148 \\ (0.002)} & \makecell[r]{0.304 \\ (0.020)} & \makecell[r]{0.491 \\ (0.032)} & \makecell[r]{0.113 \\ (0.008)} & \makecell[r]{0.443 \\ (0.013)} & \makecell[r]{1349 \\ (77)} \\
js & \makecell[r]{7.493\% \\ (0.537\%)} & \makecell[r]{0.240 \\ (0.019)} & \makecell[r]{0.397 \\ (0.023)} & \makecell[r]{0.644 \\ (0.038)} & \makecell[r]{0.156 \\ (0.011)} & \makecell[r]{0.577 \\ (0.016)} & \makecell[r]{1124 \\ (71)} \\
bl & \makecell[r]{5.180\% \\ (1.962\%)} & \makecell[r]{0.618 \\ (0.092)} & \makecell[r]{0.195 \\ (0.027)} & \makecell[r]{0.350 \\ (0.041)} & \makecell[r]{0.079 \\ (0.020)} & \makecell[r]{0.851 \\ (0.013)} & \makecell[r]{1180 \\ (70)} \\
lw & \makecell[r]{3.797\% \\ (0.226\%)} & \makecell[r]{0.154 \\ (0.002)} & \makecell[r]{0.261 \\ (0.016)} & \makecell[r]{0.421 \\ (0.026)} & \makecell[r]{0.097 \\ (0.007)} & \makecell[r]{0.456 \\ (0.013)} & \makecell[r]{1452 \\ (80)} \\
ff & \makecell[r]{7.805\% \\ (0.226\%)} & \makecell[r]{0.183 \\ (0.002)} & \makecell[r]{0.438 \\ (0.014)} & \makecell[r]{0.702 \\ (0.024)} & \makecell[r]{0.166 \\ (0.006)} & \makecell[r]{0.501 \\ (0.011)} & \makecell[r]{872 \\ (43)} \\
rp & \makecell[r]{9.462\% \\ (0.258\%)} & \makecell[r]{0.216 \\ (0.004)} & \makecell[r]{0.458 \\ (0.015)} & \makecell[r]{0.742 \\ (0.025)} & \makecell[r]{0.170 \\ (0.006)} & \makecell[r]{0.592 \\ (0.011)} & \makecell[r]{861 \\ (58)} \\
min\_v & \makecell[r]{8.935\% \\ (0.304\%)} & \makecell[r]{0.183 \\ (0.002)} & \makecell[r]{0.502 \\ (0.019)} & \makecell[r]{0.819 \\ (0.031)} & \makecell[r]{0.195 \\ (0.009)} & \makecell[r]{0.497 \\ (0.011)} & \makecell[r]{989 \\ (45)} \\
pmv & \makecell[r]{2.719\% \\ (0.161\%)} & \makecell[r]{\textbf{0.084} \\ (0.001)} & \makecell[r]{0.327 \\ (0.019)} & \makecell[r]{0.532 \\ (0.032)} & \makecell[r]{0.125 \\ (0.009)} & \makecell[r]{\textbf{0.251} \\ (0.007)} & \makecell[r]{1194 \\ (100)} \\
ctmv & \makecell[r]{-6.781\% \\ (13.863\%)} & \makecell[r]{1.223 \\ (0.173)} & \makecell[r]{0.077 \\ (0.050)} & \makecell[r]{0.306 \\ (0.090)} & \makecell[r]{-0.023 \\ (0.132)} & \makecell[r]{0.827 \\ (0.024)} & \makecell[r]{1811 \\ (112)} \\
ddpg & \makecell[r]{4.174\% \\ (0.778\%)} & \makecell[r]{0.218 \\ (0.012)} & \makecell[r]{0.217 \\ (0.031)} & \makecell[r]{0.355 \\ (0.049)} & \makecell[r]{0.101 \\ (0.012)} & \makecell[r]{0.548 \\ (0.018)} & \makecell[r]{1213 \\ (88)} \\
ppo & \makecell[r]{4.921\% \\ (0.451\%)} & \makecell[r]{0.156 \\ (0.006)} & \makecell[r]{0.295 \\ (0.027)} & \makecell[r]{0.482 \\ (0.043)} & \makecell[r]{0.122 \\ (0.010)} & \makecell[r]{0.446 \\ (0.016)} & \makecell[r]{922 \\ (82)} \\
CTRL & \makecell[r]{\textbf{12.211\%} \\ (0.219\%)} & \makecell[r]{0.232 \\ (0.003)} & \makecell[r]{\textbf{0.531} \\ (0.010)} & \makecell[r]{\textbf{0.865} \\ (0.017)} & \makecell[r]{\textbf{0.200} \\ (0.005)} & \makecell[r]{0.621 \\ (0.009)} & \makecell[r]{\textbf{521} \\ (30)} \\
\bottomrule
\end{tabular}
\caption{\textbf{Comparison of out-of-sample performance of different allocation methods from 2000 to 2020.} We report annualized return, volatility, Sharpe ratio, Sortino ratio, and Calmar ratio, together with maximum drawdown (MDD) and recovery time (RT) in days, over 100 independent experiments. Each experiment is based on 10 randomly selected stocks (except for S\&P 500 index). For each cell, the upper number is the average (over the 100 experiments) while the lower one with parentheses is the standard error. }
\label{tab:2000 to 2020}
\end{table}

While Figure \ref{fig:wealth} provides an aggregate view of the performance of different allocation methods, a more detailed evaluation using the criteria defined in Appendix~\ref{metric} is needed for a comprehensive comparison. Table~\ref{tab:2000 to 2020} reports the results for these metrics, averaged over 100 independent experiments, each with 10 randomly selected S\&P~500 constituents (except for the S\&P~500 index), during the period from 2000 to 2020. Each column corresponds to one performance metric, with the best-performing method highlighted in bold. Numbers in parentheses report the standard errors across the 100 experiments.

First, the CTRL strategy attains the highest average annualized return, achieving the closest return to the 15\% target. Moreover, in terms of risk-adjusted performance, CTRL delivers the highest Sharpe ratio among all strategies, and the outperformance  is statistically significant (Appendix~\ref{appendix:paired test}, Table~\ref{tab:wilcoxon_sharpe_pvals}). To complement the mean and standard-error summaries, Appendix~\ref{subsec:box-sharpe} further reports box plots of the Sharpe ratios across the 100 experiments, providing a clearer view of the cross-instance distribution. While the Sharpe ratio is directly aligned with the MV formulation, the Sortino and Calmar ratios are also widely used measures of risk-adjusted performance. CTRL achieves the highest values under these metrics as well. Overall, the results indicate that CTRL performs consistently strongly under different risk-adjusted criteria.

Second, in terms of MDD, which is not an explicit constraint on any methods, CTRL is somewhere in the middle, indicating its portfolios go down in a downturn market. However, consistent with the observation from Figure~\ref{fig:wealth}, CTRL has a decisively and significantly shorter recovery time (RT) than any other method, approximately half that of S\&P~500 (521 days versus 1021 days). Also,  the combination of a relatively moderate drawdown and a higher annualized return contributes to its strong Calmar ratio.

Last but probably most importantly, we observe the notable underperformance of mv and ctmv. These are derived by the classical {\it model-based}, plug-in approach, the first a (rolling horizon) static model and the other one dynamic MV model. They both need to estimate the model parameters first before optimizing. The inherent difficulty in estimating those parameters (especially the mean) and the high sensitivity of the optimal solutions with respect to the estimations have caused poor performances, as discussed earlier. In particular, the dynamic model, ctmv,  is worse than the static counterpart, mv, resulting in even bankruptcy in many instances, due to the {\it cumulative} estimation errors in a dynamic environment. By contrast, the CTRL strategy mitigates this problem by bypassing the model parameter estimation altogether, which is the fundamental reason for its outstanding performances.

\subsection{Turnover and exposure}\label{subsec:turnover}

Table~\ref{tab:trade} reports key trading and exposure statistics of all the methods (except S\&P 500) over the full period from 2000 to 2020, including bankruptcy rate (BKR), monthly turnover (TO), maximum single-asset long and short positions (MaxL and MaxS), gross (risky) exposure (GE), and the maximum gross exposure observed across all experiments (GEMax).

\begin{table}[!htbp]
\centering
\begin{tabular}{lrrrrrr}
\toprule
 & BKR & TO & MaxL & MaxS & GE & GEMax \\
\midrule
ew & 0\% & 5.798\% & 10.000\% & 0.000\% & 100.000\% & 100.000\% \\
mv & 0\% & 16.437\% & 113.971\% & 61.028\% & 113.558\% & 260.610\% \\
js & 1\% & 15.345\% & 171.726\% & 106.333\% & 140.192\% & 542.149\% \\
bl & 16\% & 15.900\% & 271.773\% & 150.030\% & 149.242\% & 1000.542\% \\
lw & 0\% & 18.745\% & 116.278\% & 82.500\% & 117.245\% & 389.493\% \\
ff & 0\% & 8.841\% & 116.479\% & 53.619\% & 96.796\% & 274.978\% \\
rp & 0\% & 11.197\% & 100.000\% & 0.000\% & 100.000\% & 100.000\% \\
min\_v & 0\% & 8.806\% & 104.477\% & 31.623\% & 126.789\% & 227.462\% \\
pmv & 0\% & 52.155\% & 75.980\% & 58.536\% & 75.156\% & 231.473\% \\
ctmv & 39\% & 39.963\% & 1532.915\% & 865.850\% & 142.307\% & 6332.816\% \\
ddpg & 1\% & 19.676\% & 77.880\% & 70.666\% & 156.801\% & 532.347\% \\
ppo & 0\% & 21.608\% & 69.267\% & 51.967\% & 95.963\% & 296.168\% \\
CTRL & 0\% & 5.679\% & 18.778\% & 1.903\% & 102.301\% & 124.034\% \\
\bottomrule
\end{tabular}
\caption{\textbf{Trading and exposure statistics over the period 2000--2020.}
BKR is the bankruptcy rate across experiments,  TO is the average monthly turnover,
MaxL and MaxS are the maximum single-asset long and short positions, respectively,
GE is the average gross exposure, and GEMax is the maximum gross exposure observed across all experiments.}
\label{tab:trade}
\end{table}

Overall, CTRL exhibits a highly conservative and stable trading profile. Its monthly turnover is the lowest across all the methods (5.679\%), substantially lower than most model-based and RL-based alternatives. At the same time, CTRL avoids concentrated  positions: the maximum long position on any single asset is below 20\%, and the maximum short position is below 2\%. By contrast, several competing methods rely on extremely  concentrated positions, with MaxL exceeding 100\% for most model-based strategies and reaching outrageous levels for ctmv and bl.

In terms of aggregate risky exposure, CTRL operates close to full investment, with an average gross exposure of approximately 102\% and a maximum exposure of 124\% across all experiments. This stands in sharp contrast to methods such as the mean--variance variants and other RL benchmarks, which frequently render substantially higher leverage and large variation in exposure across experiments. Notably, ctmv and bl suffer from extreme leverage, frequent bankruptcy instances, and explosive maximum exposure, implying they can be hardly used in practice.

As a final note, recall that we compare all the methods  {\it without} imposing additional portfolio constraints (such as no-shorting or a leverage constraint). With this freedom, CTRL still maintains more measured leverage and risky exposure, underscoring its inherent robustness.

\subsection{Transaction costs}\label{subsec:tc}
\paragraph{\textbf{Performance under transaction costs}}\label{par:tc-perf}
To compare all the strategies under transaction costs, we recompute all the performance metrics assuming a 50 bps proportional transaction cost as in \citet{demiguel2009optimal}. Table~\ref{tab:tc100bps} reports the resulting net-of-cost performance over the full sample period. Appendix \ref{subsec:comp-tc} further reports results under lower transaction costs to reflect a more realistic current market environment, and the qualitative conclusions remain unchanged.

Indeed, the lowest turnover of CTRL reported earlier already foreshadowed an even better relative performance, which is confirmed by Table~\ref{tab:tc100bps}. In particular, CTRL still achieves the highest average annualized return , Sharpe ratio, Calmar ratio and the shortest recovery time. Its performance gap  with the methods that rely on heavy trading or risky exposures, such as pmv, ctmv and the RL baselines ddpg and ppo, is further widened.

\begin{table}[!htbp]
\centering
\begin{tabular}{lrrrrrrr}
\toprule
 & Return & Volatility & Sharpe & Sortino & Calmar & MDD & RT \\
\midrule
S\&P500 & \makecell[r]{4.077\% \\ (0.000\%)} & \makecell[r]{0.189 \\ (0.000)} & \makecell[r]{0.216 \\ (0.000)} & \makecell[r]{0.341 \\ (0.000)} & \makecell[r]{0.072 \\ (0.000)} & \makecell[r]{0.568 \\ (0.000)} & \makecell[r]{1021 \\ (0)} \\
ew & \makecell[r]{9.959\% \\ (0.185\%)} & \makecell[r]{0.218 \\ (0.002)} & \makecell[r]{0.465 \\ (0.011)} & \makecell[r]{0.760 \\ (0.018)} & \makecell[r]{0.174 \\ (0.005)} & \makecell[r]{0.588 \\ (0.008)} & \makecell[r]{568 \\ (31)} \\
mv & \makecell[r]{3.203\% \\ (0.279\%)} & \makecell[r]{0.148 \\ (0.002)} & \makecell[r]{0.234 \\ (0.019)} & \makecell[r]{0.381 \\ (0.031)} & \makecell[r]{0.085 \\ (0.007)} & \makecell[r]{0.462 \\ (0.013)} & \makecell[r]{1490 \\ (72)} \\
js & \makecell[r]{6.507\% \\ (0.548\%)} & \makecell[r]{0.240 \\ (0.019)} & \makecell[r]{0.352 \\ (0.023)} & \makecell[r]{0.572 \\ (0.037)} & \makecell[r]{0.135 \\ (0.010)} & \makecell[r]{0.587 \\ (0.016)} & \makecell[r]{1199 \\ (64)} \\
bl & \makecell[r]{4.186\% \\ (1.958\%)} & \makecell[r]{0.618 \\ (0.092)} & \makecell[r]{0.172 \\ (0.027)} & \makecell[r]{0.312 \\ (0.041)} & \makecell[r]{0.066 \\ (0.020)} & \makecell[r]{0.854 \\ (0.013)} & \makecell[r]{1378 \\ (88)} \\
lw & \makecell[r]{2.635\% \\ (0.230\%)} & \makecell[r]{0.154 \\ (0.002)} & \makecell[r]{0.185 \\ (0.016)} & \makecell[r]{0.300 \\ (0.026)} & \makecell[r]{0.067 \\ (0.006)} & \makecell[r]{0.478 \\ (0.013)} & \makecell[r]{2027 \\ (120)} \\
ff & \makecell[r]{7.235\% \\ (0.230\%)} & \makecell[r]{0.183 \\ (0.002)} & \makecell[r]{0.406 \\ (0.014)} & \makecell[r]{0.653 \\ (0.024)} & \makecell[r]{0.152 \\ (0.006)} & \makecell[r]{0.510 \\ (0.011)} & \makecell[r]{929 \\ (45)} \\
rp & \makecell[r]{8.730\% \\ (0.266\%)} & \makecell[r]{0.216 \\ (0.004)} & \makecell[r]{0.426 \\ (0.016)} & \makecell[r]{0.690 \\ (0.026)} & \makecell[r]{0.155 \\ (0.006)} & \makecell[r]{0.601 \\ (0.011)} & \makecell[r]{955 \\ (69)} \\
min\_v & \makecell[r]{8.360\% \\ (0.304\%)} & \makecell[r]{0.183 \\ (0.002)} & \makecell[r]{0.471 \\ (0.018)} & \makecell[r]{0.769 \\ (0.031)} & \makecell[r]{0.181 \\ (0.009)} & \makecell[r]{0.502 \\ (0.011)} & \makecell[r]{1032 \\ (45)} \\
pmv & \makecell[r]{-0.451\% \\ (0.164\%)} & \makecell[r]{\textbf{0.085} \\ (0.001)} & \makecell[r]{-0.051 \\ (0.019)} & \makecell[r]{-0.078 \\ (0.031)} & \makecell[r]{-0.002 \\ (0.005)} & \makecell[r]{\textbf{0.370} \\ (0.011)} & \makecell[r]{2034 \\ (55)} \\
ctmv & \makecell[r]{5.844\% \\ (27.521\%)} & \makecell[r]{1.893 \\ (0.518)} & \makecell[r]{-0.021 \\ (0.053)} & \makecell[r]{0.328 \\ (0.220)} & \makecell[r]{0.071 \\ (0.246)} & \makecell[r]{0.880 \\ (0.021)} & \makecell[r]{2446 \\ (124)} \\
ddpg & \makecell[r]{2.946\% \\ (0.775\%)} & \makecell[r]{0.218 \\ (0.012)} & \makecell[r]{0.156 \\ (0.030)} & \makecell[r]{0.257 \\ (0.049)} & \makecell[r]{0.075 \\ (0.012)} & \makecell[r]{0.568 \\ (0.018)} & \makecell[r]{1318 \\ (89)} \\
ppo & \makecell[r]{3.269\% \\ (0.444\%)} & \makecell[r]{0.149 \\ (0.006)} & \makecell[r]{0.185 \\ (0.028)} & \makecell[r]{0.304 \\ (0.045)} & \makecell[r]{0.080 \\ (0.009)} & \makecell[r]{0.460 \\ (0.016)} & \makecell[r]{968 \\ (66)} \\
CTRL & \makecell[r]{\textbf{11.828\%} \\ (0.218\%)} & \makecell[r]{0.233 \\ (0.003)} & \makecell[r]{\textbf{0.514} \\ (0.010)} & \makecell[r]{\textbf{0.837} \\ (0.017)} & \makecell[r]{\textbf{0.193} \\ (0.005)} & \makecell[r]{0.623 \\ (0.009)} & \makecell[r]{\textbf{538} \\ (31)} \\
\bottomrule
\end{tabular}
\caption{\textbf{Net-of-cost performance over the full sample period (2000--2020) under 50 bps proportional transaction costs.} We report annualized return, volatility, Sharpe ratio, Sortino ratio, and Calmar ratio, together with maximum drawdown (MDD) and recovery time (RT) in days, over 100 independent experiments. Each experiment is based on 10 randomly selected stocks (except for S\&P 500 index). Numbers in parentheses are the standard errors across 100 experiments. }
\label{tab:tc100bps}
\end{table}

\paragraph{\textbf{Trading metrics under transaction costs}}\label{par:tc-trade}
Table~\ref{tab:tc100bps-trade} reports trading and exposure statistics under transaction costs over the full period. CTRL continues to have the lowest monthly turnover rates and least concentrated positions. In particular, the maximum long and short positions under CTRL remain below 20\% and 2\%, respectively, and both the average and maximum gross exposure stay close to full investment. By contrast, several model-based and RL-based methods exhibit even more extreme behaviors in terms of concentration and turnover, with the model-based continuous-time counterpart, ctmv,  in particular displaying outsized  leverage and bankruptcy rate. In summary, the results indicate that the outperformance of CTRL is even more pronounced under transaction costs.

\begin{table}[!htbp]
\centering
\begin{tabular}{lrrrrrr}
\toprule
 & BKR & TO & MaxL & MaxS & GE & GEMax \\
\midrule
ew & 0\% & 5.798\% & 10.000\% & 0.000\% & 100.000\% & 100.000\% \\
mv & 0\% & 16.432\% & 113.971\% & 61.028\% & 113.558\% & 260.610\% \\
js & 1\% & 15.341\% & 171.726\% & 106.333\% & 140.192\% & 542.149\% \\
bl & 16\% & 15.912\% & 271.773\% & 150.030\% & 149.242\% & 1000.542\% \\
lw & 0\% & 18.741\% & 116.278\% & 82.500\% & 117.245\% & 389.493\% \\
ff & 0\% & 8.840\% & 116.479\% & 53.619\% & 96.796\% & 274.978\% \\
rp & 0\% & 11.173\% & 100.000\% & 0.000\% & 100.000\% & 100.000\% \\
min\_v & 0\% & 8.804\% & 104.477\% & 31.623\% & 126.789\% & 227.462\% \\
pmv & 0\% & 52.065\% & 75.980\% & 58.536\% & 75.156\% & 231.473\% \\
ctmv & 48\% & 42.588\% & 1716.020\% & 1005.929\% & 154.106\% & 6459.914\% \\
ddpg & 1\% & 19.668\% & 77.880\% & 70.666\% & 156.801\% & 532.347\% \\
ppo & 0\% & 20.684\% & 71.224\% & 51.970\% & 94.132\% & 360.672\% \\
CTRL & 0\% & 5.680\% & 18.778\% & 1.903\% & 102.301\% & 124.034\% \\
\bottomrule
\end{tabular}
\caption{\textbf{Trading and exposure characteristics under 50 bps proportional transaction costs over the full sample period (2000--2020).}
BKR denotes the bankruptcy rate across experiments. TO is the average monthly turnover.
MaxL and MaxS report the maximum single-asset long and short positions, respectively.
GE is the average gross exposure, and GEMax is the maximum gross exposure observed across all experiments.}
\label{tab:tc100bps-trade}
\end{table}

\subsection{Market regimes}\label{subsec:regimes}
The previously reported results are based on the full sample period from 2000 to 2020, which contains both bear and bull market regimes. To examine performance across different market conditions, we split the test period into two subperiods, 2000--2010 and 2010--2020. It just so happened that the first period 2000-2010 was overall a bear market, including the dot-com bust and the 2008 financial crisis, during which the S\&P~500 had a negative annualized return of $-2.633\%$. The second subperiod was a long bull market, during which the S\&P~500 attained an annualized return of $11.070\%$. Detailed performance and trading summaries for all the strategies are reported in Appendix \ref{subsec:comp-regime}, Tables~\ref{tab:performance_bear}--\ref{tab:trade_bull}.

In the bear subperiod 2000--2010, CTRL outperforms the competing methods in terms of both annualized return and Sharpe ratio. In particular, CTRL achieves the highest annualized return ($9.788\%$) and Sharpe ratio ($0.367$), as well as  the best Sortino and Calmar ratios. Compared with the second-best method, ew, the performance gap in this subperiod is  larger than that observed over the full sample period, indicating that CTRL is an even  stronger performer during market downturns. Importantly, this outperformance is not driven by aggressive shorting or leverage. As shown by the corresponding trading summary in Table \ref{tab:trade_bear}, CTRL maintains very small short positions (MaxS around $1.9\%$), moderate single-asset long positions (MaxL below $20\%$), and gross exposure close to full investment (GE about $102\%$), along  with low turnovers. Moreover, after it plummeted, S\&P 500 had never recovered to its previous peak within this subperiod, while CTRL recovered within a short time of 264 days. By contrast, several estimation-based strategies exhibit ample instability in this market regime, including bankruptcy events for ctmv, bl, and js, consistent with their concentrated and volatile risky exposures.

During the bull subperiod 2010--2020, many strategies perform well, and differences across methods are less pronounced than in the bear period. CTRL remains among the top, achieving an annualized return of $14.539\%$ and a Sharpe ratio of $0.783$, both exceeding the strong benchmark ew. At the same time, the trading statistics show that CTRL continues to operate with low turnovers and moderate concentrations without bankruptcy events. Although several methods attain higher Sharpe ratios in this bull regime, they do so by taking substantially higher leverage and highly concentrated long or short positions.

Overall, when market conditions are favorable, most strategies generate strong performance.
A strategy therefore needs to outperform especially during adverse and volatile  periods in order to excel {\it in the long run}. It in turn calls for more robust performances, something
CTRL can provide as evident from this empirical study.

\subsection{Robustness checks for tuning parameters}
\label{sec:empirical robustness}

We further examine the robustness of the CTRL algorithm via sensitivity analysis on three groups of hyperparameters: (i) the learning rate, (ii) the temperature parameter, and (iii) the non-trainable parameters $\theta_3$ and $\phi_3$. These are the only hyperparameters in the value and policy functions. Each parameter is varied  over a wide multiplicative range, including substantial increases and decreases relative to the baseline configuration (e.g., halving or doubling their values). As reported in Tables~\ref{tab:sensitivity_gross}--\ref{tab:sensitivity_trade_tc100bps} of Appendix~\ref{appendix:sensitivity}, all performance and trading metrics remain within  narrow ranges across these variations, consistently yielding high annualized returns and strong risk-adjusted performance, with no signs of instability, excessive leverage, or abnormal risky exposure.

Importantly, this robustness holds under all the settings including no transaction costs, the 50 bps proportional transaction fees (considered in Section \ref{subsec:tc}) and lower transaction costs (reported in Appendix \ref{subsec:comp-tc}).

\section{Conclusions}
\label{sec:conclusion}

This paper presents a general data-driven RL approach for continuous-time MV portfolio selection
in markets described by observable It\^o's diffusion processes without knowing their coefficients/parameters or attempting to estimate them. It leads to an algorithm for the Black--Scholes market, whose theoretical performance guarantee including a sublinear regret is proved. Through a thorough comparative empirical study, we demonstrate the performance and robustness of the proposed CTRL strategy. This paper distinguishes itself from most existing works on applying RL to portfolio optimization in that its algorithm is based on a rigorous and explainable mathematical underpinning (relaxed control and martingality) established in \cite{wang2020reinforcement} and \cite{jia2021policy,jia2021policypg}. Moreover, it is the first to derive a {\it model-free} sublinear regret bound for dynamic MV problems, to our best knowledge.

One of the most notable insights derived from this work is the decisive outperformance of the explore-{\it and}-exploit approach of RL over the traditional estimate-{\it then}-plug-in counterparts in a dynamic market. This superiority is not due to ``big data", as our algorithm depends only on the stock price data (instead of thousands of factor data, which could be incorporated into our framework to further enhance the performance); rather it is due to a fundamentally different decision-making approach, namely, to learn the optimal policy without learning the model.

Despite the recent upsurge of interest in continuous-time RL, its study is still in the early innings, not to mention that its applications to financial decision-making remain largely uncharted territory. In particular, regret analysis is one of the most challenging problems in the {\it model-free} realm, because one can no longer rely on the well-developed theory on the accuracy of estimating model primitives as in the model-based setting. As such, we are able to derive a sublinear regret only for the Black--Scholes environment, without considering more stylized facts of asset returns such as stochastic volatility and jumps. This is a major limitation of our results; but even such a simple setup already calls for a novel yet lengthy and extremely delicate analysis on the sampling errors and policy improvement signals involved in the algorithm.  Establishing a truly model-free regret analysis—one that holds for a broader class of underlying stochastic processes (such as jump-diffusions or stochastic volatility models) using highly flexible, generalized function approximators—remains a formidable open challenge. More general dynamics may require certain conditions (e.g. \citealt[Section 4]{tang2024regret}) along with an even more involved analysis, but we hope this paper sets a starting point and a benchmark, in both methodologies and results, for future regret study with increasingly complex market environments.

Many other open questions remain.  In the MV setting, important questions include improvement of regret bound, online and/or off-policy learning, as well as large investors whose actions impact the asset prices (so counterfactuals become unobservable by mere ``paper portfolios").

\medskip

\noindent {\bf Acknowledgement.} We thank the three referees and the Associate Editor for very helpful and constructive comments that have led to a much improved manuscript. We are grateful for comments from seminar and conference participants at Fudan, University of Michigan, University of Connecticut, Humboldt-Universit\"at zu Berlin, Technische Universit\"at Berlin, CIBC, Trexquants, the 11th World Congress of the Bachelier Finance Society, the INFORMS Annual Meeting, the SIAM Conference on Financial Mathematics and Engineering, the INFORMS Conference on Financial Engineering and FinTech, the World Online Seminar on Machine Learning in Finance, the Workshop on Advances in Stochastic Analysis for Handling Risks in Finance and Insurance in Marseille, the Ninth PKU-NUS Annual International Conference on Quantitative Finance and Economics at PKU, the 2025 Workshop on Mathematical Finance at Seoul National University, the 9th Asian Quantitative Finance Conference at Peking University HSBC Business School, and the 6th ACM International Conference on AI in Finance in Singapore. Huang is supported by the Start-up Fund P0063874 at The Hong Kong Polytechnic University. Jia is supported by the Start-up Fund at The Chinese University of Hong Kong, and Hong Kong Research Grants Council (RGC) - Early Career Scheme (ECS) 24211124. 
Zhou is supported by the Nie Center for Intelligent Asset Management and  the Aurora FinTech Research Fund at Columbia University. His work is also part of a Columbia-CityU/HK collaborative project that is supported by the InnoHK Initiative, The Government of the HKSAR, and the AIFT Lab.

\bibliography{ref}
\clearpage
\appendix
\section*{Appendix}
\section{Formulation and Solutions of Exploratory Mean--Variance Problem}
\subsection{Exploratory state dynamics under stochastic policies}
\label{appendix:formulation rigorous}
We now present the precise formulation of the market environment, i.e., the asset price dynamics appearing in \eqref{classical_dynamics}, as well as the exploratory wealth dynamics under stochastic policies.

Recall that $S^0(t)$ is the price of the risk-free asset, $S^i(t)$  the price of the $i$-th risky asset, $F(t)$ represents the values of the observable covariates/factors, and $u(t)$ is the portfolio choice vector, all at time $t$. We assume $S^0$ satisfies
\begin{equation}
	\label{eq:riskfree}
	\dd S^{0}(t) = r(t,u(t),F(t)) S^{0}(t) \dd t,
\end{equation}
and $S^i$ follows
\begin{equation}
	\label{eq:stock price dynamics}
	\dd S^{i}(t)=S^{i}(t)\left[\mu^{i}(t,u(t),F(t)) \dd t+\sum_{j=1}^{m} \sigma^{i j}(t,u(t),F(t)) \dd W^{j}(t)\right], \quad i=1,2, \cdots, d,
\end{equation}
where $r(t,u(t),F(t))$ is the short rate, $\mu(t,u(t),F(t)): = (\mu^1(t,u(t),F(t)), \mu^2(t,u(t),F(t)), \cdots ,\\
\mu^d(t,u(t),F(t)))^\top \in \mathbb{R}^{d}$ and $\sigma(t,u(t),F(t)): = (\sigma^{ij}(t,u(t),F(t)))_{1\leq i\leq d,1\leq j\leq m}\in \mathbb{R}^{d\times m}$ are respectively the instantaneous expectation and volatility of the risky asset returns at $t$, and $W=(W^{1}, W^{2}, \dots, W^{m})^{\top}$ is an $m$-dimensional standard Brownian motion. We define $\Sigma(t,u(t),F(t)): = \sigma(t,u(t),F(t))\sigma(t,u(t),F(t))^\top\in \mathbb{R}^{d\times d}$ and assume it satisfies $\Sigma(t,u(t),F(t)) - \alpha I\in \mathbb{S}^d_+$ for all $t\geq 0$ with probability 1 for some constant $\alpha > 0$. We further assume that the above mentioned processes $\{r(t,u(t),F(t)),\mu(t,u(t),F(t)),\sigma(t,u(t),F(t)): 0\leq t\leq T\}$ are all well-defined and adapted in a given filtered probability space $\left(\Omega, \mathcal{F}, \p ;\left(\mathcal{F}_t\right)_{t \geq 0}\right)$ satisfying the usual conditions. Moreover, the factor process  $F$ follows
\begin{equation}
	\label{eq:factor}\dd F(t) = \iota(t,u(t),F(t))\dd t + \nu(t,u(t),F(t))\dd W(t).
\end{equation}

All the coefficients in \eqref{eq:riskfree}--\eqref{eq:factor} depend on portfolio $u$ to capture the most general scenario that a larger investor's actions may impact the values of assets and factors. They are independent of $u$ when we consider a small investor.

The wealth equation \eqref{classical_dynamics} now specializes to
\begin{equation}
	\mathrm{d} x^{u}(t) = (\mu(t,u(t),F(t)) - r(t,u(t),F(t)) \bm{e}_d)^\top u(t) \mathrm{d} t + u(t)^\top \sigma(t,u(t),F(t)) \mathrm{d} W(t), \quad 0 \leq t \leq T; \quad x_0^{u} = x_0.
	\label{classical_dynamics2}
\end{equation}

Under a stochastic policy $\bm\pi$, its ``dynamic of wealth" now describes the average of the (infinitely many) wealth processes under portfolios repeatedly sampled from $\bm\pi$; hence is different from \eqref{classical_dynamics}. Applying the notion of relaxed stochastic control, \citet{wang2020reinforcement} derive the following ``exploratory" dynamic:
\begin{equation}
	\fitdisplay{\begin{aligned}
		\dd x^{\bm\pi}(t) =& \int_{\mathbb{R}^d} [\mu(t,u,F(t)) - r(t,u,F(t))\bm{e}_d ]^\top u \bm{\pi}(u|t,x^{\bm\pi}(t),F(t)) \dd u \\
		&+ \sqrt{\int_{\mathbb{R}^d} u^\top \Sigma(t,u,F(t)) u \bm{\pi}(u|t,x^{\bm\pi}(t),F(t)) \dd u } \dd \tilde W(t).
	\end{aligned}}
	\label{rl_dynamics_general}
\end{equation}
Here, Brownian motion \(\tilde W\) is different from the one  in \eqref{eq:stock price dynamics}, see details in \citet{jia2025accuracy} for the interpretation of \eqref{rl_dynamics_general} in the weak sense. 
We emphasize that the averaged wealth process $x^{\bm\pi}$ is not observable (i.e. it is not part of the data) and \eqref{rl_dynamics_general} is used mainly for the {\it theoretical} analysis of the learning algorithms.



\subsection{Exploratory mean--variance formulation and solutions in the Black--Scholes environment}\label{appendix:solutions}

We re-state the exploratory MV problem in a frictionless, multi-stock Black--Scholes market, without any factors \(F\). The exploratory wealth equation is
\begin{equation}
	\label{rl_dynamics}
	\fitdisplay{\begin{aligned}
		\dd x^{{\bm\pi}}(t) = &(\mu - r\bm{e}_d )^\top \int_{\mathbb{R}^d} u \bm{\pi}(u|t,x^{\bm\pi}(t))\dd u \dd t + \sqrt{\int_{\mathbb{R}^d} u^\top \Sigma u \bm{\pi}(u|t,x^{\bm\pi}(t)) \dd u } \dd \tilde W(t),
	\end{aligned}}
\end{equation}
while the goal is to find the stochastic  policy $\bm\pi$ that minimizes the entropy regularized value function
\begin{equation}
	\mathbb{E}\left[\left({x}^{\bm{\pi}}(T)-w\right)^{2}+\gamma \int_{0}^{T} \int_{\mathbb{R}^d} \bm{\pi}(u|t,{x}^{\bm{\pi}}(t)) \log \bm{\pi}(u|t,{x}^{\bm{\pi}}(t)) \dd u \dd t\right]-(w-z)^{2}.
	\label{mv_rl_formulation}
\end{equation}

This problem has been solved by \cite{wang2020continuous} for the case of one stock, which can be extended readily to the multi-stock case.
The optimal value function is
\begin{equation}
	\fitdisplay{\begin{aligned}
		V^*(t, x; w) = &(x-w)^2e^{-(\mu-r)^\top (\sigma \sigma^\top)^{-1} (\mu-r)(T-t)} \\
		&+ \frac{\gamma d}{4}(\mu-r)^\top (\sigma \sigma^\top)^{-1} (\mu-r) (T^2 - t^2)\\
		&- \frac{\gamma d}{2}\bigg((\mu-r)^\top (\sigma \sigma^\top)^{-1} (\mu-r) T - \frac{1}{d} \log \frac{\det (\sigma \sigma^\top) }{ (\pi \gamma)^d} \bigg)(T-t) - (w-z)^2,\\
		&(t, x,w) \in[0, T] \times \mathbb{R}\times \mathbb{R},
		\label{eq:optimal_value_function}
	\end{aligned}}
\end{equation}
the optimal policy is
\begin{equation}
	\fitdisplay{\begin{aligned}
		\bm\pi^{*}(u \mid t, x, w)=\mathcal{N}\biggl(u \mid - (\sigma \sigma^\top)^{-1}&(\mu-r)(x-w), (\sigma \sigma^\top)^{-1} \frac{\gamma}{2}e^{(\mu-r)^\top (\sigma \sigma^\top)^{-1} (\mu-r) (T-t)}\biggl)\\
		&(u,t, x,w) \in\mathbb{R}^d\times [0, T] \times \mathbb{R}\times \mathbb{R},
		\label{eq:optimal_policy}
	\end{aligned}}
\end{equation}
and the corresponding  Lagrange multiplier is
\begin{equation}
	\label{eq:optimal_w}
	w^* = \frac{z e^{(\mu-r)^\top (\sigma \sigma^\top)^{-1} (\mu-r) T} - x_0}{e^{(\mu-r)^\top (\sigma \sigma^\top)^{-1} (\mu-r) T} - 1}.
\end{equation}

Once again, these analytical expressions are not used to compute the solutions (because the problem primitives are unknown); rather they are employed to {\it parameterize} the policies and value functions for learning.

\section{Details of Implementations of the Algorithm}\label{appendix:algo}
\subsection{The algorithm}\label{appendix:baseline algo}

To begin, consider the moment conditions in \eqref{eq:joint moment} under the absence of the factor \(F(t)\) and with the stochastic policy \(\bm\pi^{\bm\phi}\) parameterized as \eqref{eq:policy parameterize}. The moment conditions can be reformulated as:

\begin{equation}
	\label{eq:joint moment_new}
	\left\{
	\fitdisplay{\begin{aligned}
		&  \E\left[ \int_0^T \frac{\partial J}{\partial \theta}\left(t, x^{\bm\pi}(t) ; w ; \bm \theta\right) \left[ \dd J(t,x^{\bm\pi}(t);w;\bm\theta)+ \gamma \hat{p}(t,\bm\phi)\dd t \right] \right] = 0,  \\
		& \E\bigg[\int_0^T  \left[ \frac{\partial }{\partial \bm\phi} \log \bm\pi(u(t)|t,x^{\bm\pi}(t);w;\bm\phi ) \right]  \Big[ \dd J(t,x^{\bm\pi}(t);w;\bm\theta)+ \gamma \hat{p}(t,\bm\phi)\dd t  \bigg]  + \gamma \frac{\partial \hat{p}}{\partial \bm\phi}(t, \bm\phi) \biggl ] = 0, \\
		& \E\left[ x^{\bm\pi}(T) - z \right] = 0,
	\end{aligned}} \right.
\end{equation}
where \(\hat{p}(t, \bm\phi)\) represents the differential entropy of the policy $\bm\pi(\cdot \mid t, x ; w;\bm\phi)$, which can be explicitly calculated as
\begin{equation*}
	\fitdisplay{\begin{aligned}
		\hat{p}(t, \bm\phi)&=  -\frac{d}{2} \log{(2 \pi e)} + \frac{1}{2} \log (\det \phi_2 ^{-1}) - \frac{d}{2} \phi_3 (T-t),
	\end{aligned}}
\end{equation*}
revealing its independence of $x,w$ and $\phi_1$. To see the equivalence between conditions \eqref{eq:joint moment} and \eqref{eq:joint moment_new}, we note that
$
\E[\log \bm\pi(u(t)|t,x^{\bm\pi}(t);w;\bm\phi )] = \hat{p}(t,\bm\phi)$, and hence,
\[\fitdisplay{\begin{aligned}
	\frac{\partial \hat{p}}{\partial \bm\phi}(t, \bm\phi) = & \E\left[\int_{\mathbb{R}^d} \log \bm\pi(u|t,x^{\bm\pi}(t);w;\bm\phi) \frac{\partial }{\partial \bm\phi} \bm\pi(u|t,x^{\bm\pi}(t);w;\bm\phi) \dd u \right] + \E\left[\int_{\mathbb{R}^d} \frac{\partial }{\partial \bm\phi} \bm\pi(u|t,x^{\bm\pi}(t);w;\bm\phi) \dd u \right] \\
	= & \E\left[\int_{\mathbb{R}^d} \log \bm\pi(u|t,x^{\bm\pi}(t);w;\bm\phi) \frac{\partial }{\partial \bm\phi} \bm\pi(u|t,x^{\bm\pi}(t);w;\bm\phi) \dd u \right].
\end{aligned}}  \]	
Therefore,	
\[
\fitdisplay{\begin{aligned}
	&\E\biggl[ \frac{\partial }{\partial \bm\phi} \log \bm\pi(u(t)|t,x^{\bm\pi}(t);w;\bm\phi) \log \bm\pi(u(t)|t,x^{\bm\pi}(t);w;\bm\phi )\biggl]\\
	= & \E\left[\int_{\mathbb{R}^d} \log \bm\pi(u|t,x^{\bm\pi}(t);w;\bm\phi) \frac{\partial }{\partial \bm\phi} \bm\pi(u|t,x^{\bm\pi}(t);w;\bm\phi) \dd u \right] \\
	= & \frac{\partial \hat{p}}{\partial \bm\phi}(t, \bm\phi)
	= \E\biggl[ \frac{\partial }{\partial \bm\phi} \log \bm\pi(u(t)|t,x^{\bm\pi}(t);w;\bm\phi) \biggl] \hat{p}(t,\bm\phi)  + \frac{\partial \hat{p}}{\partial \bm\phi}(t, \bm\phi) .
\end{aligned}}
\]

To design the baseline algorithm, we first calculate the relevant gradients given the parameterization in \eqref{eq:value function parameterize} and \eqref{eq:policy parameterize}. Indeed, we have
\[
\frac{\partial J}{\partial \theta_{1}}(t, x ; w; \bm\theta)=t-T, \;
\frac{\partial J}{\partial \theta_{2}}(t, x ; w; \bm\theta)=t^{2}-T^{2}, \;
\]
\[
\frac{\partial \hat{p}}{\partial \phi_{1}}(t, \bm\phi)=0, \;
\frac{\partial \hat{p}}{\partial \phi_{2}^{-1}}(t, \bm\phi)=\frac{\phi_2}{2}, \;
\]
and
\[
\fitdisplay{\begin{aligned}
	&\frac{\partial \log \bm\pi(u \mid t, x ; w; \bm\phi)}{\partial \phi_{1}} =  -e^{-\phi_3(T-t)} \left[(x-w) \phi_2^{-1}u + (x-w)^2 \phi_2^{-1} \phi_1 \right], \\
	&\frac{\partial \log \bm\pi(u \mid t, x ; w; \bm\phi)}{\partial \phi_{2}^{-1}} = \frac{1}{2} \phi_2 - \frac{1}{2} e^{-\phi_3(T-t)} (u + \phi_1(x-w)) (u + \phi_1(x-w))^\top .
\end{aligned}}
\]

Recall the (theoretical) updating rules for $\bm\theta$, $\phi_1$, $\phi_2$ in \eqref{eq:theta_update1}--\eqref{eq:phi2_update1} involve integrals. For actual implementation we use discretized summations to approximate those integrals: we discretize $[0,T]$ into small time intervals with an equal length of $\Delta t$. 
Then the updating rules are modified to
\begin{equation}
	\bm\theta \leftarrow \Pi_{K_{\bm\theta, n}} \biggl(\bm\theta+a_n \sum_{k=0}^{\left\lfloor\frac{T}{\Delta t}\right\rfloor-1} \frac{\partial J}{\partial \bm\theta}\left(t_k, x_{t_k} ; w; \bm\theta \right)\left[J\left(t_{k+1}, x_{t_{k+1}} ; w; \bm\theta\right) - J\left(t_k, x_{t_k} ; w; \bm\theta\right)+\gamma \hat{p}(t_k, \bm\phi
	) \Delta t\right] \biggl),
	\label{eq:theta_update_discrete}
\end{equation}

\begin{equation}
	\begin{split}
		\phi_1 \leftarrow \Pi_{K_{1, n}} \biggl(\phi_1-a_n \sum_{k=0}^{\left\lfloor\frac{T}{\Delta t}\right\rfloor-1}\biggl\{\frac{\partial \log {\bm\pi}}{\partial \phi_1}\left(u_{t_k} \mid t_k, x_{t_k} ; w; \bm\phi\right)
		&\biggl[J\left(t_{k+1}, x_{t_{k+1}} ; w; \bm\theta\right) - J\left(t_k, x_{t_k} ; w; \bm\theta\right)\\
		&+\gamma \hat{p}(t_k, \bm\phi) \Delta t\biggl]+\gamma \frac{\partial \hat{p}}{\partial \phi_1}\left(t_k, \bm\phi\right) \Delta t \biggl\}\biggl),
		\label{eq:phi1_update_discrete}
	\end{split}
\end{equation}

\begin{equation}
	\begin{split}
		\phi_2 \leftarrow \Pi_{K_{2, n}} \biggl(\phi_2+a_n \sum_{k=0}^{\left\lfloor\frac{T}{\Delta t}\right\rfloor-1}\biggl\{\frac{\partial \log {\bm\pi}}{\partial \phi_2^{-1}}\left(u_{t_k} \mid t_k, x_{t_k} ; w; \bm\phi\right)
		&\biggl[J\left(t_{k+1}, x_{t_{k+1}} ; w; \bm\theta\right) - J\left(t_k, x_{t_k} ; w; \bm\theta\right)\\
		&+\gamma \hat{p}(t_k, \bm\phi) \Delta t\biggl]+\gamma \frac{\partial \hat{p}}{\partial \phi_2^{-1}}\left(t_k, \bm\phi\right) \Delta t \biggl\} \biggl).
		\label{eq:phi2_update_discrete}
	\end{split}
\end{equation}

For each iteration, the algorithm starts with time 0 and initial wealth $x_0$. At each discretized timestep $t$, $t=0, \Delta t,2\Delta t,..., \left\lfloor\frac{T}{\Delta t}\right\rfloor-1$, it samples an action $u(t)$ from the Gaussian policy in \eqref{eq:policy parameterize}, and calculates the new wealth at next timestep  based on the current wealth, action and  the assets price movement. At the final timestep $\left\lfloor\frac{T}{\Delta t}\right\rfloor$, the algorithm then  uses the whole wealth trajectory to update parameters $\bm\theta$ and $\bm\phi$.

\subsection{Algorithm implementation}\label{appendix:algo empirical}

We now describe the detailed choice of hyperparameters for the implementation of the CTRL algorithm.

\begin{itemize}
	\item Learning rate for the Lagrange multiplier $w$: 0.0001.
	\item Learning rates for $\bm\theta$ and $\bm\phi$: 0.0001.
	\item Initial wealth is set to be \$1, with a target set to $(1.15)^{\frac{1}{12}}$.
	\item Investment horizon $T$: 1 month, with a time step size $\Delta t=\frac{1}{21}$.
	\item Temperature parameter $\gamma$: 0.1.
	\item Total number of training iterations: 100,000 for pretrain and 50 for finetuning each month, updating the Lagrange multiplier every 10 iterations.
	\item Batch size: 64.
\end{itemize}

\section{Alternative Asset Allocation Methods}\label{alternative}
In this section, we briefly describe the other  portfolio selection methods to be compared with  our CTRL strategy. They include computation-free approaches, risk-based strategies, other RL methods and, predominantly, those based on both static and dynamic MV frameworks  with different statistical techniques to estimate the mean and covariance matrix of asset returns.  To dynamically implement all the static models, we rebalance monthly with the following month for single-period optimization, taking a rolling window of the immediately prior 10 years for estimating the model parameters.

For all the methods involved,  we define $R(t) = (R^1(t),\cdots,R^d(t))^\top$ to be the vector of monthly excess returns of the  $d$ assets in the $t$-th month,  and $w(t) = (w^1(t),\cdots,w^d(t))$ the portfolio in the $t$-th month, where $w^i(t)$ is the fraction of total wealth allocated to the $i$-th asset at $t$, $1\leq t\leq 240$. The various portfolio choice methods are used to determine these weights.

Table~\ref{tab:strategies} gives an overview of all the allocation methods under comparison.

\begin{table}[h!]
    \centering
    \begin{tabular}{lll}
        \toprule
        Strategy & Symbol & Reference / Description\\
        \midrule
        Buy-and-hold S\&P 500 index & S\&P 500 & Capitalization‑weighted U.S. market index \\
        Equally weighted allocation & ew & \cite{demiguel2009optimal} \\
        Mean--variance (single‑period) & mv & \cite{markowitz1952portfolio} \\
        Minimum variance & min\_v & \cite{markowitz1952portfolio} \\
        James--Stein shrinkage (mean) & js & \cite{jorion1986bayes} \\
        Ledoit--Wolf shrinkage (covariance) & lw & \cite{ledoit2003improved} \\
        Black--Litterman model & bl & \cite{black1990asset} \\
        Fama--French three‑factor model & ff & \cite{fama1993common} \\
        Risk parity & rp & \cite{maillard2010properties} \\
        Continuous-time mean--variance  & ctmv & \cite{zhou2000continuous} \\
        Predictive mean--variance & pmv & \cite{bali2016empirical} \\
        Deep deterministic policy gradient & ddpg & \cite{lillicrap2016continuous} \\
        Proximal policy optimization & ppo & \cite{schulman2017proximal} \\
        \bottomrule
    \end{tabular}
    \caption{Alternative asset allocation strategies compared in this study.}
    \label{tab:strategies}
\end{table}

\subsection{Buy-and-hold market index (S\&P 500)}
The S\&P 500 index is capitalization weighted with dynamically adjusted constituents (\url{https://www.spglobal.com/spdji/en/index-family/equity/us-equity/us-market-cap/\#overview}). It serves as a natural barometer of the overall market performance, a proxy of the market portfolio, and a benchmark many funds compare against. The buy-and-hold strategy of the S\&P 500 index does not require any computation -- its return over any given period is calculated based on the index's values on the first and last days of the period.

\subsection{Equally weighted allocation (ew)}
\label{ew}
Another straightforward allocation method is the equally weighted allocation where $w^i(t)=\frac{1}{d}$ for $1\leq i\leq d$. This strategy does not depend on any  data, nor does it require any statistical estimation. Despite its  simplicity and disregard of information, \cite{demiguel2009optimal} find that it exhibits admirable performance and remarkable robustness. Indeed, none of the 14 alternative allocation methods they tested consistently outperformed the equally weighted portfolio on real market data. As such, we take it as another important benchmark for comparison in our study.

\subsection{Sample-based (single-period) mean--variance (mv)}
Many portfolio selection methods are based on the one-period MV problem \citep{markowitz1952portfolio}:
\begin{equation}
	\label{eq:mv problem}
	\begin{array}{ll}
		\underset{w}{\operatorname{min}} & w^\top \Sigma w \\
		\text {subject to } & w^\top \mu  \geq {\mu}^{*},
	\end{array}
\end{equation}
where $\mu$ and $\Sigma$ are the mean vector and covariance matrix of asset \textit{excess} returns respectively, and ${\mu}^*$ is the investor's target expected excess return.  The solution to this problem can be found explicitly as
\begin{equation}
	\label{eq:optimal mv}
w^* = \frac{\mu^*}{\mu^\top\Sigma^{-1}\mu}\Sigma^{-1} \mu.
\end{equation}

Various plug-in methods are differentiated by the ways to  estimate the unknown mean and covariance. Among them, the sample-based method estimates them
using sample mean and covariance
based on the most recent 120-month data:
\begin{equation}
	\label{eq:sample mean}
	\hat{\mu}(t)\equiv (\hat{\mu}_1(t),\cdots,\hat{\mu}_d(t))^\top = \frac{1}{M}\sum_{\tau=1}^M R_{t-\tau},\;\;\hat{\Sigma}(t)\equiv (\hat{\Sigma}_{ij}(t))_{d\times d} =  \frac{1}{M-1}\sum_{\tau=1}^M (R_{t-\tau} - \hat{\mu})(R_{t-\tau} - \hat{\mu})^\top ,
\end{equation}
and then plugs them into  the formula \eqref{eq:optimal mv} to compute the portfolio weights.

\subsection{Sample-based minimum variance (min\_v)}
\label{minv}
A minimum variance  portfolio achieves the lowest variance with a set of risky assets,  without setting any expected return target.
Mathematically, the minimum variance portfolio is obtained by solving
$$
\begin{array}{ll}
	\underset{w}{\operatorname{min}} & w^\top \Sigma w \\
	\text { subject to } & w^\top \bm{e}_d = 1.
\end{array}
$$
The solution is
\begin{equation}
	\label{eq:minv}
	w^* =  \frac{1}{\bm{e}_d^\top\Sigma^{-1}\bm{e}_d}\Sigma^{-1}\bm{e}_d.
\end{equation}

An advantage of the minimum variance portfolio is that it does not involve the mean returns of the stocks, which are significantly harder to estimate to a workable accuracy compared with the covariances. In our experiments we plug the sample covariance in \eqref{eq:sample mean} into \eqref{eq:minv} to obtain the minimum variance portfolio.

\subsection{James--Stein shrinkage estimator for mean (js)}
\cite{jorion1986bayes} proposes a James--Stein type of shrinkage estimator \citep{james1992estimation}  to shrink the estimates for the mean returns  towards those of the sample-based minimum variance portfolio:
\[ \tilde{\mu}(t) = \frac{\hat{\mu}(t)^\top\hat{\Sigma}(t)^{-1}\bm{e}_d}{\bm{e}_d^\top\hat{\Sigma}(t)^{-1}\bm{e}_d} \bm{e}_d, \]
where $\hat{\mu},\hat{\Sigma}$ are the sample estimators given by \eqref{eq:sample mean}. The James--Stein shrinkage estimator for the mean is then
\[ \hat{\mu}^{\text{js}}(t) = (1-{\alpha}(t)) \hat{\mu}(t) + {\alpha}(t)  \tilde{\mu}(t), \]
where ${\alpha}(t) = \frac{d+2}{d+2+(M-d-2) (\hat{\mu}(t) - \tilde{\mu}(t))^\top \hat{\Sigma}(t)^{-1}  (\hat{\mu}(t) - \tilde{\mu}(t)) }$,
to be plugged into the solution of the MV problem, \eqref{eq:optimal mv}.

\subsection{Ledoit--Wolf shrinkage estimator for covariance matrix (lw)}
\citet{ledoit2003improved} propose a shrinkage estimator for the covariance matrix.
It starts with the single-index model for stock returns \citep{sharpe1963simplified} at the $t$-th month:
$$
R^i(t)=a_{i}+b_{i} R^m(t)+\varepsilon^i(t),\;\;i=1,2,\cdots,d,
$$
where $R^m(t)$ is the excess return of the market and $\varepsilon^i(t)$, $i=1,2,\cdots,d$, are
the
residuals that are uncorrelated to the market and to one another. Then the sample estimator of the covariance matrix of this model is:
$$
\hat {F}(t)\equiv (\hat F_{i j}(t))_{d\times d}=b b^{\top}\hat\sigma_m^2(t)+\hat {D}(t)
$$
where $\hat\sigma_m^2(t)$ is the sample variance of the market return, $b$ is the vector of the slopes $b_i$ and $\hat D(t)$ is the diagonal matrix containing the residual sample variances estimates. Denote by $\hat{\mu}_{m}(t)$ the sample mean of the market, and by $\hat\sigma_{im}(t)$ be the sample covariance between stock $i$ and the market, both at time $t$.

Set $k_{ij}(t)=\frac{p_{ij}-r_{ij}}{c_{ij}(t)}$ to be the shrinkage estimator, where
$$
p_{i j}=\frac{1}{M} \sum_{t=1}^{M}\left\{\left( R^i(t)-\hat{\mu}_{i}(t)\right)\left(R^j(t)-\hat{\mu}_{j}(t)\right)-\hat\Sigma_{i j}(t)\right\}^{2},\;\;
c_{i j}(t)=\left(\hat F_{i j}(t)-\hat\Sigma_{i j}(t)\right)^{2},
$$
and $r_{ii} = p_{ii}$ while $r_{ij} = \frac{\sum_{t=1}^M r_{ij}(t)}{M}$ for $i \neq j$ where
$$
\fitdisplay{\begin{aligned}
	r_{i j}(t)=& \frac{\hat\sigma_{jm}(t) \hat\sigma_m(t)\left(R^i(t)-\hat{\mu}_{i}(t)\right)+\hat\sigma_{im}(t) \hat\sigma_m(t)\left(R^j(t)-\hat{\mu}_{j}(t)\right)-\hat\sigma_{im}(t)\hat\sigma_{jm}(t) \left(R^m(t)-\hat{\mu}_{m}(t)\right)}{\hat\sigma_m^2(t)} \\
	&\times\left(R^m(t)-\hat{\mu}_{m}(t)\right)\left(R^i(t)-\hat{\mu}_{i}(t)\right) \left(R^j(t)-\hat{\mu}_{j}(t)\right)-\hat F_{i j} \hat\Sigma_{i j}(t).
\end{aligned}}
$$

Then the Ledoit--Wolf shrinkage estimator for the covariance matrix is
$\hat{\Sigma}^{lw}(t)\equiv (\hat{\Sigma}^{lw}_{ij}(t))$ where
\begin{equation}
	\hat{\Sigma}^{lw}_{ij}(t)=\frac{{k}_{ij}(t)}{M} \hat F_{ij}(t)+\left(1-\frac{k_{ij}(t)}{M}\right) \hat{\Sigma}_{ij}(t).
\end{equation}

This estimator, along with any  estimated mean returns, is to be plugged into the solution of the MV problem, \eqref{eq:optimal mv}.

\subsection{Black--Litterman model (bl)}
\label{bl}
Premised upon the CAPM \citep{sharpe1964capital}, \cite{black1990asset} propose  to use the market portfolio to infer mean returns of individual stocks.
More precisely, at time $t$, we take the sample covariance matrix $\hat{\Sigma}^{all}(t)$  of all the 300 stocks in our stock universe, and compute the corresponding market portfolio (of these 300 stocks) $ w^{all}(t)$ based on their market capitalizations. Then the implied stock mean return vector  $\mu^{all}(t)$ and the market portfolio  have the relation:
\[ \mu^{all}(t)= \gamma(t) \hat{\Sigma}^{all}(t) w^{all}(t)\]
for some risk-adjusted coefficient $\gamma(t)$. This parameter is estimated using $\hat\gamma(t) = \frac{\hat{\mu}_{m}(t)}{\hat{\sigma}^2_{m}(t)}$, where $\hat{\mu}_{m}(t)$ and $\hat{\sigma}^2_{m}(t)$ are the sample mean and variance of the S\&P 500 index respectively at $t$. 

Then we extract the corresponding entries in $\mu^{all}(t)$ as our estimated expected returns for the $d$ selected stocks, denoted by $\hat{\mu}^{bl}(t)\in \mathbb{R}^{d}$, and  feed them along with any estimate of the covariance matrix into the solution  \eqref{eq:optimal mv}.

\subsection{Fama--French three factor model (ff)}
The celebrated Fama--French three factor model \citep{fama1993common}  provides a decomposition for asset returns in the following form:
\[ R(t) = \bm{\alpha} + \bm{B} F(t) + \bm{\epsilon}(t) , \]
where $F(t)\in \mathbb{R}^3$ is a vector of  mean-zero factors (``MKT", ``SMB", and ``HML"; see \url{https://mba.tuck.dartmouth.edu/pages/faculty/ken.french/data_library.html}) and $\bm\epsilon(t)$ consists of i.i.d. idiosyncratic noise terms for the stocks.
Then the model parameters can be estimated by running linear regression on each individual stock against the centered factor values. Specifically,  we first centralize the factors by
\[\fitdisplay{ \tilde F(s) = \left(MKT(s) - \frac{1}{M}\sum_{\tau=1}^M MKT({t-\tau}), SMB(s) - \frac{1}{M}\sum_{\tau=1}^M SMB({t-\tau}), HML(s) - \frac{1}{M}\sum_{\tau=1}^M HML({t-\tau})\right)^\top , }\]
for $s=t-M,\cdots,t-1$, where $MKT(s),SMB(s),HML(s)$ are the observed factor values at time $s$.
Then we use the least square to estimate the linear regression:
\[ R^i(s) = \bm{\alpha}^i + \bm{B}[i,]\tilde F(s) + \bm{\epsilon}^i(s) , \]
for each individual asset $i$, where $\bm{B}[i,]$ stands for the $i$-th row of the matrix $\bm B$.

This procedure produces estimates $\hat{\bm{\alpha}}^i, \hat{\bm{B}}[i,]$, and the residual $\hat{\bm{\epsilon}}^i(s)$ for each $1\leq i\leq d$ and each time instant $t-M\leq s\leq t-1$. The first two items lead to the estimators $\hat{\bm\alpha}$ and $\hat{\bm B}$. Moreover, we obtain the sample covariance matrix of the factors by $\hat{\Sigma}_{F}(t) = \frac{1}{M-1}\sum_{\tau=1}^{M} \tilde F({t-\tau})\tilde F({t-\tau})^\top$, as well as  a diagonal residual matrix $\hat{\Sigma}_{\epsilon}(t) = \frac{1}{M-1}diag\{ \sum_{\tau=1}^M \hat{\bm{\epsilon}}^1(t-\tau)^{2}, \cdots, \sum_{\tau=1}^M \hat{\bm{\epsilon}}^d(t-\tau)^{2} \}$. Finally, we set
\[ \hat{\mu}^{\text{ff}}(t) = \hat{\bm{\alpha}},\ \hat{\bm \Sigma}^{\text{ff}}(t) = \hat{\bm B} \hat{\Sigma}_{F}(t) \hat{\bm B}^\top + \hat{\Sigma}_{\epsilon}(t) \]
to be plugged into the solution \eqref{eq:optimal mv}.

\subsection{Risk parity (rp)}
\label{rp}
Risk parity is a volatility based portfolio allocation strategy that equalizes risk contribution of individual stocks to the whole portfolio. Mathematically, the volatility (standard deviation) of a portfolio $w=(w_1,\cdots,w_d)^{\top}$ is
$$
C(w)=\sqrt{w^{\top} \Sigma w}=\sum_{i=1}^{d} C_{i}(w)
$$
where
\[ C_{i}(w)=w_{i} \frac{\partial C(w)}{\partial{w_{i}}}=\frac{w_{i}(\Sigma w)_{i}}{\sqrt{w^{\top} \Sigma w}}
\]
is the risk contribution of the asset $i$. A risk parity portfolio $w$ requires $C_{i}(w)=\frac{C(w)}{d}$, which can in turn be determined by the following system of equations:
$$
w_{i}=\frac{C(w)^2}{(\Sigma w)_{i} d},\;\;i=1,\cdots,d.
$$
Alternatively, it can be derived by solving the following optimization problem
$$
\begin{array}{ll}
	\underset{w}{\operatorname{min}} & \sum_{i=1}^{d}\left[w_{i}-\frac{C(w)^2}{(\Sigma w)_{i} d}\right]^{2} \\
	\text { subject to } & w^\top \bm{e}_d = 1.
\end{array}
$$
In our implementation, we use a standard solver in Python package SciPy to solve the above optimization problem. Since it is a nonlinear, non-convex optimization, there may be numerical errors in the optimization routine, and the results may vary with the configuration of the algorithm. Our configuration is to set the maximum number of iteration limit to be 1000, and early terminate the solver when the maximum absolute deviation of the normalized asset-level risk contributions from the equal-risk target, $1/d$, falls below $10^{-10}$, i.e., when $\max_{1\le i\le d}\left|w_i(\widehat{\Sigma}w)_i/
	(w^\top \widehat{\Sigma}w)-1/d\right| < 10^{-10}$.

\subsection{Sample-based continuous-time mean--variance (ctmv)}
All the methods described so far in this section are for static models, while implemented dynamically on a rolling horizon basis. Our CTRL algorithms are  inherently for dynamic optimization; so we also include the classical model-based  continuous-time MV  method \citep{zhou2000continuous} for comparison purpose.
As with all the plug-in approaches,  this method does not explore nor update policy parameters. Instead, it estimates the mean vector  $\mu(t)$ and covariance matrix $\sigma(t)$ using  the 10-year historical stock data immediately prior to $t$, and then plug into the following formula for optimal policy \citep{zhou2000continuous}:
\begin{equation}
	\fitdisplay{\begin{aligned}
		\bm u^{*}(t, x; w^*)&=-\Sigma(t)^{-1}\left(\mu(t)-r(t)\right)(x-w^*)
		\label{optimal_policy2}
	\end{aligned}}
\end{equation}
where
\begin{equation}
	w^* = \frac{z e^{\left(\mu(t)-r(t)\right)^\top \Sigma(t)^{-1} \left(\mu(t)-r(t)\right) T} - x_0}{e^{\left(\mu(t)-r(t)\right)^\top \Sigma(t)^{-1} \left(\mu(t)-r(t)\right) T} - 1} .
\end{equation}

In our experiments, all the model coefficients are re-estimated at a rebalancing time point (monthly) using the most recent 10 years of stock price data.

\subsection{Predictive mean--variance (pmv)}

None of the classical methods described above employs any predictive models for expected returns. However, the empirical asset pricing literature highlights the predictive power of certain factors in forecasting stock returns \citep{lewellen2014cross}. Among the hundreds of factors documented in the literature \citep{cochrane2011presidential}, we focus on two of the most prominent ones as advocated by \citet{bali2016empirical}: the short-term reversal factor \citep{jegadeesh1990evidence, lehmann1990fads} and the medium-term momentum factor \citep{jegadeesh1993returns}.

The short-term reversal factor is among the strongest and most straightforward in empirical asset pricing \citep{bali2016empirical}. It is based on the empirical observation that top performers in a given month tend to underperform in the following month, while underperforming stocks often rebound. This reversal factor is defined as, simply,
\[
F_{rev}^i(t) = R^i(t),
\]
for stock \(i\) in month \(t\).

The medium-term momentum factor is based on investors’ often delayed responses and overreactions to information. The momentum of a stock \(i\) in month \(t\) is defined as the return of the stock during the 11-month period from months \(t-11\) to \(t-1\):
\[
F_{mom}^i(t) = \prod_{s=t-11}^{t-1}(1 + R^i(s)) - 1.
\]

We then employ a predictive linear regression model to estimate expected stock returns:
\[
R^{i}(t+1) = \alpha + \beta_{rev}^i F_{rev}^i(t) + \beta_{mom}^i F_{mom}^i(t) + \epsilon^i(t+1),
\]
for \(i = 1, 2, \ldots, d\). The coefficients \(\alpha\), \(\beta_{rev}^i\), and \(\beta_{mom}^i\) are estimated using the method of least squares over a 10-year rolling window of historical data, resulting in parameter estimates \(\hat{\alpha}\), \(\hat{\beta}_{rev}\), and \(\hat{\beta}_{mom}\).

To enhance accuracy of predictions and ensure alignment with established economic theory and empirical evidence, we impose constraints on the estimated coefficients based on their anticipated economic behaviors, as suggested by \citet{campbell2008predicting}. Specifically, the momentum coefficient \(\beta_{mom}^i\) is anticipated to be positive, indicating a persistence in returns, while the short-term reversal coefficient \(\beta_{rev}^i\) is expected to be negative, capturing the mean-reversion. Therefore, we modify \(\tilde{\beta}_{mom}:=\max\{ \hat{\beta}_{mom},0\}\) and  \(\tilde {\beta}_{rev}:=\min\{\hat{\beta}_{rev},0\}\). This adjustment results in the final predictive model:
\begin{equation}
	\label{eq_pmv_pred}
	R^{i}(t+1) = \hat{\alpha} + \tilde{\beta}_{rev}^i F_{rev}^i(t) + \tilde{\beta}_{mom}^i F_{mom}^i(t).
\end{equation}

Finally, the predicted returns obtained from \eqref{eq_pmv_pred} are plugged into the MV solution in \eqref{eq:optimal mv}.

\subsection{Two existing reinforcement learning algorithms}
In this subsection, we introduce  two existing state-of-the-art  RL algorithms: deep deterministic policy gradient and proximal policy optimization  which we will use to compare with our CTRL algorithm.

\paragraph{Deep deterministic policy gradient (DDPG)}
DDPG is a cutting-edge actor--critic algorithm designed for problems with continuous action spaces. It integrates the benefits of both DPG (deterministic policy gradient) and DQN (deep Q-network); see e.g. \cite{lillicrap2016continuous, mnih2015human}.
DDPG has the following important features:

\begin{itemize}
	\item \textbf{Architecture.} It employs two separate networks: the actor network that proposes an action given the current state, and the critic network that evaluates the proposed action by estimating the value function. The two networks are trained simultaneously.
	\item \textbf{Exploration strategy.} It carries out exploration by adding noises to  action output, often using Ornstein--Uhlenbeck processes, to facilitate efficient exploration of the action space.
	\item \textbf{Experience replay.} It utilizes experience replay, where a replay buffer stores past states, actions, and rewards. This technique improves sample efficiency and breaks correlations between consecutive learning steps.
	\item \textbf{Advantages for financial applications.} DDPG's ability to handle high-dimensional and continuous action spaces makes it particularly well-suited for financial applications including dynamic portfolio choice.
\end{itemize}

\paragraph{Proximal policy optimization (PPO)}
PPO has emerged as a popular choice in RL for its balance between performance and ease of implementation. It modifies traditional policy gradient approaches for improved stability and efficiency \citep{schulman2017proximal}.
PPO has the following important features:

\begin{itemize}
	\item \textbf{Objective function.} It introduces a clipped objective function that limits the size of policy updates. This approach reduces the likelihood of destructive large policy updates, ensuring more stable training.
	\item \textbf{Policy update.} It uses a policy update rule that keeps the new policy  not too far away from the old one (hence the term ``proximal").
	\item \textbf{Advantage estimation.} It often employs generalized advantage estimation (GAE) for calculating the advantage function, which helps reduce the variance of policy gradient estimates while retaining a bias.
	\item \textbf{Advantages for financial applications.} PPO's robustness and adaptability to various environments make it suitable for modeling complex financial systems, including   optimal portfolio strategies over a range of market conditions.
\end{itemize}


\section{Performance Metrics}
\label{appendix:performance metric}
\paragraph{Annualized return, volatility, and Sharpe ratio}
\label{metric}
We use $r_p$ to denote the return of the constructed portfolio. The annualized mean return rate $\mu_{p}=\E[r_p]$ and annualized volatility (standard deviation) $${\sigma}_{p}=\sqrt{\E[(r_p - \mu_p)^2]}$$ are two fundamental measures of portfolio performance. In Figure~\ref{fig:wealth} and all tables, whenever a wealth process becomes negative, we treat it as a bankruptcy event: the remaining wealth is set to zero, and the corresponding annualized return is set to $-100\%$.

The (annualized) Sharpe ratio is defined as
$$
\text{Sharpe Ratio} =\frac{\mu_{p}-r}{{\sigma}_{p}},
$$
which is a widely-used risk-adjusted return measure.

\paragraph{Sortino ratio}
The Sharpe ratio equally penalizes upside and downside volatilities, while investors often take upside volatility positively. The Sortino ratio addresses this by focusing on downside risk. It is defined as:
$$
\text{Sortino Ratio} =\frac{\mu_{p}-r}{{\sigma}_{downside}},
$$
where ${\sigma}_{\text{downside}}=\sqrt{\E[ \left( (r_p - \mu_p)^{-} \right)^2]} = \sqrt{\E[ \min\{0, r_p - \mu_p\}^2 ]}$ is the downside semi-deviation. The Sortino ratio offers a more nuanced evaluation of  risk-adjusted return.

\paragraph{Maximum drawdown (MDD), Calmar ratio, and recovery time (RT)}
Maximum drawdown (MDD) is another key metric for  downside risk. It measures the loss from the peak to the trough during a given period, relative to the peak value, and is defined as:
$$
\text{MDD}=\frac{\text{Peak Value} -\text{Trough Value}}{\text{Peak Value}}.
$$
The Calmar ratio provides a risk-adjusted return measure but uses MDD as the risk denominator:
$$
\text{Calmar Ratio} =\frac{\mu_{p}-r}{\text{MDD}}.
$$
Lastly, recovery time is the time (in days), {\it within a given testing period},  spent by a portfolio to rebound from its lowest point back to its previous peak. In our empirical results presented in all tables, for each strategy except the S\&P 500 index, if a wealth trajectory among the 100 independent simulations does not fully recover, we use the highest observed RT from the other trajectories as a substitute when computing the average RT.

\section{Trading and Exposure Statistics}
\label{appendix:trading metric}

\paragraph{Bankruptcy rate (BKR)}
The bankruptcy rate measures the proportion of experiments in which the wealth processes become non-positive during the testing period. As described in the main paper, once a wealth trajectory reaches zero, the remaining wealth is set to zero for the rest of the period. The bankruptcy rate is computed as the percentage of such events across the 100 independent experiments.

\paragraph{Turnover (TO)}
Turnover measures a portfolio's trading activity level. Let $\hat w_{j,t-}$ and $\hat w_{j,t+}$  denote respectively the portfolio weights in asset $j$ immediately before and after rebalancing at time $t$. The turnover at $t$ is defined as 
$$
\text{TO}_t = \sum_{j=1}^{d}\left|\hat w_{j,t+}-\hat w_{j,t-}\right|.
$$
We report the average monthly turnover over the testing period.

\paragraph{Position concentration (MaxL and MaxS)}
To evaluate concentration risk, we report the maximum single-asset long and short positions over the testing period. Specifically,
$$
\text{MaxL} = \max_{t,j} \hat w_{j,t+}^{+},
\qquad
\text{MaxS} = \max_{t,j} (-\hat w_{j,t+})^{+},
$$
where $x^{+}=\max(x,0)$. 

\paragraph{Gross exposure (GE and GEMax)}
Gross exposure measures the total positions defined as
$$
\text{GE}_t = \sum_{j=1}^d |\hat w_{j,t+}|.
$$
We report the time-averaged gross exposure (GE) as well as the maximum gross exposure observed over the testing period across all the 100 experiments (GEMax). Gross exposure reflects the overall leverage of a portfolio and its variability over time.

\section{Results Based on Simulated Data}\label{appendix:simulation-results}

This section presents the results of a simulation study on the baseline CTRL strategy shown in Algorithm~\ref{alg:offline} for which there are theoretically proved performance guarantees, and meanwhile one knows the ``ground truth" in a simulation (as opposed to an empirical study). As such, the purpose of this study is to demonstrate that the convergence of the related parameters, its speed and the regret bound closely match the theoretical results.

\subsection{Experiment setup}
\label{subsec:experimental-setup}

Our experiment simulates a two-stock market environment, with each stock's price following a geometric Brownian motion. The model parameters are set as follows: drift vector $\mu = (0.2, 0.3)^\top$, marginal volatilities $0.3$ and $0.4$ with a correlation coefficient of $0.1$, risk-free rate $r = 0.02$, initial wealth $x_0= 1$, investment horizon $T = 1$, target expected terminal wealth $z = 1.4$, and temperature parameter $\gamma = 0.1$. The time discretization is set to be $\Delta t = 0.004$, and the total number of episodes is $10^5$.

\subsection{Convergence rate and regret bound}

\begin{figure}[htbp]
\centering
\includegraphics[trim=0 9.25cm 0 9.5cm, clip, width=0.9\textwidth]{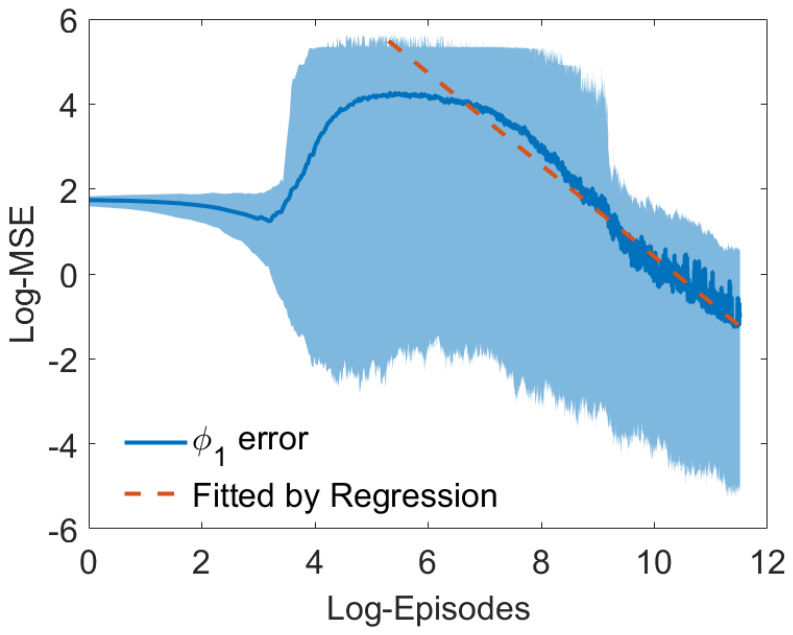}
\caption{\textbf{Error of parameter $\phi_1$} The solid curves and the upper and lower boundaries of the shaded regions represent the average, 2.5\% and 97.5\% percentile of the error over 1000 independent simulation runs, respectively. The slope for $\phi_1$ by least squares regression is -1.09.  The vertical and horizontal axes are on natural log-scale.}
\label{fig:phi_1}
\end{figure}

\begin{figure}[htbp]
\centering
\includegraphics[trim=0 9.25cm 0 9.5cm, clip, width=0.9\textwidth]{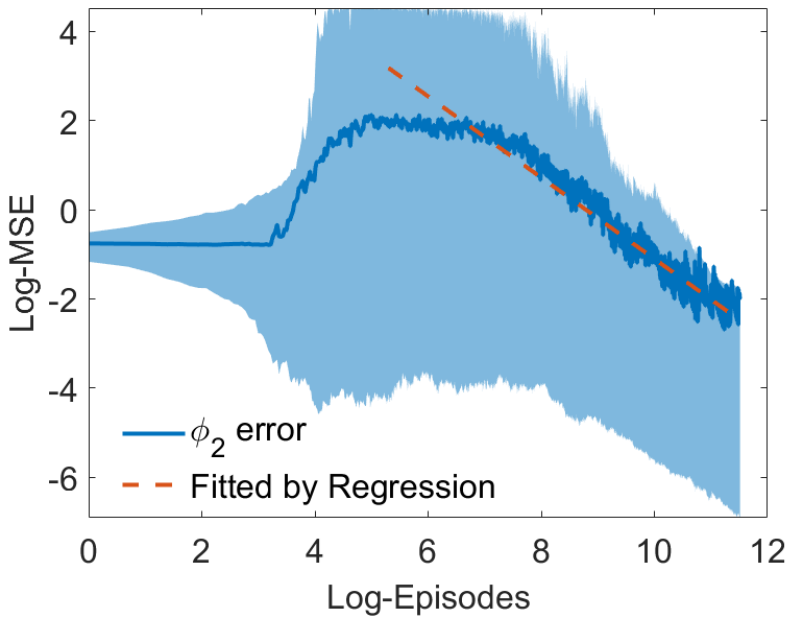}
\caption{\textbf{Error of parameter $\phi_2$} The solid curves and the upper and lower boundaries of the shaded regions represent the average, 2.5\% and 97.5\% percentile of the error over 1000 independent simulation runs, respectively. The slope for $\phi_2$ by least squares regression is -0.91.  The vertical and horizontal axes are on natural log-scale.}
\label{fig:phi_2}
\end{figure}

\begin{figure}[htbp]
\centering
\includegraphics[trim=0 9.25cm 0 9.5cm, clip, width=0.9\textwidth]{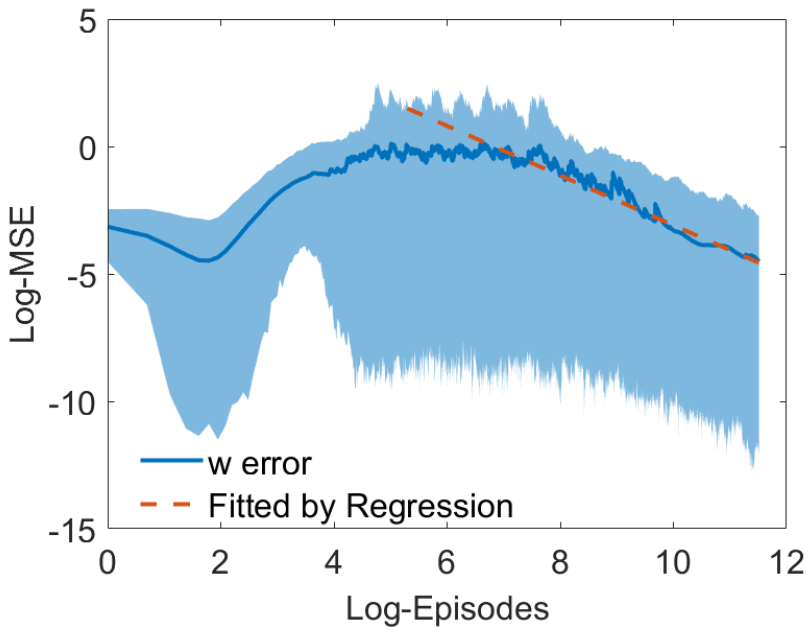}
\caption{\textbf{Error of parameter $w$} The solid curves and the upper and lower boundaries of the shaded regions represent the average, 2.5\% and 97.5\% percentile of the error over 1000 independent simulation runs, respectively. The slope for $w$ by least squares regression is -0.97.  The vertical and horizontal axes are on natural log-scale.}
\label{fig:w}
\end{figure}

Algorithm \ref{alg:offline} is initialized with $\bm\theta = (0,0)^\top$, $\phi_1 = (0, 0)^\top$, $\phi_2 = I$, and $w = 1.5$. A total of 1000 independent simulation runs are conducted independently. 
As we know the oracle values of $\phi_1$, $\phi_2$, and $w$, we can compute the mean-squared errors (MSEs) of these learned parameters against number of episodes, both in log scale.
Figures \ref{fig:phi_1}, \ref{fig:phi_2}, and \ref{fig:w} indicate  that the learned  parameters $\phi_1$, $\phi_2$, and $w$ all converge, and converge rapidly  after certain numbers of episodes.
Moreover, by Theorem \ref{thm:convergence_both}, the theoretical convergence rate of $\phi_{1,n}$ is $\frac{(\log n)^p \log \log n}{n}$ under the configuration specified in Remark \ref{remark1}. On a log scale, this corresponds to a slope close to -1, because $\log{\frac{(\log n)^p \log \log n}{n}} = -\log n + p\log\log n + \log\log\log n$, of which the first term dominates when $n$ is large.  Figure \ref{fig:phi_1} shows that the fitted slope of the log average error against log number of episodes for $\phi_1$ is -1.09, closely approximating the theoretical one. While theoretical convergence bounds for the iterates $\phi_{2,n}$ and $w_n$ are not yet available, Figures \ref{fig:phi_2} and \ref{fig:w} show fitted slopes of -0.91 and -0.97 respectively, yielding  convergence rates of these two parameters of close to $1/n$.

\begin{figure}[htbp]
\centering
\includegraphics[trim=0 9.25cm 0 9.5cm, clip, width=1\textwidth]{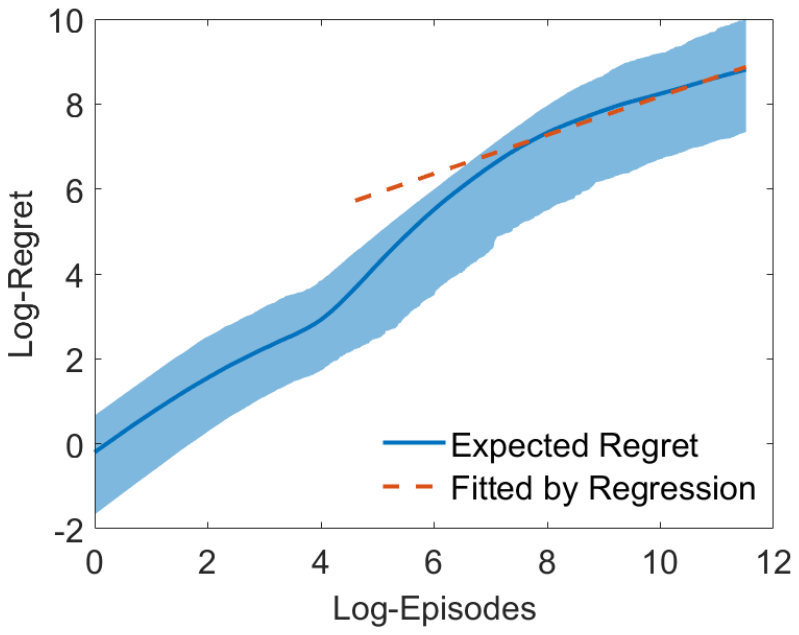}
\caption{\textbf{Cumulative regret rate in number of episodes.} The solid blue curve and the upper and lower boundary of the shaded region represent the mean, 2.5\% and 97.5\% percentile of the regret over 1000 independent simulation runs, respectively. The red dashed line is the fitted value by linearly regressing the log average regret against the logarithm of the number of episodes starting from the 100th episode. The fitted slope by least squares regression is $0.456$.  The vertical and horizontal axes are on natural log-scale.}
\label{fig:regret}
\end{figure}

On the other hand, Theorem \ref{thm:regret0} stipulates that the theoretical regret bound of Algorithm \ref{alg:offline} is $\sqrt{N (\log N)^p \log\log N}$ under the setting of Remark \ref{remark1}. On a log scale, this corresponds approximately to a slope close to 0.5, because $\log{\sqrt{N (\log N)^p \log\log N}} = \frac{1}{2}(\log N + p\log\log N + \log\log\log N)$. Figure \ref{fig:regret} shows that the fitted slope of regret is 0.520 (on log scale), again very close to the theoretical one.

\subsection{Numerical Illustration of Exploration–-Exploitation Tradeoff}
\label{appendix:tradeoff}

To numerically investigate the exploration–-exploitation tradeoff, we examine how the variance of the updating direction  \(Z_1\) of the parameter \(\phi_1\), varies with the exploration parameter \(\phi_2\). With the same experimental setup as outlined  in Section \ref{subsec:experimental-setup}, we plot \(\log(|\operatorname{Var}(Z_1)|)\) against \(\log(|\phi_2^{-1}|)\) to visualize the interplay between the two.

\begin{figure}[h!]
\centering
\includegraphics[width=0.8\textwidth]{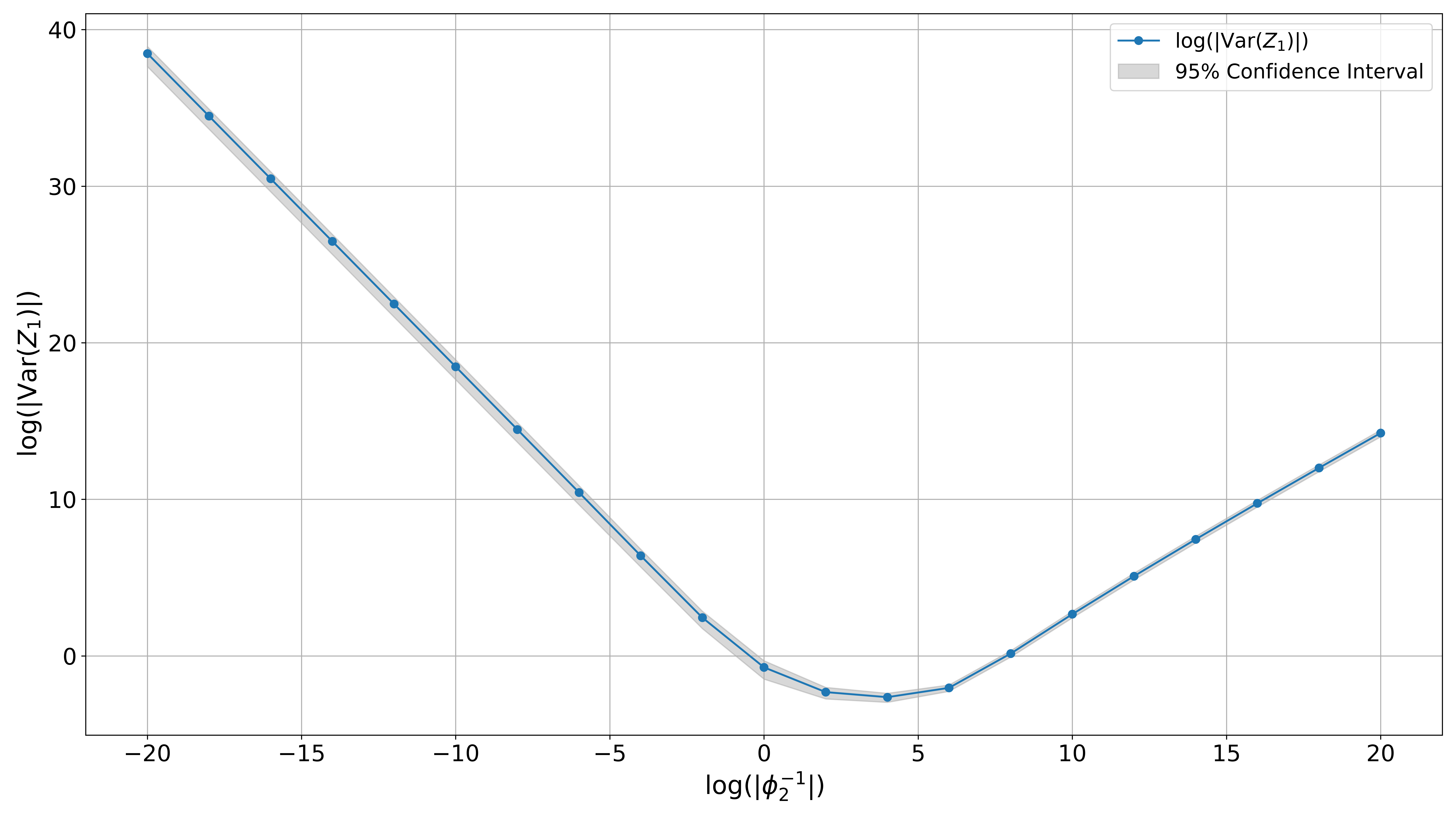}
\caption{Log–log plot of \(|\operatorname{Var}(Z_1)|\) versus \(|\phi_2^{-1}|\), showing how the exploration level \(\phi_2\) affects the stability of the update for \(\phi_1\). The shaded area represents the 95\% confidence interval.}
\label{fig:tradeoff}
\end{figure}

Figure~\ref{fig:tradeoff} shows a U-shaped relationship between \(\log(|\operatorname{Var}(Z_1)|)\) and \(\log(|\phi_2^{-1}|)\). The variance increases when \(\phi_2\) is either too small or too large. A very small \(\phi_2\) renders a highly deterministic policy, leading  to large variances in the updates due to inadequacy of data. An excessively large \(\phi_2\) introduces too much noise, also causing high variances. Both cases hamper learning efficiency. This highlights the need of a careful exploration--exploitation balance for optimal learning.

\section{Proofs of Statements}\label{appendix:proof}


\subsection{Proof of Theorem \ref{thm:tradeoff}}
\label{appendix:tradeoff_thm}

In this subsection, we present the proof of Theorem~\ref{thm:tradeoff}, which characterizes the mean and variance of the update direction for the parameter \(\phi_1\). We also carry out a similar analysis for \(\phi_2\) and \(w\). Understanding the update behavior of all three parameters is essential for establishing the convergence and regret results that follow.

The full expression of \eqref{eq:mean for update phi 1} in Theorem \ref{thm:tradeoff} is
\[
R(\phi_{1}, \phi_{2}, w)= 2 \left[\frac{(x_0-w)^2 e^{-\phi_{3} T}(e^{Q(\phi_{1})T} - 1)}{Q(\phi_1)} + \frac{\langle \sigma \sigma^\top, \phi_{2} \rangle (e^{Q(\phi_{1})T} - 1-Q(\phi_{1})T)}{Q(\phi_{1})^2}\right],
\]
where
\[
Q(\phi_{1}) = -2(\mu-r)^\top \phi_{1} + \langle \sigma \sigma^\top, \phi_{1}\phi_1^\top \rangle +\phi_{3}.
\]

We start by examining the agent's discounted wealth process \eqref{classical_dynamics} in the $n$-th iteration satisfies the wealth equation
\begin{equation}
\label{eq:classical_SDE}
\dd x_n(t) = (\mu - r\bm{e}_d)^\top u_n(t) \dd t + u_n(t)^\top \sigma \dd W_n(t), ~~~~ 0 \leq t \leq T;\;\;x_n(0)=x_0,
\end{equation}
where $W_n$ is a Brownian motion in the $n$-th iteration, and (with a slight abuse of notation) $u_n(t)=u_n(t,x_n(t))$ while  $u_n(t,x) \sim \mathcal{N}\left( -\phi_{1,n} (x - w_n) , \phi_{2,n} e^{\phi_3(T-t)} \right)$ independent of $W_n$.

Recall that $\theta_3=\phi_3$ is fixed and not updated in our algorithm, and that $Z_{1,n}(T)$ and $Z_{2,n}(T)$ are defined in \eqref{eq:z1_def1} and \eqref{eq:z2_def1}.
Denote by $\xi_n = (\xi_{1, n}, \xi_{2, n})^{\top}$ the ``noise'' parts of these random variables, namely,
\[
\xi_{1,n+1} = Z_{1,n}(T) - h_1(\phi_{1,n}, \phi_{2,n}, w_n) \;\mbox{ where } h_1(\phi_{1,n}, \phi_{2,n}, w_n) = \E\left[ Z_{1,n}(T) \big| \bm\theta_n, \bm\phi_n, w_n \right],
\]
\[
\xi_{2,n+1} = Z_{2,n}(T) - h_2(\phi_{1,n}, \phi_{2,n}, w_n) \;\mbox{ where } h_2(\phi_{1,n}, \phi_{2,n}, w_n) = \E\left[ Z_{2,n}(T) \big| \bm\theta_n, \bm\phi_n, w_n \right].
\]
Similarly, define $\xi_{w, n}\in \mathbb{R}$ as the noise counterpart in updating $w$:
\[
\xi_{w,n+1} = x_{n}(T)-z - h_{w}(\phi_{1,n}, \phi_{2,n}, w_n)  \;\mbox{ where } h_w(\phi_{1,n}, \phi_{2,n}, w_n) = \E\left[x_{n}(T)-z \big| \bm\theta_n, \bm\phi_n, w_n \right].
\]
Then the updating rules for $\bm\phi$ and $w$ can be rewritten as
\begin{equation}
\fitdisplay{\begin{aligned}
	\label{eq:projected}
	\phi_{1,n+1} &= \Pi_{K_{1, n+1}}\left( \phi_{1,n} - a_n [h_1(\phi_{1,n},\phi_{2,n},w_n;\phi_3) + \xi_{1,n+1}] \right),\\
	\phi_{2,n+1} &= \Pi_{K_{2, n+1}}\left( \phi_{2,n} + a_n [h_2(\phi_{1,n},\phi_{2,n},w_n;\phi_3) + \xi_{2,n+1}] \right),\\
	w_{n+1} &= \Pi_{K_{w, n+1}}\left( w_n - a_{w, n} [h_w(\phi_{1,n},\phi_{2,n},w_n;\phi_3) + \xi_{w, n+1}] \right).
\end{aligned}}
\end{equation}


We divide the rest of the proof into several steps.

\subsubsection*{Moment estimates.}
First we establish the moment expressions and estimates for the wealth trajectory under the policy \eqref{eq:policy parameterize}.
\begin{lemma}
\label{lemma_xt-w}
Let $\{\tilde x^{\bm\phi}(t):0\leq t\leq T\}$ be the wealth trajectory under the policy \eqref{eq:policy parameterize}. Then we have
\begin{equation}\label{eq}
	\fitdisplay{\begin{aligned}
		\E[\tilde x^{\bm\phi}(t)-w] = &(x_0-w)e^{-(\mu-r)^\top \phi_1 t},\\
		\E [(\tilde x^{\bm\phi}(t) - w)^2] =&\left[(x_0-w)^2 + \frac{\langle \Sigma, \phi_{2}\rangle e^{\phi_3T}}{-2 (\mu-r)^\top \phi_{1} + \langle \Sigma , \phi_{1} \phi_{1}^\top\rangle + \phi_3}\right]e^{(-2 (\mu-r)^\top \phi_{1} + \langle \Sigma , \phi_{1} \phi_{1}^\top\rangle)t} \\
		&- \frac{\langle \Sigma, \phi_{2}\rangle e^{\phi_3(T-t)}}{-2 (\mu-r)^\top \phi_{1} + \langle \Sigma , \phi_{1} \phi_{1}^\top\rangle +\phi_3}.
	\end{aligned}}
\end{equation}
Moreover, there exists a constant $C > 0$ that only depends on $\mu,r,x_0,T$ and $\Sigma$ such that we have
\begin{equation}\label{ineq}
	\fitdisplay{\begin{aligned}
		&\E[(\tilde x^{\bm\phi}(t) - w)^2] \leq C  (1 + |w|^2 + |\phi_2|) \exp{(C |\phi_1|^2 t)},\\
		&\operatorname{Var}\left( \tilde x^{\bm\phi}(t)\right) \leq C  (1 + |w|^2 + |\phi_2|) \exp{(C |\phi_1|^2 t)},\\
		&	\E[(\tilde x^{\bm\phi}(t) - w)^4] \leq C  (1 + |w|^4 + |\phi_2|^2) \exp{(C |\phi_1|^4 t)},\\
		&		\E[(\tilde x^{\bm\phi}(t) - w)^8] \leq C(1 + |w|^8 + |\phi_2|^4)\exp{(C |\phi_1|^8 t)}.
	\end{aligned}}
\end{equation}
\end{lemma}
\begin{proof}

Denote $\hat{x}(t) = \tilde x^{\bm\phi}(t) - w$. It follows from \eqref{rl_dynamics} that
\begin{equation}
	\label{eq:ito1}
	\hat{x}(t) =x_0 - w + \int_0^t -(\mu - r)^\top\phi_1 \hat{x}(s) \dd s +\int_0^t \sqrt{\hat{x}(s)^{2}\phi_1^\top \Sigma\phi_1 + \langle \Sigma, \phi_2 e^{\phi_3(T-s)} \rangle} \dd W(s) .
\end{equation}
Taking  expectation on both sides and solving the resulting ODE, we obtain the first equation of \eqref{eq}.
%
%
%
%

Apply It\^o's lemma to $\hat{x}^2(t)$ in \eqref{eq:ito1} and take expectation on both sides to obtain

\[
\E[\hat{x}^2(t)] = (x_0-w)^2 + \E \int_0^t [-2(\mu-r)^\top \hat{x}^2(s) + \langle \Sigma, \phi_1\phi_1^\top \hat{x}^2(s) + \phi_2e^{\phi_3(T-s)} \rangle] \dd s .
\]
Solving the above ODE in $\E[\hat{x}^2(\cdot)]$ we obtain the second equation of
\eqref{eq}.

%
%
%

Next, we take the eighth power and then apply expectation on both sides of \eqref{eq:ito1}. By H\"older's inequality, we have
\begin{equation}
	\fitdisplay{\begin{aligned}
		\E [\hat{x}(t)^8] = &\E\left[ \biggl(x_0 - w + \int_0^t -(\mu - r)^\top\phi_1 \hat{x}(s) \dd s + \int_0^t \sqrt{\hat{x}(s)^{2}\phi_1 \Sigma\phi_1 + \langle \Sigma, \phi_2 e^{\phi_3(T-s)} \rangle} \dd W(s) \biggl)^8 \right] \\
		\leq &C|x_0 - w|^8 +  C  \E\left[ \biggl(\int_0^t -(\mu - r)^\top\phi_1 \hat{x}(s) \dd s\biggl)^8 \right] + C \E \left[ \biggl(\int_0^t \sqrt{\hat{x}(s)^{2}\phi_1 \Sigma\phi_1 + \langle \Sigma, \phi_2 e^{\phi_3(T-s)} \rangle} \dd W(s) \biggl)^8\right] \\
		\leq &C|x_0 - w|^8 + C  ((\mu - r)^\top\phi_1)^8 \E\left[ (\int_0^t \hat{x}(s) \dd s)^8 \right] +
		C  \E\left[ (\int_0^t\hat{x}(s)^{2}\phi_1 \Sigma\phi_1 + \langle \Sigma, \phi_2 e^{\phi_3(T-s)} \rangle \dd s)^4 \right]\\
		\leq  &C|x_0 - w|^8 +  C |\phi_1|^8  \E\left[ \int_0^t \hat{x}(s)^8 \dd s \right] + C \E\left[ \int_0^t \hat{x}(s)^8|\phi_1|^8 + |\phi_2|^4 \dd s\right] .
	\end{aligned}}
\end{equation}

Gronwall's inequality thus leads to the fourth inequality of \eqref{ineq}.
The similar argument can be applied to prove the remaining inequalities of \eqref{ineq} for the moment estimate of the second and fourth orders. In particular, $	\operatorname{Var}\left( \tilde x^{\bm\phi}(t)\right) \leq \E[(\tilde x^{\bm\phi}(t) - w)^2]$.

\end{proof}
Next, we estimate the variances of the increments $Z_{1,n}(T)$ and $Z_{2,n}(T)$ defined  in \eqref{eq:z1_def1} and \eqref{eq:z2_def1} respectively.

\begin{lemma}
\label{lemma_xi_upper_z}

There exists a constant $C>0$ such that
\begin{equation}
	\label{eq:noise_upper}
	\fitdisplay{\begin{aligned}
		|\operatorname{Var}\left( Z_{1,n}(T) \Big| \bm\theta_n, \bm\phi_n, w_n \right)| & \leq C \bigg( 1 + |w_n|^{16} + |\phi_{1,n}|^8 + |\phi_{2,n}|^8 + |b_n|^8 \bigg) e^{C|\phi_{1,n}|^8}.\\
		|\operatorname{Var}\left( Z_{2,n}(T) \Big| \bm\theta_n, \bm\phi_n, w_n \right)| &\leq C  \bigg( 1 + |w_n|^{16} + |\phi_{1,n}|^8 + |\phi_{2,n}|^8\bigg) e^{C|\phi_{1,n}|^8}.
	\end{aligned}}
\end{equation}
\end{lemma}

\begin{proof}
We first derive the dynamics of  $\{(Z_{1, n}(t),Z_{2, n}(t)):0\leq t\leq T\}$.
Applying It\^o's lemma  we obtain
\begin{equation}
	\fitdisplay{\begin{aligned}
		d J(t, x_n(t); w_n; \bm\theta_n) = &\biggl( \frac{\partial J(t, x_n(t); w_n; \bm\theta_n)}{\partial t} + (\mu - re_d)^\top u_n(t) \frac{\partial J(t, x_n(t); w_n; \bm\theta_n)}{\partial x} \\
		+& \frac{u_n(t)^\top \Sigma u_n(t)}{2} \frac{\partial^2 J(t, x_n(t); w_n; \bm\theta_n)}{\partial x^2} \biggl)dt + u_n(t)^\top \sigma \frac{\partial J(t, x_n(t); w_n; \bm\theta_n)}{\partial x} dW_n(t).
	\end{aligned}}
\end{equation}
Noting the explicit forms \eqref{eq:value function parameterize} and \eqref{eq:policy parameterize}, we deduce from \eqref{eq:z1_def1} and \eqref{eq:z2_def1} that
{\small
	\begin{equation}
		\label{eq:dz1}
		\fitdisplay{\begin{aligned}
			&\dd Z_{1,n}(t)\\ = &\frac{\partial \log \pi (u_n(t) | t, x_n(t); w_n; \phi_n)}{\partial \phi_{1}} \biggl[ \biggl(\theta_3(x_n(t)-w_n)^2 e^{-\theta_3(T-t)} + 2(x_n(t)-w_n)e^{-\theta_3(T-t)}(\mu -re_d)^\top u_n(t) + \\
			&e^{-\theta_3(T-t)} u_n(t)^\top \sigma \sigma^\top u_n(t) + 2\theta_{2,n} t + \theta_{1,n} \biggl)\dd t + 2(x_n(t)-w_n) u_n(t)^\top \sigma e^{-\theta_3(T-t)} \dd W(t)  + \gamma p^\phi (t) \dd t\biggl] + \gamma \frac{\partial p^\phi (t)}{\partial \phi_{1}}\dd t\\
			=&-e^{-\phi_3(T-t)}\phi_{2,n}^{-1} \left[(x_n(t)-w_n) u_n(t) + (x_n(t)-w_n)^2  \phi_{1,n} \right] \\
			&\times \biggl[ \biggl(\theta_3(x_n(t)-w_n)^2 e^{-\theta_3(T-t)} + 2(x_n(t)-w_n)e^{-\theta_3(T-t)}(\mu -re_d)^\top u_n(t) + e^{-\theta_3(T-t)} u_n(t)^\top \sigma \sigma^\top u_n(t) + 2\theta_{2,n} t + \theta_{1,n} \biggl)\dd t\\
			&+ 2(x_n(t)-w_n) u_n(t)^\top \sigma e^{-\theta_3(T-t)} \dd W(t)  + \gamma (-\frac{d}{2} \log{(2 \pi e)} + \frac{1}{2} \log (\det( \phi_{2,n} ^{-1})) \dd t - \frac{d}{2} \phi_3 (T-t)) \dd t\biggl]\\
			=&-e^{-\phi_3(T-t)}\phi_{2,n}^{-1} \left[(x_n(t)-w_n) u_n(t) + (x_n(t)-w_n)^2  \phi_{1,n} \right]\\
			&\times \biggl[ \biggl(\theta_3(x_n(t)-w_n)^2 e^{-\theta_3(T-t)} + 2(x_n(t)-w_n)e^{-\theta_3(T-t)}(\mu -re_d)^\top u_n(t) + e^{-\theta_3(T-t)} u_n(t)^\top \sigma \sigma^\top u_n(t) + 2\theta_{2,n} t + \theta_{1,n} \biggl)\\
			& + \gamma (-\frac{d}{2} \log{(2 \pi e)} + \frac{1}{2} \log (\det( \phi_{2,n} ^{-1})) - \frac{d}{2} \phi_3 (T-t))\biggl ] \dd t\\
			& -2e^{-\phi_3(T-t)}\phi_{2,n}^{-1} \left[(x_n(t)-w_n) u_n(t) + (x_n(t)-w_n)^2  \phi_{1,n} \right] (x_n(t)-w_n) u_n(t)^\top \sigma e^{-\theta_3(T-t)} \dd W_n(t)\\
			\triangleq &Z_{1,n}^{(1)}(t) \dd t + Z_{1,n}^{(2)}(t) \dd W_n(t),
		\end{aligned}}
\end{equation}}
and
{\small
	\begin{equation}
		\label{eq:dz2}
		\fitdisplay{\begin{aligned}
			&\dd Z_{2,n}(t)\\ = &\frac{\partial \log \pi (u_n(t)|t,x_n(t);w_n;\phi_n)}{\partial \phi_{2}^{-1}} \biggl[ \biggl(\theta_3(x_n(t)-w_n)^2 e^{-\theta_3(T-t)} + 2(x_n(t)-w_n)e^{-\theta_3(T-t)}(\mu -re_d)^\top u_n(t)  \\
			&+e^{-\theta_3(T-t)} u_n(t)^\top \sigma \sigma^\top u_n(t) + 2\theta_{2,n} t + \theta_{1,n} \biggl)\dd t + 2(x_n(t)-w_n) u_n(t)^\top \sigma e^{-\theta_3(T-t)} \dd W(t)  + \gamma p^\phi (t) \dd t\biggl] + \gamma \frac{\partial p^\phi (t)}{\partial \phi_{2}^{-1}}\dd t\\
			= &\left[\frac{1}{2} \phi_{2,n} - \frac{1}{2} e^{-\phi_3(T-t)} (u_n(t) + \phi_{1,n}(x_n(t)-w_n)) (u_n(t) + \phi_{1,n}(x_n(t)-w_n))^\top\right] \\
			&\times \biggl[ \biggl(\theta_3(x_n(t)-w_n)^2 e^{-\theta_3(T-t)} + 2(x_n(t)-w_n)e^{-\theta_3(T-t)}(\mu -re_d)^\top u_n(t) + e^{-\theta_3(T-t)} u_n(t)^\top \sigma \sigma^\top u_n(t) + 2\theta_{2,n} t + \theta_{1,n} \biggl)\dd t\\
			&+ 2(x_n(t)-w_n) u_n(t)^\top \sigma e^{-\theta_3(T-t)} \dd W(t)  + \gamma (-\frac{d}{2} \log{(2 \pi e)} + \frac{1}{2} \log (\det( \phi_{2,n} ^{-1})) \dd t - \frac{d}{2} \phi_3 (T-t)) \dd t\biggl] + \gamma \frac{\phi_{2,n}}{2} \dd t\\
			=&\left[\frac{1}{2} \phi_{2,n} - \frac{1}{2} e^{-\phi_3(T-t)} (u_n(t) + \phi_{1,n}(x_n(t)-w_n)) (u_n(t) + \phi_{1,n}(x_n(t)-w_n))^\top\right]\\
			& \times \biggl\{\biggl[ \biggl(\theta_3(x_n(t)-w_n)^2 e^{-\theta_3(T-t)} + 2(x_n(t)-w_n)e^{-\theta_3(T-t)}(\mu -re_d)^\top u_n(t) + e^{-\theta_3(T-t)} u_n(t)^\top \sigma \sigma^\top u_n(t) + 2\theta_{2,n} t + \theta_{1,n} \biggl)\\
			& + \gamma (-\frac{d}{2} \log{(2 \pi e)} + \frac{1}{2} \log (\det( \phi_{2,n} ^{-1})) - \frac{d}{2} \phi_3 (T-t))\biggl ] + \gamma \frac{\phi_{2,n}}{2} \biggl\} \dd t\\
			& +2\left\{\frac{1}{2} \phi_{2,n} - \frac{1}{2} e^{-\phi_3(T-t)} [u_n(t) + \phi_{1,n}(x_n(t)-w_n)] [u_n(t) + \phi_{1,n}(x_n(t)-w_n)]^\top\right\}) (x_n(t)-w_n) u_n(t)^\top \sigma e^{-\theta_3(T-t)} \dd W_n(t)\\
			\triangleq & Z_{2,n}^{(1)}(t) \dd t + Z_{2,n}^{(2)}(t) \dd W_n(t).
		\end{aligned}}
\end{equation}}

Noting that $u_n(t)\equiv u_n(t,x_n(t))$ while  $u_n(t,x) \sim \mathcal{N}\left( -\phi_{1,n} (x - w_n) , \phi_{2,n} e^{\phi_3(T-t)} \right)$, we can easily upper bound $|Z_{1,n}^{(1)}|^2$ and $|Z_{1,n}^{(2)}|^2$ by
\[
\fitdisplay{\begin{aligned}
	\E[|Z_{1,n}^{(1)}(t)|^2 | \bm\theta_n,\bm\phi_n,w_n,x_n(t)]  \leq & C \bigg[1 + (x_n(t)-w_n)^4 |\phi_{2,n}^{-1}|^2 + (x_n(t)-w_n)^8 |\phi_{1,n}|^2|\phi_{2,n}^{-1}|^2 \\
	&+ (x_n(t)-w_n)^8 + (x_n(t)-w_n)^8 |\phi_{1,n}|^4 + (\log{\det(\phi_{2,n}^{-1})})^4\bigg],
\end{aligned}}
\]
\[
\fitdisplay{\begin{aligned}
	\E[|Z_{1,n}^{(2)}(t)|^2 | \bm\theta_n,\bm\phi_n,w_n,x_n(t)] \leq &C \bigg[1 + (x_n(t)-w_n)^4 |\phi_{2,n}^{-1}|^2 + (x_n(t)-w_n)^8 |\phi_{1,n}|^2|\phi_{2,n}^{-1}|^2 \\
	&+ (x_n(t)-w_n)^4 |\phi_{1,n}|^4 + (x_n(t)-w_n)^4 |\phi_{2,n}|^2\bigg].
\end{aligned}}
\]


Then we conclude from  Lemma \ref{lemma_xt-w} that
\[	\fitdisplay{\begin{aligned}
&\E[(|Z_{1,n}^{(1)}(t)|^2 + |Z_{1,n}^{(2)}(t)|^2) | \bm\theta_n,\bm\phi_n,w_n]\\
\leq & C \bigg[1 + (1 + |w_n|^4 + |\phi_{2,n}|^2) \exp{(C |\phi_{1,n}|^4t)}(|\phi_{2,n}^{-1}|^2 + |\phi_{2,n}|^2 + |\phi_{1,n}|^4) \\
&+ (1 + |w_n|^8 + |\phi_{2,n}|^4)\exp(C|\phi_{1,n}|^8t)(1 + |\phi_{2,n}^{-1}|^4 + |\phi_{1,n}|^4)) + (d \log{|\phi_{2,n}^{-1}|})^4\bigg]\\
\leq& C \bigg( 1 + |w_n|^{16} + |\phi_{1,n}|^8 + |\phi_{2,n}|^8 + |\phi_{2,n}^{-1}|^8 \bigg) e^{C|\phi_{1,n}|^8}.
\end{aligned}}
\]
This leads to the second part of Theorem~\ref{thm:tradeoff}.

By virtue of  the projection $\phi_{2,n} \succeq \frac{1}{b_n} I$, or $|\phi_{2,n}^{-1}| \leq b_n$, we further obtain
\[	\fitdisplay{\begin{aligned}
\E[(|Z_{1,n}^{(1)}(t)|^2 + |Z_{1,n}^{(2)}(t)|^2) | \bm\theta_n,\bm\phi_n,w_n] \leq C \bigg( 1 + |w_n|^{16} + |\phi_{1,n}|^8 + |\phi_{2,n}|^8 + |b_n|^8 \bigg) e^{C|\phi_{1,n}|^8},
\end{aligned}}
\]
leading to the first inequality of \eqref{eq:noise_upper}. The second inequality of \eqref{eq:noise_upper} can be proved similarly.
\end{proof}


\subsubsection*{Explicit expressions of mean increments}
Next, we derive the analytical forms of the functions $h_1,h_2,h_w$, which are the means of the increments in the algorithms approximating $\phi_1,\phi_2$ and $\phi_w$ respectively. 

To start, note that $\{Z_{1,n}^{(2)}(t):0\leq t\leq T\}$ and $\{Z_{2,n}^{(2)}(t):0\leq t\leq T\}$ are both square integrable based on the proof of  Lemma \ref{lemma_xi_upper_z} along with the moment estimates in Lemma \ref{lemma_xt-w}. Thus, when we integrate
\eqref{eq:dz1} and \eqref{eq:dz2} and take expectation, the It\^o integrals vanish. Denote
\[
\fitdisplay{\begin{aligned}
&A_n = (\mu-r)^\top \phi_{1,n},\;\;B_n=\langle \sigma \sigma^\top , \phi_{1,n} \phi_{1,n}^\top\rangle,\;\;E_n = \langle \sigma \sigma^\top, \phi_{2,n} e^{\phi_3(T-t)} \rangle, \\
&G=e^{-\theta_3(T-t)},\;\;H_n = -\frac{d}{2} \log{(2 \pi e)} + \frac{1}{2} \log ( \det(\phi_{2,n} ^{-1}) ),\;\;P_n = 2 \theta_{2,n} t - \frac{\gamma d}{2}\phi_3(T-t) + \theta_{1,n} + \gamma H_n.
\end{aligned}}
\]
Then it follows from \eqref{eq:dz1} and \eqref{eq:dz2} that
{\small \[
\fitdisplay{\begin{aligned}
	\dd \E [Z_{1, n}(t)] =& \E\biggl \{ \frac{\partial \log \pi\left(u_n(t) \mid t, x_n(t) ; w_n; \bm\phi_n\right)}{\partial \phi_1}
	\left[\mathrm{d} J\left(t, x_n(t) ; w_n ; \bm\theta_n \right)+\gamma \hat{p}\left(t, x_n(t), \bm\phi_n\right) \mathrm{d} t\right]+
	\gamma \frac{\partial \hat{p}}{\partial \phi_1}\left(t, x_n(t), \bm\phi_n \right) \mathrm{d} t  \biggl \}\\
	=& \E \biggl \{ -G [(x_n(t)-w_n) \phi_{2, n}^{-1} u_n(t) + (x_n(t) - w_n)^2 \phi_{2, n}^{-1} \phi_{1, n}] \\
	&\times [\theta_{3} G (x_n(t)-w_n)^2 + 2 G (x_n(t) - w) (\mu - r)^{\top} u_n(t) + G (x_n(t) - w) u_n(t) \langle \sigma \sigma^\top, u_n(t)u_n(t)^\top \rangle + P_n ] \dd t\biggl \}\\
	=&\E \biggl\{ \{-G\phi_{2,n}^{-1}[\theta_3G(x_n(t)-w_n)^3u_n(t) + 2G(x_n(t)-w_n)^2u_n(t)u_n(t)^\top(\mu-r)\\
	&+ G(x_n(t)-w_n)u_n(t)\langle \Sigma, u_n(t)u_n(t)^\top\rangle + P_n(x_n(t)-w_n)u_n(t)]\}\dd t \\
	&+\{-G\phi_{2,n}^{-1}\phi_{1,n} [\theta_3G(x_n(t)-w_n)^4 + 2G(x_n(t)-w_n)^3(\mu-r)^\top u_n(t)\\
	&+G(x_n(t)-w_n)^2\langle \sigma \sigma^\top, u_n(t)u_n(t)^\top \rangle + P_n(x_n(t)-w_n)^2]\} \dd t\biggl\}\\
	=&\E (x_n(t) - w_n)^2 [2 G (-(\mu-r) + \sigma\sigma^\top \phi_{1,n})] \dd t,
\end{aligned}}
\]}
and
{\small
\[
\fitdisplay{\begin{aligned}
	\dd \E [Z_{2, n}(t)] =& \E\biggl \{ \frac{\partial \log \pi\left(u_n(t) \mid t, x_n(t) ; w_n, \bm \phi_n\right)}{\partial \phi_2^{-1}}
	\left[\mathrm{d} J\left(t, x_n(t) ; w_n; \bm \theta_n\right)+\gamma \hat{p}\left(t, x_n(t), \bm\phi_n\right) \mathrm{d} t\right]+
	\gamma \frac{\partial \hat{p}}{\partial \phi_2^{-1}}\left(t, x_n(t), \phi\right) \dd t \biggl \}\\
	=&\E\biggl \{ [\frac{1}{2} \phi_{2,n} - \frac{1}{2} G (u_n(t)u_n(t)^\top  + u_n(t) \phi_{1,n}^\top (x_n(t)-w_n) + \phi_{1,n}u_n(t)^\top(x_n(t)-w_n)+ \phi_{1,n}\phi_{1,n}^\top(x_n(t)-w_n)^2)] \\
	& \times [\theta_{3} G (x_n(t)-w_n)^2 + 2 G (x_n(t) - w) (\mu - r)^{\top} u_n(t)+ G \langle \sigma \sigma^\top, u_n(t)u_n(t)^\top \rangle + P_n ] \dd t + \gamma \frac{\phi_{2,n}}{2} \dd t  \biggl \}\\
	=&\frac{1}{2}\phi_{2,n}[(\theta_3-2A_n+B_n)G\E((x_n(t)-w_n)^2) + GE_n + P_n + \gamma]\dd t\\
	&-\frac{1}{2}G\E \biggl\{ \theta_3G(x_n(t)-w_n)^2u_n(t)u_n(t)^\top + 2G(x_n(t)-w_n)u_n(t)u_n(t)^\top (\mu-r)^\top u_n(t) \\
	&+ Gu_n(t)u_n(t)^\top\langle\sigma\sigma^\top,u_n(t)u_n(t)^\top\rangle + P_nu_n(t)u_n(t)^\top\\
	&+\theta_3G(x_n(t)-w_n)^3u_n(t)\phi_{1,n}^\top + 2G(x_n(t)-w_n)^2u_n(t)\phi_{1,n}^\top (\mu-r)^\top u_n(t) \\
	&+G(x_n(t)-w_n)u_n(t)\phi_{1,n}^\top \langle\sigma\sigma^\top,u_n(t)u_n(t)^\top\rangle + P_n(x_n(t)-w_n)u_n(t)\phi_{1,n}^\top\\
	&+\theta_3G(x_n(t)-w_n)^3\phi_{1,n}u_n(t)^\top + 2G(x_n(t)-w_n)^2\phi_{1,n}u_n(t)^\top (\mu-r)^\top u_n(t) \\
	&+G(x_n(t)-w_n)\phi_{1,n}u_n(t)^\top \langle\sigma\sigma^\top,u_n(t)u_n(t)^\top\rangle + P_n(x_n(t)-w_n)\phi_{1,n}u_n(t)^\top\\
	&+\theta_3G(x_n(t)-w_n)^4\phi_{1,n}\phi_{1,n}^\top + 2G(x_n(t)-w_n)^3\phi_{1,n}\phi_{1,n}^\top (\mu-r)^\top u_n(t) \\
	&+G(x_n(t)-w_n)^2\phi_{1,n}\phi_{1,n}^\top \langle\sigma\sigma^\top,u_n(t)u_n(t)^\top\rangle + P_n(x_n(t)-w_n)^2\phi_{1,n}\phi_{1,n}^\top \biggl\}\dd t\\
	=&\frac{1}{2} [\gamma \phi_{2,n} - 2 \phi_{2,n}\sigma\sigma^\top\phi_{2,n}] \dd t.
\end{aligned}}
\]}
However, Lemma \ref{lemma_xt-w} yields
\[
\fitdisplay{\begin{aligned}
\E [(x_n(t) - w_n)^2] &= [(x_0-w_n)^2 + \frac{\langle \Sigma, \phi_{2,n}\rangle e^{\phi_3T}}{-2 (\mu-r)^\top \phi_{1,n} + \langle \Sigma , \phi_{1,n} \phi_{1,n}^\top\rangle + \phi_3}]e^{(-2 (\mu-r)^\top \phi_{1,n} + \langle \Sigma , \phi_{1,n} \phi_{1,n}^\top\rangle)t} \\
&- \frac{\langle \Sigma, \phi_{2,n}\rangle e^{\phi_3(T-t)}}{-2 (\mu-r)^\top \phi_{1,n} + \langle \Sigma , \phi_{1,n} \phi_{1,n}^\top\rangle +\phi_3}.
\end{aligned}}
\]
Integrating $\dd \E [Z_{1, n}(t)]$ from 0 to $T$ and plugging in the above expression of $\E (x_n(t) - w_n)^2$, we obtain
\begin{equation}
\label{eq:h1}
h_1(\phi_{1,n}, \phi_{2,n}, w_n) = -R(\phi_{1,n}, \phi_{2,n}, w_n)(\mu-r-\Sigma \phi_{1,n}),
\end{equation}
where the function $R$ is defined by
\begin{equation}\label{R}
R(\phi_{1}, \phi_{2}, w)= 2 \left[\frac{(x_0-w)^2 e^{-\phi_{3} T}(e^{Q(\phi_{1})T} - 1)}{Q(\phi_1)} + \frac{\langle \sigma \sigma^\top, \phi_{2} \rangle (e^{Q(\phi_{1})T} - 1-Q(\phi_{1})T)}{Q(\phi_{1})^2}\right],
\end{equation}
while
\begin{equation}
Q(\phi_{1}) = -2(\mu-r)^\top \phi_{1} + \langle \sigma \sigma^\top, \phi_{1}\phi_1^\top \rangle +\phi_{3}.
\end{equation}

This completes the proof of the first part of Theorem~\ref{thm:tradeoff}. Similarly (and more easily), we have
\begin{equation}
\label{eq:h2}
h_2(\phi_{1,n}, \phi_{2,n}, w_n) = -\left( \phi_{2,n} \Sigma \phi_{2,n} - \frac{\gamma}{2}\phi_{2,n}\right) T,
\end{equation}
which is quadratic in $\phi_{2,n}$.
Moreover, Lemma \ref{lemma_xt-w} implies
\begin{equation}
\label{eq:hw}
h_w(\phi_{1,n}, \phi_{2,n}, w_n) = \left(1 -e^{-(\mu-r)^\top \phi_{1,n} T} \right) w_n + (x_0 e^{-(\mu-r)^\top \phi_{1,n}T}-z),
\end{equation}
which is linear in $w_n$.

\subsection{Proof of Theorem \ref{thm:convergence_both}}
\label{appendix:proof covergence}
The proof  will also be carried out  through several steps. It will apply some general stochastic approximation results including those in \citet{andradottir1995stochastic} and \citet{broadie2011general}. However, we need to verify several assumptions for our specific  problem  and to overcome difficulties arising from those that are not satisfied by our problem.


\subsubsection*{Properties of mean increments}
With the explicit expressions of $h_1$, $h_2$ and $h_w$ in \eqref{eq:h1}, \eqref{eq:h2} and \eqref{eq:hw} respectively, we further investigate  properties of these functions, which will be useful in the sequel.  Recall that the function $R$ defined in \eqref{R} depends on $\phi_3$. We first show that this function   has a positive lower bound when $\phi_3$ is sufficiently large. Indeed, noting  that $\sigma\sigma^\top$ is positive definite we have
\begin{equation}
\fitdisplay{\begin{aligned}
	Q(\phi_{1}) &= [ \phi_{1} - (\sigma \sigma^\top)^{-1}(\mu-r)]^\top(\sigma \sigma^\top) [\phi_{1} - (\sigma \sigma^\top)^{-1}(\mu-r) ] + \phi_{3} - (\mu-r)^\top(\sigma \sigma^\top)^{-1} (\mu-r) \\
	&> \phi_{3} - (\mu-r)^\top(\sigma \sigma^\top)^{-1} (\mu-r) =: C_Q>0
\end{aligned}}
\end{equation}
when $\phi_{3}$ is sufficiently large. Hence,
\begin{equation}
\label{eq:R_lower}
\fitdisplay{\begin{aligned}
	R(\phi_{1}, \phi_{2}, w )&= 2 \left[\frac{(x_0-w)^2 e^{-\phi_{3} T}(e^{Q(\phi_{1})T} - 1)}{Q(\phi_{1})} + \frac{\langle \sigma \sigma^\top, \phi_{2} \rangle (e^{Q(\phi_{1})T} - 1-Q(\phi_{1})T)}{Q(\phi_{1})^2}\right]\\
	& \geq 2[(x_0-w)^2 e^{-\phi_{3} T} T + \frac{1}{2}\langle \sigma \sigma^\top, \phi_{2} \rangle T^2] =: C_R>0,
\end{aligned}}
\end{equation}
where the inequality follows from the familiar general result $e^x - 1 - x - \frac{1}{2}x^2 \geq 0$  $\forall x \geq 0$.

On the other hand, $Q$ is a quadratic function in $\phi_1$; hence there exist constants $C_{Q_0},C_{Q_1} > 0$ such that $Q(\phi_{1} ) \leq C_{Q_0} + C_{Q_1} |\phi_{1}|^2$. As a consequence,
\[
\fitdisplay{\begin{aligned}
R(\phi_{1}, \phi_{2}, w) &\leq 2 \left[\frac{C(1+|w|^2)e^{C_{Q_0} + C_{Q_1} |\phi_{1}|^2}}{C_Q} + \frac{C |\phi_{2}| e^{C_{Q_0} + C_{Q_1} |\phi_{1}|^2}}{C_Q^2}\right]\\
&\leq C_{R_0} (1 + |w| ^2+ |\phi_{2}|) \exp{(C_{Q_0} + C_{Q_1} |\phi_{1}|^2)},
\end{aligned}}
\]
where $C_{R_0}>0$ is some constant.

Next, we derive  the upper bounds for $h_1$, $h_2$ and $h_w$. We have
\begin{equation}
\label{eq:h1_upper}
\fitdisplay{\begin{aligned}
	\left| h_1(\phi_{1,n},\phi_{2,n},w_n) \right| =& R(\phi_{1,n}, \phi_{2,n}, w_n)|\mu-r-\Sigma \phi_{1,n}|\\
	\leq & \biggl(C_{R_0} (1 + |w_n| ^2+ |\phi_{2,n}|) \exp{(C_{Q_0} + C_{Q_1} |\phi_{1,n}|^2)}  \biggl) |\mu-r-\Sigma \phi_{1,n}|\\
	\leq  & C \biggl( 1 + |\phi_{1,n}| + |\phi_{1,n}| |w_{n}|^2 e^{C|\phi_{1,n}|^2} + |\phi_{1,n}||\phi_{2,n}|e^{|\phi_{1,n}|^2} \biggl),
\end{aligned}}
\end{equation}
and
\begin{equation}
\label{eq:h2_upper}
\fitdisplay{\begin{aligned}
	\left| h_{2}(\phi_{1,n},\phi_{2,n},w_n) \right| = & T \left| \phi_{2,n} \Sigma \phi_{2,n} - \frac{\gamma}{2} \phi_{2,n} \right| \leq C (1 + |\phi_{2,n}|^{2}),
\end{aligned}}
\end{equation}
where the constant $C$ only depends on $\Sigma$, $\gamma$ and $\phi_3$.
Denoting $\bm h(\phi_1, \phi_2, w; \phi_3) = (h_1(\phi_{1,n},\phi_{2,n},w_n;\phi_3),\allowbreak h_2(\phi_1, \phi_2, w; \phi_3))^\top$, we conclude by \eqref{eq:h1_upper} and \eqref{eq:h2_upper} that
\begin{equation}
\label{eq:h_upper}
\fitdisplay{\begin{aligned}
	|\bm h(\phi_{1,n},\phi_{2,n},w_n)|^2 \leq & |h_1(\phi_{1,n},\phi_{2,n},w_n)|^2 + |h_2(\phi_{1,n},\phi_{2,n},w_n)|^2 \\
	\leq & \biggl( C ( 1 + |\phi_{1,n}| + |\phi_{1,n}| |w_{n}|^2 e^{C|\phi_{1,n}|^2} + |\phi_{1,n}||\phi_{2,n}|e^{|\phi_{1,n}|^2} ) \biggl) ^2 + \biggl( C (1 + |\phi_{2,n}|^{2})  \biggl)^2\\
	\leq & C \biggl(1 + |\phi_{1,n}|^{2} + |\phi_{2,n}|^{4}  +  |\phi_{1,n}|^{2}|w_{n}|^{4} e^{C |\phi_{1,n}|^{2}} + |\phi_{1,n}|^{2}|\phi_{2,n}|^{2}  e^{C |\phi_{1,n}|^{2}} \biggl).
\end{aligned}}
\end{equation}

Furthermore, it follows from \eqref{eq:hw}
that
\begin{equation}
\label{eq:hw_upper}
|h_w(\phi_{1,n},\phi_{2,n},w_n)|^2 \leq C(1 + e^{C |\phi_{1,n}|}) |w_{n}|.
\end{equation}

\subsubsection*{Almost sure convergence of $\bm\phi_n$}
We now prove the almost sure convergence of $\bm\phi_n$. Indeed, we present  a more general result of  such convergence, of which Theorem \ref{thm:convergence_both}-(a) is a special case.

\begin{theorem}
\label{thm:convergence_phi}
Let $\phi_{3}$ be a sufficiently large constant, while $\bm\phi_n = (\phi_{1,n}, \phi_{2,n})^\top$ and $w$ be updated according to \eqref{eq:projected}.
Assume that the noise vector $\xi_n = (\xi_{1, n}, \xi_{2, n})^{\top}$ satisfies $\E\left[ \xi_{i, n+1} \Big|\g_n\right] = \beta_{i, n}$ for $i=1,2$ and $\E\left[ \xi_{w, n+1} \Big|\g_n \right] = \beta_{w, n}$, where $\g_n$ are the filtration generated by $\{\bm\theta_m, \bm\phi_m, w_m, m=0,1,2,...,n\}$, with the following upper bounds:
\begin{equation} \label{eq:thm1_assumption0}
	\fitdisplay{\begin{aligned}
		\E\left[ \left|\xi_{1, n+1} - \beta_{1, n} \right|^2 \Big| \g_n \right] \leq & C \bigg( 1 + |w_n|^{16} + |\phi_{1,n}|^8 + |\phi_{2,n}|^8 + |b_n|^8 \bigg) e^{C|\phi_{1,n}|^8}, \\
		\E\left[ \left|\xi_{2, n+1} - \beta_{2, n} \right|^2 \Big| \g_n\right] \leq & C \bigg( 1 + |w_n|^{16} + |\phi_{1,n}|^8 + |\phi_{2,n}|^8 \bigg) e^{C|\phi_{1,n}|^8} ,
\end{aligned}}\end{equation}
where $C > 0$ is a constant independent of $n$.
Moreover, assume

\begin{equation} \label{eq:thm1_assumption}
\fitdisplay{\begin{aligned}
	(i)& \quad \sum_{n} a_n = \infty, \quad \sum_{n}  a_n |\beta_{i, n}| < \infty, \quad \text{for } i=1,2; \\
	(ii)& \quad c_{1, n}\uparrow \infty, \quad c_{2, n}\uparrow \infty, \quad c_{w, n}\uparrow \infty, \quad \sum_{n}  a_n^2 b_n^{8} c_{2, n} ^ {8} c_{w, n} ^ {16} e^{c_{1, n}^{8}} < \infty; \\
	(iii)& \quad b_n\uparrow \infty, \quad \sum_{n}  \frac{a_n}{b_n} = \infty.
\end{aligned}}
\end{equation}
Then $\bm\phi_n = (\phi_{1,n},\phi_{2,n})^\top$  almost surely converges to the unique equilibrium point $\bm\phi^*=(\phi_1^*, \phi_2^*)^\top$ where $\phi_{1}^{*} = \Sigma^{-1} (\mu - r)$ and $\phi_{2}^{*} = \frac{\gamma}{2} \Sigma^{-1}$.
\end{theorem}


\begin{proof}
The main idea is to derive inductive upper bound of  $|\bm\phi_{n} - \bm\phi^*|^2$, namely, to bound $|\bm\phi_{n+1} - \bm\phi^*|^2$ in terms of  $|\bm\phi_n - \bm\phi^*|^2$.

First, for any closed, convex set $K\subset \mathbb{S}^d_+$ and $x\in K, y\in \mathbb{S}^d$, it follows from a property of projection that the function $f(t)=|t\Pi_K(y) + (1-t) x - y|^2$, $t\in\mathbb{R}$,  achieves minimum at $t=1$. However,
\[ f(t) = t^2 |\Pi_K(y) - y|^2 + (1-t)^2|x-y|^2 + 2t(1-t) \langle \Pi_K(y) - y, x - y \rangle. \]
The first-order condition at $t = 1$ yields
\[ 2 |\Pi_K(y) - y|^2 - 2\langle \Pi_K(y) - y, x - y \rangle = 0 .\]
Therefore,
\[\fitdisplay{ |\Pi_K(y) - x|^2 = |\Pi_K(y) - y + y - x |^2 = |y-x|^2 + | \Pi_K(y) - y|^2 + 2\langle  \Pi_K(y) - y, y - x \rangle = |y - x|^2 - |\Pi_K(y) - y|^2 \leq |y-x|^2 . }\]

Now, consider $n$ sufficiently large such that $\bm\phi^*\in K_{1,n+1}\times K_{2,n+1}$ and denote
\[\bm h(\phi_1, \phi_2, w) = (h_1(\phi_1, \phi_2, w), h_2(\phi_1, \phi_2, w))^\top.\]
By the above general projection inequality, we have
\[ |\bm\phi_{n+1} - \bm\phi^*|^2 \leq \left|\bm\phi_n - a_n[\bm h(\phi_{1,n}, \phi_{2,n}, w_n) + \xi_{n+1}] - \bm\phi^*\right|^2.\]

Denoting $U_n = \bm\phi_n - \bm\phi^*$ and $\bm\beta_n=(\beta_{1,n},\beta_{2,n})^\top$, we have
\[\fitdisplay{\begin{aligned}
& \E\left[|U_{n+1}|^2 \Big| \bm\phi_n, w_n \right] \\
\leq & \E\left[| U_n -  a_n[\bm h(\phi_{1,n}, \phi_{2,n}, w_n) + \xi_{n+1}] |^2 \Big| \bm\phi_n, w_n \right] \\
= & |U_n|^2 - 2a_n \langle U_n, \bm h(\phi_{1,n}, \phi_{2,n}, w_n) + \bm \beta_n \rangle + a_n^2 \E\left[ |\bm h(\phi_{1,n}, \phi_{2,n}, w_n) + \xi_{n+1}|^2  \Big| \bm\phi_n, w_n \right] \\
= & |U_n|^2 - 2a_n \langle U_n, \bm h(\phi_{1,n}, \phi_{2,n}, w_n) + \bm \beta_n \rangle + a_n^2 \E\left[ |\bm h(\phi_{1,n}, \phi_{2,n}, w_n) + (\xi_{n+1} - \bm \beta_n) + \bm \beta_n|^2  \Big| \bm\phi_n, w_n \right] \\
\leq & |U_n|^2 - 2a_n \langle U_n, \bm h(\phi_{1,n}, \phi_{2,n}, w_n) \rangle + 2 a_n |\bm \beta_n| |U_n| \\
&+ 3a_n^2 \left( |\bm h(\phi_{1,n}, \phi_{2,n}, w_n)|^2 + |\bm \beta_n|^2 +  \E\left[ \left|\xi_{n+1} - \bm \beta_n \right|^2 \Big| \bm\phi_n, w_n\right] \right)  \\
\leq & |U_n|^2 - 2a_n \langle U_n, \bm h(\phi_{1,n}, \phi_{2,n}, w_n) \rangle + a_n |\bm \beta_n| (1 +|U_n|^2) \\
&+ 3a_n^2 \left( |\bm h(\phi_{1,n}, \phi_{2,n}, w_n)|^2 + |\bm \beta_n|^2 +  \E\left[ \left|\xi_{n+1} - \bm \beta_n \right|^2 \Big| \bm\phi_n, w_n\right] \right) .
\end{aligned}}
\]

Recall that $|\phi_{1, n}| \leq c_{1, n}$, $|\phi_{2, n}| \leq c_{2, n}$, $|w_n| \leq c_{w, n}$ almost surely. By the estimate \eqref{eq:h_upper},
\begin{equation}
\fitdisplay{\begin{aligned}
	|\bm h(\phi_{1,n}, \phi_{2,n}, w_n)|^2 \leq C (1 + c_{1, n}^{2} + c_{2, n}^{4}  +  c_{1, n}^{2}c_{w, n}^{4} e^{C c_{1, n}^{2}} + c_{1, n}^{2}c_{2, n}^{2}  e^{C c_{1, n}^{2}} ).
\end{aligned}}
\end{equation}

However, the assumption \eqref{eq:thm1_assumption0} yields
\[\fitdisplay{\begin{aligned}
\E\left[ \left|\xi_{n+1} - \bm \beta_n \right|^2 \Big| \bm\phi_n, w_n\right] \leq & \E\left[ \left|\xi_{1,n+1} - \beta_{1,n} \right|^2 \Big| \bm\phi_n, w_n\right] + \E\left[ \left|\xi_{2,n+1} -  \beta_{2,n} \right|^2 \Big| \bm\phi_n, w_n\right] \\
\leq &C \bigg(1 + (1 + |w_n|^4 + |\phi_{2, n}|^2) \exp{(C |\phi_{1, n}|^4)}(b_n^2 + |\phi_{2, n}|^2 + |\phi_{1, n}|^4)\\
&+ (1 + |w_n|^8 + |\phi_{2, n}|^4)\exp( C|\phi_{1, n}|^8)(1 + b_n^4 + |\phi_{1, n}|^4) + (d \log{b_n})^4\bigg)\\
&+ C \left(1 + |\phi_{2, n}|^4 + (1 + |w_n|^8 + |\phi_{2, n}|^4)\exp(C|\phi_{1, n}|^8)(1 + |\phi_{1, n}|^4)\right)\\
\leq &C \bigg(1 + |\phi_{2,n}|^4 + (d \log b_n)^4 + \exp (C|\phi_{1,n}|^4) (1+|w_n|^8|+|\phi_{2,n}|^4+b_n^4+|\phi_{1,n}|^8)\\
&+\exp (C|\phi_{1,n}|^8) (1+|w_n|^{16}|+|\phi_{2,n}|^8+b_n^8+|\phi_{1,n}|^8)\bigg)\\
\leq &C \bigg(1 + c_{2,n}^4 + (d \log b_n)^4 +\exp (Cc_{1,n}^8) (1+c_{w,n}^{16}+c_{2,n}^8+b_n^8+c_{1,n}^8)\bigg)
\end{aligned}}\]
almost surely, for some positive constant $C$ that only depends on the model primitives $\mu,\Sigma,d$.

Therefore,
\begin{equation}\label{unpo}
\fitdisplay{\begin{aligned}
	& \E\left[|U_{n+1}|^2 \Big| \bm\phi_n, w_n \right] \\
	\leq & |U_n|^2 - 2a_n \langle U_n, \bm h(\phi_{1,n}, \phi_{2,n}, w_n) \rangle + a_n |\bm \beta_n| |U_n|^2 + a_n |\bm \beta_n| \\
	&+ 3a_n^2 \left( |\bm h(\phi_{1,n}, \phi_{2,n}, w_n)|^2 + |\bm \beta_n|^2 +  \E\left[ \left|\xi_{n+1} - \bm \beta_n \right|^2 \Big| \bm\phi_n, w_n\right] \right) \\
	\leq & |U_n|^2 - 2a_n \langle U_n, \bm h(\phi_{1,n}, \phi_{2,n}, w_n) \rangle + a_n |\bm \beta_n| |U_n|^2 + a_n |\bm \beta_n| \\
	&+3a_n^2 \biggl( C (1 + c_{1, n}^{2} + c_{2, n}^{4}  +  c_{1, n}^{2}c_{w, n}^{4} e^{C c_{1, n}^{2}} + c_{1, n}^{2}c_{2, n}^{2}  e^{C c_{1, n}^{2}} ) + |\bm \beta_n|^2 \\
	& + C (1 + c_{2,n}^4 + (d \log b_n)^4 + e^{Cc_{1,n}^4} (1+c_{w,n}^8+c_{2,n}^4+b_n^4+c_{1,n}^8)+e^{Cc_{1,n}^8} (1+c_{w,n}^{16}+c_{2,n}^8+b_n^8+c_{1,n}^8)) \biggl) \\
	= & (1 + a_n |\bm \beta_n|)|U_n|^2 -  2a_n \langle U_n, \bm h(\phi_{1,n}, \phi_{2,n}, w_n) \rangle + a_n|\bm \beta_n| + \\
	&+3a_n^2 \biggl( C (1 + c_{1, n}^{2} + c_{2, n}^{4}  +  c_{1, n}^{2}c_{w, n}^{4} e^{C c_{1, n}^{2}} + c_{1, n}^{2}c_{2, n}^{2}  e^{C c_{1, n}^{2}} ) + |\bm \beta_n|^2 \\
	& + C (1 + (d \log b_n)^4 + e^{Cc_{1,n}^4} (1+c_{w,n}^8+c_{2,n}^4+b_n^4+c_{1,n}^8)+e^{Cc_{1,n}^8} (1+c_{w,n}^{16}+c_{2,n}^8+b_n^8+c_{1,n}^8)) \biggl) \\
	=&: (1 + \gamma_n) |U_n|^2  - \zeta_n + \eta_n,
\end{aligned}}\end{equation}
where $\gamma_n = a_n |\beta_n|$, $\zeta_n = 2a_n \langle U_n, h(\phi_{1,n}, \phi_{2,n}, w_n) \rangle $, and
\begin{equation}
\label{eq:eta_n}
\fitdisplay{\begin{aligned}
	\eta_n =& a_n|\bm \beta_n| + 3a_n^2 |\bm \beta_n|^2 + 3a_n^2 \biggl( C (1 + c_{1, n}^{2} + c_{2, n}^{4}  +  c_{1, n}^{2}c_{w, n}^{4} e^{C c_{1, n}^{2}} + c_{1, n}^{2}c_{2, n}^{2}  e^{C c_{1, n}^{2}} ) + |\bm \beta_n|^2 \\
	& + C (1 + (d \log b_n)^4 + e^{Cc_{1,n}^4} (1+c_{w,n}^8+c_{2,n}^4+b_n^4+c_{1,n}^8)+e^{Cc_{1,n}^8} (1+c_{w,n}^{16}+c_{2,n}^8+b_n^8+c_{1,n}^8)) \biggl).
\end{aligned}}
\end{equation}

By Assumptions (i)--(ii), we know $\sum_{n} \gamma_n<\infty$ and $\sum_{n} \eta_n<\infty$. It then follows from \citet[Theorem 1]{robbins1971convergence} that $\left|U_n\right|^2$ converges to a finite limit and $\sum_{n} \zeta_n<\infty$ almost surely.

It remains to show $|U_n| \to 0$ almost surely.
Consider the term
\[\fitdisplay{\begin{aligned}
&\langle \bm \phi - \bm \phi^*, \bm h(\phi_{1}, \phi_{2}, w) \rangle \\ =&\langle \phi_{1} - \phi_{1}^{*}, h_1(\phi_{1}, \phi_{2}, w) \rangle + \langle \phi_{2} - \phi_{2}^{*}, h_2(\phi_{2}) \rangle \\
= &\langle \phi_{1} - \phi_{1}^{*}, R(\phi_{1}, \phi_{2}, w) \Sigma (\phi_{1}-\phi_{1}^{*})  \rangle + \langle \phi_{2} - \phi_{2}^{*}, \phi_{2,n}^{\top} \Sigma (\phi_{2}-\phi_{2}^{*})  \rangle \\
= & R(\phi_{1}, \phi_{2}, w) \langle \Sigma, (\phi_{1}-\phi_{1}^{*}) (\phi_{1} - \phi_{1}^{*})^{\top}  \rangle + \langle \Sigma \phi_{2},  (\phi_{2} - \phi_{2}^{*})(\phi_{2} - \phi_{2}^{*})^\top \rangle.
\end{aligned}}\]
Note that $\langle  \Sigma,  (\phi_{1} - \phi_{1}^{*})(\phi_{1} - \phi_{1}^{*})^\top \rangle \geq 0$ because
$\Sigma \in \mathbb{S}^d_{++}$ and $(\phi_{1}-\phi_{1}^{*}) (\phi_{1} - \phi_{1}^{*})^{\top} \in \mathbb{S}^d_+$.

To proceed, let us first consider a special case when $\Sigma = I$ to get the main idea of the rest of the proof. Indeed, when $\Sigma = I$,
\[ \langle \Sigma, (\phi_{1}-\phi_{1}^{*}) (\phi_{1} - \phi_{1}^{*})^{\top}  \rangle = \langle I, (\phi_{1}-\phi_{1}^{*}) (\phi_{1} - \phi_{1}^{*})^{\top}  \rangle  \geq  |\phi_{1}-\phi_{1}^{*}|^2 \geq \delta^2,
\]
whenever $|\phi_{1} - \phi_{1}^{*}| \geq \delta > 0$. In this case,
$$R(\phi_{1}, \phi_{2}, w) \langle I, (\phi_{1}-\phi_{1}^{*}) (\phi_{1} - \phi_{1}^{*})^{\top}  \rangle \geq C_R \delta^2$$
because $R(\phi_{1}, \phi_{2}, w; \phi_3) \geq C_R > 0$ due to \eqref{eq:R_lower}.
Moreover, $\langle  \phi_{2},  (\phi_{2} - \phi_{2}^{*})(\phi_{2} - \phi_{2}^{*})^\top \rangle \geq 0$ because $\phi_{2}\in \mathbb{S}^d_+$ and $(\phi_{2} - \phi_{2}^{*})(\phi_{2} - \phi_{2}^{*})^\top \in \mathbb{S}^d_+$. In particular, when $|\phi_{2} - \phi_{2}^{*}| \geq \delta > 0$ and $\phi_{2} - \frac{1}{b_n}I\in \mathbb{S}^d_+$, we have
\[ \fitdisplay{\begin{aligned}
\langle \phi_{2},  (\phi_{2} - \phi_{2}^{*})(\phi_{2} - \phi_{2}^{*})^\top \rangle =  &\langle  \phi_{2} - \frac{1}{b_n}I,  (\phi_{2} - \phi_{2}^{*})(\phi_{2} - \phi_{2}^{*})^\top \rangle + \frac{1}{b_n} \langle I, (\phi_{2} - \phi_{2}^{*})(\phi_{2} - \phi_{2}^{*})^\top \rangle \\
\geq & \frac{1}{b_n}|\phi_{2}-\phi_{2}^{*}|^2 \geq \frac{\delta^2}{b_n} .
\end{aligned}}\]

Now, suppose $|U_n| \nrightarrow 0$ almost surely. Then there exists a set $Z\in \f$ with $\p(Z) =1$ so that for every $\omega\in Z$, there is $\delta(\omega) > 0$ such that for all $n$ sufficiently large, at least one of the following two cases holds true: (a) $|\phi_{1}(\omega) - \phi_{1}^{*}| \geq \delta(\omega) > 0$; (b) $|\phi_{2}(\omega) - \phi_{2}^{*}| \geq \delta(\omega) > 0$.

Recall that $\bm\phi_n(\omega)\in K_{1,n}\times K_{2,n}$. If (a) is true, then the above analysis yields
\[ \langle U_n(\omega), \bm h(\bm \phi_n(\omega), w_n(\omega))\rangle \geq \delta(\omega)^2. \]
Thus, by Assumption (iii), we have
\[ \sum_{n} \zeta_n(\omega) = 2 \sum_{n} a_n \langle U_n(\omega), \bm h( \bm \phi_n(\omega), w_n(\omega))\rangle \geq 2C_R\delta(\omega)^2 \sum_{n} a_n = \infty . \]
This is a contradiction.

If (b) is true, then
\[ \langle U_n(\omega), \bm h(\bm \phi_n(\omega), w_n(\omega))\rangle \geq \frac{\delta(\omega)^2}{b_n} , \]
and hence, by Assumption-(iii),
\[ \sum_{n} \zeta_n(\omega) = 2 \sum_{n} a_n \langle U_n(\omega), \bm h( \bm \phi_n(\omega), w_n(\omega))\rangle \geq 2\delta(\omega)^2 \sum_{n} \frac{a_n}{b_n} = \infty . \]
This is again a contradiction.

Now let us consider the general case when $\Sigma \neq  I$. Introduce a different inner product and norm on $\mathbb{R}^{d}\times  \mathbb{R}^{d\times d}$ induced by $\Sigma\in \mathbb{S}^d_{++}$:
\[ \langle (A_1, A_2)^\top, (B_1, B_2)^\top \rangle_{\Sigma} := \langle A_1, B_1 \rangle + \langle\Sigma A_2 \Sigma, B_2 \rangle,\]
\[ |(A_1, A_2)^\top|_{\Sigma} :=  |A_1| + \sqrt{\langle A, A \rangle_{\Sigma} } = |A_1| + | \Sigma^{1/2} A_2 \Sigma^{1/2} | .   \]
It is straightforward to verify that $\langle \cdot, \cdot \rangle_{\Sigma}$ is indeed an inner product and $|\cdot|_{\Sigma}$ is the associated norm. Moreover, since all norms on a finite dimensional space are equivalent, there exist constants $\overline{C} > \underline{C} > 0$ depending only on $\Sigma$ and the dimension $d$ such that
\[ \underline{C} |(A_1, A_2)^\top| \leq |(A_1, A_2)^\top|_{\Sigma} \leq \overline{C} |(A_1, A_2)^\top| ,   \]
for any $A_1 \in \mathbb{R}^{d}$, $A_2 \in \mathbb{R}^{d\times d}$.

When $n$ is sufficiently large such that $\bm \phi^*\in K_{n+1}$,
\[ |\bm \phi_{n+1} - \bm \phi^*|^2 \leq |\bm \phi_n - a_n[\bm h(\phi_{1,n}, \phi_{2,n}, w_n) + \xi_{n+1}] - \bm \phi^*|^2,\]
or $|\bm \phi_{n+1} - \bm \phi^*|^2_{\Sigma} \leq \frac{\overline{C}^2}{\underline{C}^2} |\bm \phi_n - a_n[\bm h(\phi_{1,n}, \phi_{2,n}, w_n) + \xi_{n+1}] - \bm \phi^*|^2_{\Sigma} $.
Hence the estimate \eqref{unpo} for $U_{n+1}$ still holds true under  the new norm $|\cdot|_{\Sigma}$. It follows that
\[ \sum a_n \langle U_n, \bm h(\phi_{1,n}, \phi_{2,n}, w_n) \rangle_{\Sigma} < \infty  \]
and $|U_{n}|^2_{\Sigma}$ converges to a finite limit almost surely.

Consider the term
\[\fitdisplay{\begin{aligned}
&\langle \bm \phi - \bm \phi^*, \bm h(\phi_{1}, \phi_{2}, w) \rangle_{\Sigma}\\
= & R(\phi_{1}, \phi_{2}, w) \langle \Sigma, (\phi_{1}-\phi_{1}^{*}) (\phi_{1} - \phi_{1}^{*})^{\top}  \rangle
+  \langle \Sigma (\phi_{2} - \phi_{2}^{*})\Sigma, \phi_{2,n}^{\top} \Sigma (\phi_{2}-\phi_{2}^{*})  \rangle \\
= &  R(\phi_{1}, \phi_{2}, w) \langle \Sigma, (\phi_{1}-\phi_{1}^{*}) (\phi_{1} - \phi_{1}^{*})^{\top}  \rangle+\langle \Sigma^{1/2} (\phi_{2} - \phi_{2}^{*})\Sigma^{1/2}, \Sigma^{1/2}\phi_{2,n}^{\top} \Sigma^{1/2} \Sigma^{1/2} (\phi_{2}-\phi_{2}^{*}) \Sigma^{1/2} \rangle  \\
= & R(\phi_{1}, \phi_{2}, w) \langle \Sigma, (\phi_{1}-\phi_{1}^{*}) (\phi_{1} - \phi_{1}^{*})^{\top}  \rangle + \langle \tilde{\phi_{2}} - \tilde{\phi_{2}^{*}} , \tilde{\phi_{2,n}^{\top}} (\tilde{\phi_{2}} - \tilde{\phi_{2}^{*}})\rangle \\
= & R(\phi_{1}, \phi_{2}, w) \langle \Sigma, (\phi_{1}-\phi_{1}^{*}) (\phi_{1} - \phi_{1}^{*})^{\top}  \rangle + \langle \tilde{\phi_{2}}, (\tilde{\phi_{2}} - \tilde{\phi_{2}^{*}})(\tilde{\phi_{2}} - \tilde{\phi^{(2)*\top}}) \rangle,
\end{aligned}} \]
where $\tilde{\phi_{2}} = \Sigma^{1/2}\phi_{2}\Sigma^{1/2}$ and $\tilde{\phi_{2}^{*}} = \Sigma^{1/2}\phi_{2}^{*}\Sigma^{1/2}$.
As before, we need to prove $|U_n|_{\Sigma} \to 0$ almost surely.
If not, then there exists a set $Z\in \f$ with $\p(Z) =1$ so that for every $\omega\in Z$, there is $\delta(\omega) > 0$ such that for all $n$ sufficiently large, at least one of the following two cases are true: (a) $|\phi_{1}(\omega) - \phi_{1}^{*}| \geq \delta(\omega) > 0$; (b) $|\phi_{2}(\omega) - \phi_{2}^{*}|_{\Sigma} \geq \delta(\omega) > 0$.

If (a) is true, then there is a contradiction based on the same argument before.
If (b) is true, then when $|\phi_{2}(\omega)  - \phi_{2}^{*} |_{\Sigma} \geq \delta(\omega) > 0$ and $\phi_{2}(\omega) - \frac{1}{b_n}I\in \mathbb{S}^d_+$, we have $\Sigma^{1/2}(\phi_{2}(\omega) - \frac{1}{b_n}I)\Sigma^{1/2} \in \mathbb{S}^d_+$, and
\[ \fitdisplay{\begin{aligned}
\langle \bm \phi - \bm \phi^*, \bm h(\phi_{1}, \phi_{2}, w) \rangle_{\Sigma} \geq & \langle  \Sigma^{1/2}\phi_{2} \Sigma^{1/2},  (\tilde{\phi_{2} } - \tilde{\phi_{2}^{*}})(\tilde{\phi_{2}} - \tilde{\phi_{2}^{*}})^\top \rangle \\
= & \langle  \Sigma^{1/2} (\phi_{2} - \frac{1}{b_n}I)\Sigma^{1/2},  (\tilde{\phi_{2}} - \tilde{\phi_{2}^{*}})(\tilde{\phi_{2}} - \tilde{\phi_{2}^{*}})^\top \rangle + \frac{1}{b_n} \langle \Sigma,(\tilde{\phi_{2}} - \tilde{\phi_{2}^{*}})(\tilde{\phi_{2}} - \tilde{\phi_{2}^{*\top}}) \rangle \\
\geq & \frac{1}{b_n} \langle \Sigma,(\tilde{\phi_{2}} - \tilde{\phi_{2}^{*}})(\tilde{\phi_{2}} - \tilde{\phi^{(2)*\top}}) \rangle \\
\geq & \frac{\lambda_{min}}{b_n}  \langle I,(\tilde{\phi_{2}} - \tilde{\phi_{2}^{*}})(\tilde{\phi_{2}^{*}} - \tilde{\phi_{2}^{*}})^\top \rangle \\
= & \frac{\lambda_{min}}{b_n} |\tilde{\phi_{2}} - \tilde{\phi_{2}^{*}}|^2 = \frac{\lambda_{min}}{b_n} |\Sigma^{1/2} (\phi_{2} - \phi_{2}^{*}) \Sigma^{1/2}|^2\\
\geq & \frac{\lambda_{min} \delta^2}{b_n},
\end{aligned}}\]
where $\lambda_{min} > 0$ is the smallest eigenvalue of $\Sigma$.
Hence
\[ \langle U_n(\omega), \bm h( \bm \phi_n(\omega), w_n(\omega))\rangle_\Sigma \geq \frac{\lambda_{min} \delta^2}{b_n} . \]
Thus, Assumption-(iii) implies
\[\sum_{n} a_n \langle U_n(\omega), \bm h(\bm \phi_n(\omega), w_n(\omega))\rangle \geq \lambda_{min} \delta(\omega)^2 \sum_{n} \frac{a_n}{b_n} = \infty , \]
which is a contradiction.
The proof is now complete.
\end{proof}

\begin{remark}
\label{remark1}
When $\beta_{1, n} = 0$, $\beta_{2, n} = 0$, $\beta_{w, n} = 0$ for all $n$, which holds true in our mean--variance problem, a typical choice of the sequences satisfying Assumptions (i)--(iii) is $a_n =\frac{\alpha}{n+\beta}$ with constants $\alpha>0$ and $\beta>0$,  $b_n = 1 \vee (\log \log n)^{\frac{1}{8}}, c_{1, n} = 1\vee (\log \log n)^{\frac{1}{8}}, c_{2, n} = 1\vee (\log \log n)^{\frac{1}{8}}$ and  $c_{w, n} = 1\vee (\log \log n)^{\frac{1}{16}}$. This is because  $\sum \frac{1}{n (\log\log n)^{\kappa}} = \infty$ and $\sum \frac{(\log n)^{\kappa_1} (\log \log n)^{\kappa_2}}{n^2} <\infty$, for any $\kappa, \kappa_1, \kappa_2 > 0$. 
\end{remark}

\subsubsection*{Mean--squared error of $\phi_{1,n}-\phi_{1}^*$}
Now we move forward to  derive the error bound of $\phi_{1,n}-\phi_{1}^*$ in the mean-squared sense, which is Theorem \ref{thm:convergence_both}-(b).
Note that this result is also necessary for subsequently proving  the almost sure convergence of $w_n$, because $h_w$ not only depends on $w$, but also on $\phi_{1}$. Moreover, the error bound of $\phi_{1,n} - \phi_{1}^*$ affects the property of $h_w$.

We first need a general recursive relation satisfied by a typical learning rate sequence.
\begin{lemma}
\label{lemma:a_n}
For any $A > 0$, there exist positive numbers $\alpha > \frac{1}{A}$ and $\beta \geq \frac{1}{A \alpha -1}$ such that the learning rate sequence  $a_n = \frac{\alpha}{n + \beta}$, $n\geq 0$, satisfies
$a_n \leq a_{n+1}(1 + A a_{n+1})$ for any $n\geq0$.
\end{lemma}
\begin{proof} It is clear that $a_n  \leq a_{n+1}(1 + A a_{n+1})$ is equivalent to
$n + 1 + \beta \leq A \alpha n + A \alpha \beta$.
However, the latter holds true when $\alpha > \frac{1}{A}$, $\beta \geq \frac{1}{A \alpha -1}$.
\end{proof}

With  Lemma \ref{lemma:a_n}, we present the following result for the mean-squared error of $\phi_{1,n}$.

\begin{theorem}
\label{phi_1_rate}
Under the assumptions of Theorem \ref{thm:convergence_phi}, if the sequence $\{a_n\}$ further satisfies
\[a_n \leq a_{n+1}(1 + A a_{n+1}),\] for some sufficiently small constant $A>0$  and $|\beta_n| = O(a_n^{\frac{1}{2}})$,
then there exists an increasing sequence $\{\hat{\eta}_{n}\}$ 
and a constant ${C}'>0$ such that
\[
\E [|\phi_{1, n+1} - \phi_1^*|^2] \leq {C}' a_n \hat{\eta}_{1,n}.
\]
In particular, if we set the parameters $a_n, b_n, c_{1,n}, \beta_{1,n}$ as in Remark \ref{remark1}, then
\[
\E [|\phi_{1, n+1} - \phi_1^*|^2] \leq C \frac{(\log n)^{p} (\log \log n)}{n}
\]
for any $n$, where $C$ and $p$ are positive constants that only depend on model primitives.
\end{theorem}



\begin{proof}
Denote $n_0=\inf \{n\geq0:\bm\phi^*\in K_{1,n+1}\times K_{2,n+1}\}$ and $U_{1, n} = \phi_{1, n} - \phi_1^*$. It follows from  \eqref{eq:h1} and \eqref{eq:R_lower} that
\[
\langle U_{1, n}, h_1(\phi_{1,n},\phi_{2,n},w_n;\phi_3) \rangle \geq C_R^\prime|\phi_{1, n} - \phi_1^*|^2 = C_R^\prime |U_{1, n}|^2
\]
with some constant $C_R^\prime > 0$. When $n \geq n_0$, this together with a similar argument to the proof of Theorem \ref{thm:convergence_phi} yields
\begin{equation}
\label{eq:phi1_rate(t)emp}
\fitdisplay{\begin{aligned}
	& \E\left[|U_{1, n+1}|^2 \Big| \bm\phi_n, w_n \right] \\
	\leq & |U_{1, n}|^2 - 2a_n \langle U_{1, n}, h_1(\phi_{1,n},\phi_{2,n},w_n;\phi_3) \rangle + 2a_n |\beta_{1, n}| |U_{1, n}| + 3a_n^2 \biggl( |h_1(\phi_{1,n},\phi_{2,n},w_n;\phi_3)|^2 + |\beta_{1, n}|^2  \\
	&+\E\left[ \left|\xi_{1, n+1} - \beta_{1, n} \right|^2 \Big| \bm\phi_n, w_n\right] \biggl)\\
	\leq & |U_{1, n}|^2 - 2a_n \langle U_{1, n}, h_1(\phi_{1,n},\phi_{2,n},w_n;\phi_3) \rangle + a_n\left( \frac{1}{C_R'}|\beta_{1,n}|^2 + C_R'|U_{1,n}|^2  \right) \\
	& + 3a_n^2 \biggl( |h_1(\phi_{1,n},\phi_{2,n},w_n;\phi_3)|^2 + |\beta_{1, n}|^2 +  \E\left[ \left|\xi_{1, n+1} - \beta_{1, n} \right|^2 \Big| \bm\phi_n, w_n\right] \biggl) \\
	\leq & (1-a_n C_R^\prime)|U_{1, n}|^2 + 3a_n^2 \hat{\eta}_n .
\end{aligned}}
\end{equation}
Now, by the proof of Theorem \ref{thm:convergence_phi},
\begin{equation}
\label{eq:combined_h1_noise1}
\fitdisplay{\begin{aligned}
	&\left| h_1(\phi_{1,n},\phi_{2,n},w_n;\phi_3) \right|^2 + \E\left[ \left|\xi_{1, n+1} - \beta_{1, n} \right|^2 \Big| \bm\theta_n, \bm\phi_n, w_n\right] \\
	\leq & C \biggl(1 + c_{1, n}^{2} +  c_{1, n}^{2}c_{w, n}^{4} e^{C c_{1, n}^{2}} + c_{1, n}^{2}c_{2, n}^{2}  e^{C c_{1, n}^{2}} \\
	&+ \exp{\{C c_{1, n}^4\}}(1 + c_{w, n}^8 + c_{2, n}^4 + b_n^4 + c_{2, n}^4 + c_{1, n}^8) \\
	&+ \exp\{ Cc_{1, n}^8\}(1 + c_{w, n}^{16} + c_{2, n}^8 + b_n^8 + c_{1, n}^8) + (d \log{b_n})^8 \bigg).\\
\end{aligned}}
\end{equation}
Moreover, the assumption $|\beta_n| = O(a_n^{\frac{1}{2}})$ imply that $\frac{|\beta_n|^2}{a_n} \leq c$, where $c>0$ is a constant. When $n \geq n_0$, it follows from \eqref{eq:phi1_rate(t)emp} that
\[
\E\left[|U_{1, n+1}|^2 \Big| \bm\phi_n, w_n \right] \leq (1-a_n C_R^\prime)|U_{1, n}|^2 + 3a_n^2 \hat{\eta}_n,
\]
where
\begin{equation}
\label{eq:eta_hat}
\fitdisplay{\begin{aligned}
	\hat{\eta}_n =& C \biggl(1 + c_{1, n}^{2} +  c_{1, n}^{2}c_{w, n}^{4} e^{C c_{1, n}^{2}} + c_{1, n}^{2}c_{2, n}^{2}  e^{C c_{1, n}^{2}} \\
	&+ \exp{\{C c_{1, n}^4\}}(1 + c_{w, n}^8 + c_{2, n}^4 + b_n^4 + c_{2, n}^4 + c_{1, n}^8) \\
	&+ \exp\{ Cc_{1, n}^8\}(1 + c_{w, n}^{16} + c_{2, n}^8 + b_n^8 + c_{1, n}^8) + (d \log{b_n})^8 \bigg),
\end{aligned}}
\end{equation}
which is monotonically increasing because so are $c_{1,n}$, $c_{2,n}$, $c_{w,n}$, $b_n$ by the assumptions. Taking expectation on both sides of the above and
denoting $\rho_n = \E[|U_{1, n}|^2]$, we get
\begin{equation}
\label{eq:phi1_rate_induction}
\rho_{n+1} \leq (1-a_n C_R^\prime)\rho_n + 3a_n^2 \hat{\eta}_n,
\end{equation}
where $n \geq n_0$.

Next, we show $\rho_{n+1} \leq C^\prime a_n\hat{\eta}_{n}$ for all $n \geq 0$ by induction, where
$C^\prime = max\{ \frac{\rho_1}{a_0 \hat{\eta}_{0}},  \frac{\rho_2}{a_1 \hat{\eta}_{1}}, \cdots,  \frac{\rho_{n_0+1}}{a_{n_0} \hat{\eta}_{n_0}}, \frac{3}{C_R^\prime} \} + 1$.
Indeed, it is true when $n\leq n_0$.  Assume that $\rho_{k+1} \leq c'a_k \hat{\eta}_{1,k}$ is true for $n_0 \leq k \leq n-1$. 
Then \eqref{eq:phi1_rate_induction} yields

\[
\fitdisplay{\begin{aligned}
\rho_{n+1} & \leq  (1-a_n C_R^\prime)\rho_n + 3a_n^2 \hat{\eta}_n\\
&\leq (1-a_n C_R^\prime)C^\prime a_{n-1} \hat{\eta}_{n-1} + 3a_n^2 \hat{\eta}_n \\
&\leq (1-a_n C_R^\prime)C^\prime a_{n}(1 + A a_n) \hat{\eta}_{n-1} + 3a_n^2 \hat{\eta}_n \\
&\leq (1-a_n C_R^\prime)C^\prime a_{n}(1 + A a_n) \hat{\eta}_{n} + 3a_n^2 \hat{\eta}_n \\
&=C^\prime a_n\hat{\eta}_n + C^\prime \hat{\eta}_n a_n^2 \biggl(-AC_R^\prime a_n + (A-C_R^\prime) + \frac{3}{C^\prime} \biggl).
\end{aligned}}
\]
Consider the function
\[
f(x) = C^\prime \hat{\eta}_nx^2 \biggl(-AC_R^\prime x + (A-C_R^\prime) + \frac{3}{C^\prime} \biggl),
\]
which has two roots at $x_{1,2}=0$ and one root at $x_3=\frac{A-C_R^\prime + \frac{3}{C^\prime}}{AC_R^\prime}$.
Because  $C_R^\prime - \frac{3}{C'} > 0$, we can  choose $0<A < C_R^\prime - \frac{3}{C'}$ so that $x_3 < 0$.
So ${f}(x) < 0$ when $x > 0$, leading to
\[
C^\prime \hat{\eta}_n a_n^2 \biggl(-AC_R^\prime a_n + (A-C_R^\prime) + \frac{3}{C^\prime} \biggl) < 0, \;\;\forall n,
\]
since $a_n>0$. We have now proved $\E [|U_{1, n+1}|^2] \leq C^\prime a_n \hat{\eta}_n$.

In particular, under the settings of Remark \ref{remark1}, it is straightforward to verify that $|\beta_n| = O(a_n^{\frac{1}{2}})$. Then
\begin{equation}
\label{eq:eta_hat_bound}
\fitdisplay{\begin{aligned}
	\hat{\eta}_n =& C \biggl(1 + c_{1, n}^{2} +  c_{1, n}^{2}c_{w, n}^{4} e^{C c_{1, n}^{2}} + c_{1, n}^{2}c_{2, n}^{2}  e^{C c_{1, n}^{2}} \\
	&+ \exp{\{C c_{1, n}^4\}}(1 + c_{w, n}^8 + c_{2, n}^4 + b_n^4 + c_{2, n}^4 + c_{1, n}^8) \\
	&+ \exp\{ Cc_{1, n}^8\}(1 + c_{w, n}^{16} + c_{2, n}^8 + b_n^8 + c_{1, n}^8) + (d \log{b_n})^8 \bigg)\\
	\leq & C\biggl(1 + \log\log n + \log\log n(\log n)^p + (\log n)^p(1 + \log\log n)\biggl) \\
	\leq & C (\log n)^p(\log \log n),
\end{aligned}}
\end{equation}
where $C$ and $p$ are positive constants independent of $n$. The proof is now complete.

\end{proof}

\subsubsection*{Almost sure convergence of $w_n$}
We finally prove the almost sure convergence of $w_n$.
\begin{theorem}
\label{thm_w_convergence}
Let $w_n$ be updated following \eqref{eq:projected}, and
the assumptions \eqref{eq:thm1_assumption}  and the following additional assumptions be satisfied:	
\begin{equation} \label{eq:thm3_assumption}
\fitdisplay{\begin{aligned}
	(i)& \quad \sum_{n} a_{w, n} = \infty, \quad \sum_{n} a_{w, n} |\beta_{w, n}| < \infty; \\
	(ii)& \quad c_{1, n}\uparrow \infty, \quad c_{2, n}\uparrow \infty, \quad c_{w, n}\uparrow \infty, \quad \sum_{n} a_{w,n}^2 c_{2, n} c_{w, n} ^ {2} e^{c_{1, n}^{2}} < \infty ;\\
	(iii)& \quad \sum_{n} a_{w,n} a_n c_{1, n}^{8} c_{2, n}^{8} b_{n}^{8} c_{w, n}^{16} e^{c_{1, n}^{8}} < \infty.
\end{aligned}}
\end{equation}
Then $w_n\rightarrow w^*=\frac{ze^k - x_0}{e^k - 1}$ almost surely as $n\rightarrow \infty$, where $k=(\mu-r)^\top \Sigma^{-1}(\mu-r) T$.
\end{theorem}


\begin{proof}
Recall from Lemma \ref{lemma_xt-w} that	
$\E\left[ \left|\xi_{w, n+1} - \beta_{w, n} \right|^2 \Big| \phi_n\right]  \leq C (1 + |w_n|^2 + |\phi_{2, n})|e^{C|\phi_{1, n}|^2}\leq C (1 + c_{w,n}^2 + c_{2, n})e^{Cc_{1, n}^2}$ and $\beta_{w,n}=0$ in our case.

Also, we can estimate the upper bound of function $|h_w|$ as
\[
|h_w(\phi_{1, n}, w_n)| \leq C (1 + |w_n|)e^{C|\phi_{1, n}|}.
\]

Then,
\[
|h_w(\phi_{1, n}, w_n)|^2 \leq C (1 + |w_n|^2)e^{C|\phi_{1, n}|} \leq C (1 + c_{w,n}^2)e^{C c_{1, n}}.
\]

Denote $U_{w, n} = w_n - w^*$. Then similarly as in the proof of Theorem \ref{thm:convergence_phi}, we have

\begin{equation}
\label{eq:w_long}
\fitdisplay{\begin{aligned}
	& \E\left[|U_{w, n+1}|^2 \Big| \bm\phi_n, w_n \right] \\
	\leq & |U_{w, n}|^2 - 2a_{w,n}  U_{w, n} h_w(\phi_{1, n}, w_n) + a_{w,n} |\beta_{w, n}| (1 +|U_{w, n}|^2) \\
	&+ 3a_{w,n}^2 \biggl( |h_w(\phi_{1, n}, w_n)|^2 + |\beta_{w, n}|^2 + \E\left[ \left|\xi_{w, n+1} - \beta_{w, n} \right|^2 \Big| \bm\phi_{n}, w_n\right] \biggl)\\
	\leq & |U_{w, n}|^2 - 2a_{w,n}  U_{w, n} h_w(\phi_{1, n}, w_n)  + a_{w,n} |\beta_{w, n}| (1 +|U_{w, n}|^2) \\
	&+ 3a_{w,n}^2 \biggl( C (1 + c_{w,n}^2)e^{C c_{1, n}} + |\beta_{w, n}|^2 + C   (1 + c_{w,n}^2 + c_{2, n})e^{Cc_{1, n}^2} \biggl)\\
	\leq & \left[ 1+a_{w,n}|\beta_{w,n}| \right] |U_{w, n}|^2 - 2a_{w,n} U_{w, n} h_w(\phi_{1, n}, w_n)  + a_{w,n} |\beta_{w, n}|\\
	&+ 3a_{w,n}^2 \biggl( C (1 + c_{w,n}^2)e^{C c_{1, n}} + |\beta_{w, n}|^2 + C   (1 + c_{w,n}^2 + c_{2, n})e^{Cc_{1, n}^2} \biggl)\\
	= & \left[ 1+a_{w,n}|\beta_{w,n}|+4a_{w,n}(1-e^{-(\mu-r)^\top \phi_{1,n} T })^-\right] |U_{w, n}|^2 \\
	&-2a_{w,n} \biggl(U_{w, n} h_w(\phi_{1, n}, w_n)  +2 (1-e^{-(\mu-r)^\top \phi_{1,n} T})^- |U_{w,n}|^2+ M(\phi_{1,n})\biggl) \\
	&+ a_{w,n} |\beta_{w, n}|+ 3a_{w,n}^2 \biggl( C (1 + c_{w,n}^2)e^{C c_{1, n}} + |\beta_{w, n}|^2 + C   (1 + c_{w,n}^2 + c_{2, n})e^{Cc_{1, n}^2} \biggl) + 2a_{w,n} M(\phi_{1,n}) \\
	=&: (1 + \gamma_n) |U_{w,n}|^2  - \zeta_n + \eta_n,
\end{aligned}}
\end{equation}
where $f^- = \max(-f, 0)$, $\gamma_n = a_{w,n} |\beta_{w,n}|+4a_{w,n}(1-e^{-(\mu-r)^\top \phi_{1,n} T})^-$, $\zeta_n = 2a_{w,n} \bigg[ U_{w, n}  h_w(\phi_{1, n}, w_n) +2 (1-e^{-(\mu-r)^\top \phi_{1,n} T})^- U_{w,n}^2+ M(\phi_{1,n}) \bigg],$
$	\eta_n = a_{w,n} |\beta_{w, n}| + 3a_{w,n}^2 \biggl( C (1 + c_{w,n}^2)e^{C c_{1, n}} + |\beta_{w, n}|^2 + C  (1 + c_{w,n}^2 + c_{2, n})e^{Cc_{1, n}^2} \biggl)+ 2a_{w,n} M(\phi_{1,n})$, while
\[
M(\phi_{1,n}) = \frac{(z-x_0)^2}{4(e^k- 1)^2} \frac{(e^{-(\mu-r)^\top (\phi_{1,n}-\phi_{1}^*) T} - 1)^2}{|1-e^{-(\mu-r)^\top \phi_{1,n} T}|}\geq 0.
\]

First, we consider the term $\gamma_n$. By Theorem \ref{thm:convergence_phi}, almost surely, $\phi_{1,n} \to \phi_1^*=\Sigma^{-1} (\mu - r)$ as $n \to \infty$. Hence $(\mu-r)^\top \phi_{1, n} \to (\mu-r)^\top \Sigma^{-1} (\mu - r) >0$ since $\Sigma \in \mathbb{S}^d_{++}$ and $\mu\neq r$. Then for any $\epsilon>0$ and any $\omega \in \Omega$ except for a zero-probability set, there exists $0<N_1(\epsilon, \omega)<\infty$ such that when $n\geq N_1(\epsilon, \omega)$, $1-e^{-(\mu-r)^\top \phi_{1, n}(\omega) T}>\epsilon>0$. Then,
\[
\fitdisplay{\begin{aligned}
\sum_{n=1}^\infty \gamma_n &= 4\sum_{n=1}^\infty a_{w,n}(1-e^{-(\mu-r)^\top \phi_{1,n}(\omega) T})^-\\
&= 4\sum_{n=1}^{N_1(\epsilon, \omega)} a_{w,n}(1-e^{-(\mu-r)^\top \phi_{1,n}(\omega) T})^- <\infty .
\end{aligned}}
\]

Next, we consider the term $ U_{w, n} h_w(\phi_{1, n}, w_n) \rangle +2 \left(1-e^{-(\mu-r)^\top \phi_{1,n}T }\right)^- U_{w,n}^2$.
When $1-e^{-(\mu-r)^\top \phi_{1,n} T} \geq 0$,
\[
\fitdisplay{\begin{aligned}
&U_{w, n} h_w(\phi_{1, n}, w_n)  +2 \left(1-e^{-(\mu-r)^\top \phi_{1,n} T}\right)^- U_{w,n}^2 \\
=& (w_n - w^*)\left[ \left( 1-e^{-(\mu-r)^\top \phi_{1, n} T} \right)w_n + \left( x_0e^{-(\mu-r)^\top \phi_{1, n} T}-z \right) \right]\\
=& \left( 1-e^{-(\mu-r)^\top \phi_{1, n}T}\right)w_n^2 + \frac{1}{e^k - 1}\left\{ \left[ e^k(x_0 + z) -2x_0 \right] e^{-(\mu-r)^\top \phi_{1, n} T} - 2ze^k + z+x_0  \right\} w_n  \\
&- \frac{1}{e^k - 1}(ze^k - x_0)\left[ x_0e^{-(\mu-r)^\top \phi_{1, n} T}-z \right],
\end{aligned}}
\]
which is a convex quadratic function of $w_n$ with the minimum value
$$-\frac{(z-x_0)^2}{4(e^k-1)^2} \frac{(e^{-(\mu-r)^\top (\phi_{1,n}-\phi_{1}^*) T} - 1)^2}{1-e^{-(\mu-r)^\top \phi_{1,n} T}}  \leq0.$$
When $1-e^{-(\mu-r)^\top \phi_{1,n} T} < 0$,
\[
\fitdisplay{\begin{aligned}
&  U_{w, n} h_w(\phi_{1, n}, w_n)  +2 \left( 1-e^{-(\mu-r)^\top \phi_{1,n} T} \right)^- U_{w,n}^2\\
= & (w_n - w^*)\left[ \left( 1-e^{-(\mu-r)^\top \phi_{1, n} T} \right)w_n + \left( x_0e^{-(\mu-r)^\top \phi_{1, n} T}-z \right) \right] +2 \left( e^{-(\mu-r)^\top \phi_{1,n} T} - 1\right) U_{w,n}^2 \\
=& \left( e^{-(\mu-r)^\top \phi_{1, n} T}-1 \right) w_n^2 + \frac{1}{e^k-1}\left\{ e^{-(\mu-r)^\top \phi_{1,n} T} \left[ (-3z + x_0)e^k + 2x_0 \right] + 2ze^k - 3x_0 + z \right\} w_n  \\
&+ \frac{ze^k - x_0}{(e^k - 1)^2}\left\{ e^{-(\mu-r)^\top \phi_{1,n}T }\left[ (2z-x_0)e^k - x_0 \right] + 2x_0 - z-ze^k \right\},
\end{aligned}}
\]
which is also a convex quadratic function of $w_n$ with the minimum value of
$$-\frac{(z-x_0)^2}{4(e^k-1)^2} \frac{(e^{-(\mu-r)^\top (\phi_{1,n}-\phi_{1}^*) T} - 1)^2}{e^{-(\mu-r)^\top \phi_{1,n} T} - 1}\leq0.$$
To sum up, in both cases $ U_{w, n} h_w(\phi_{1, n}, w_n) +2 (1-e^{-(\mu-r)^\top \phi_{1,n}})^- U_{w,n}^2$ is a convex quadratic function of $w_n$ with the minimum value of $-M(\phi_{1,n})$. This implies that
$$\zeta_n = 2a_{w,n} \biggl(U_{w, n} h_w(\phi_{1, n}, w_n)  +2 (1-e^{-(\mu-r)^\top \phi_{1,n}})^- U_{w,n}^2+ M(\phi_{1,n})\biggl)\geq 0$$
is always true for any $n$.

Third, we aim to prove $\sum \eta_n < \infty$ almost surely. By Theorem \ref{thm:convergence_phi}, $\phi_{1,n} \to \phi_1^*$; hence, there exists $0 < N_2(\omega) < \infty$ such that $-1 < (\mu-r)^\top (\phi_{1,n}-\phi_1^*) T < 1$ for all $n \geq N_2(\omega)$. Additionally,  for any $\delta > 0$ there exists $N_3(\epsilon, \delta, \omega) > 0$ such that $|\frac{z-x_0}{e^k-1}(1-e^{-(\mu-r)^\top (\phi_{1, n} - \phi_1^*) T})| < \frac{\epsilon \delta}{2}$ when $n \geq N_3(\epsilon, \delta, \omega)$.

Choose $N(\epsilon, \delta, \omega) = \max\{ N_1(\epsilon, \omega), N_2(\omega), N_3(\epsilon, \delta, \omega)\}$. Notice that $(e^{-x}-1)^2 \leq 4x^2$ when $-1 \leq x \leq 1$. So when $n \geq N(\epsilon, \delta, \omega)$,
\[ (e^{-(\mu-r)^\top (\phi_{1,n}-\phi_{1}^*) T} - 1)^2 \leq 4 |(\mu-r)^\top (\phi_{1,n}-\phi_{1}^*)|^2 T^2 \leq  4T^2 |\mu-r|^2 |\phi_{1,n} - \phi_1^*|^2. \]
Furthermore, when $n \geq N(\epsilon, \delta, \omega)$, we have
\[
M(\phi_{1,n}) \leq \frac{(z-x_0)^2}{(e^k-1)^2} \frac{T^2 |\mu-r|^2 |\phi_{1,n} - \phi_1^*|^2 }{\epsilon} \leq C_{\epsilon} |\phi_{1,n}-\phi_{1}^*|^2.
\]


By Theorem \ref{phi_1_rate}, the definition of $\hat{\eta}_n$ in \eqref{eq:eta_hat} and the assumption \eqref{eq:thm3_assumption} on $\{a_{w,n}\}$, we know \[\sum_{n=1}^\infty a_{w,n} \E [|\phi_{1, n}-\phi_1^*|^2] \leq C^\prime \sum_{n=1}^{\infty} a_{w,n} a_n \hat{\eta}_n < \infty.\]
Consider the sequence $S_m = \sum_{n=1}^m a_{w,n} |\phi_{1, n}-\phi_1^*|^2$, which is a monotone increasing sequence and $S_m \to S=\sum_{n=1}^\infty a_{w,n} |\phi_{1, n}-\phi_1^*|^2$. By the monotone convergence theorem, we have $\E[S_m] \to \E[S] = \sum_{n=1}^\infty a_{w,n} \E [|\phi_{1, n}-\phi_1^*|^2] < \infty$. It follows that $S = \sum_{n=1}^\infty a_{w,n} |\phi_{1, n}-\phi_1^*|^2 < \infty$ almost surely. This implies $\sum_{n=1}^\infty a_{w,n} M(\phi_{1,n}) \leq \sum_{n=1}^{N(\epsilon,\delta,\omega) - 1}a_{w,n} M(\phi_{1,n}) + C_{\epsilon} \sum_{n=N(\epsilon,\delta,\omega)}^\infty a_{w,n} |\phi_{1, n}-\phi_1^*|^2 < \infty$ almost surely. Furthermore, if assumptions in \eqref{eq:thm1_assumption} in Theorem \ref{thm:convergence_phi} and assumptions in \eqref{eq:thm3_assumption} in Theorem \ref{thm_w_convergence} are satisfied, then we have $\sum \eta_n<\infty$.

The above analysis yields $\sum \gamma_n<\infty$, $\sum \eta_n<\infty$ and $\zeta_n$ is non-negative. It follows from  \citet[Theorem 1]{robbins1971convergence} that $\left|U_{w,n}\right|^2$ converges to a finite limit and $\sum \zeta_n<\infty$ almost surely.

Finally, we show $|U_{w,n}| \to 0$. Otherwise, there exists a set $Z\in \f$ with $\p(Z) > 0$, for every $\omega\in Z$, there exists $\delta(\omega) > 0$ such that for all $n$ sufficiently large, $|w_{n}(\omega) - w^{*}| \geq \delta(\omega) > 0$.
Consider the following function:
\[
f(\phi_{1, n},  w_n)= U_{w, n} h_w(\phi_{1, n}, w_n)  = (w_n - w^*)\left[ \left( 1-e^{-(\mu-r)^\top \phi_{1, n} T}\right) w_n + \left(x_0 e^{-(\mu-r)^\top \phi_{1, n} T}-z \right) \right].
\]
When $n>N(\epsilon, \delta, \omega)$, we have
\[
\fitdisplay{\begin{aligned}
f(\phi_{1, n}(\omega),  w^*+\delta(\omega))=& \delta(\omega)\left[ \left( 1-e^{-(\mu-r)^\top \phi_{1, n}(\omega) T} \right) \delta(\omega) + \frac{z-x_0}{e^k-1}\left( 1-e^{-(\mu-r)^\top (\phi_{1, n}(\omega) - \phi_1^*) T} \right) \right],
\end{aligned}}
\]
and
\[
\fitdisplay{\begin{aligned}
f(\phi_{1, n}(\omega),  w^*-\delta(\omega))=& -\delta(\omega)\left[ -\left( 1-e^{-(\mu-r)^\top \phi_{1, n}(\omega)}\right) \delta(\omega) + \frac{z-x_0}{e^k-1}\left( 1-e^{-(\mu-r)^\top (\phi_{1, n}(\omega) - \phi_1^*) T} \right) \right].
\end{aligned}}
\]
Recall that for $n > N(\epsilon, \delta, \omega)$, $|\frac{z-x_0}{e^k-1}(1-e^{-(\mu-r)^\top (\phi_{1, n}(\omega) - \phi_1^*)T})| < \frac{\epsilon \delta(\omega)}{2}$ holds true, and $f$ is a convex quadratic function of $w_n$ with one root to be $w^*$.
Then we have $f(\phi_{1, n}(\omega),  w^*+\delta(\omega)) \geq \frac{\epsilon \delta(\omega)^2}{2} > 0$ and $f(\phi_{1, n}(\omega),  w^*-\delta(\omega)) \geq \frac{\epsilon \delta(\omega)^2}{2} > 0$. Moreover, by the property of quadratic functions, we obtain $f(\phi_{1,n}(\omega), w) > \frac{\epsilon \delta(\omega)^2}{2} > 0$ for all $w\in (-\infty, w^*-\delta(\omega)] \cup [w^*+\delta(\omega), \infty)$. Thus, if $|w_n(\omega) - w^*| > \delta(\omega)$ for any $n > N(\epsilon, \delta, \omega)$,
\[ \fitdisplay{\begin{aligned}
\zeta_n(\omega) = & 2a_{w,n} U_{w,n}(\omega) h_w(\phi_{1, n}(\omega), w_n(\omega))  + 2a_{w,n}M(\phi_{1,n}(\omega)) \\
\geq & 2a_{w,n}  U_{w,n}(\omega) h_w(\phi_{1, n}(\omega) w_n(\omega))  \geq a_{w,n} \epsilon \delta(\omega)^2.
\end{aligned}}\]

Then
$$\sum_{n=1}^\infty \zeta_n(\omega) = \sum_{n=1}^{N(\epsilon, \delta, \omega) -1} \zeta_n(\omega) + \sum_{n=N(\epsilon, \delta, \omega)}^\infty \zeta_n(\omega) \geq \sum_{n=1}^{N(\epsilon, \delta, \omega) -1} \zeta_n(\omega)+\sum_{n=N(\epsilon, \delta, \omega)}^\infty a_{w,n} \epsilon \delta(\omega)^2 = \infty,$$
which contradicts the fact that $\sum_{n=1}^\infty \zeta_n<\infty$ almost surely. Therefore, $w_n \to w^*$ almost surely.

\end{proof}


Now,  Theorem \ref{thm:convergence_both} follows from combining Theorems \ref{thm:convergence_phi}, \ref{phi_1_rate}, \ref{thm_w_convergence}, and Remark \ref{remark1}.
\subsection{Proof of Theorem \ref{thm: more random higher variance}}
\label{appendix:proof variance}
We first recall a simple result regarding the inner product between two positive semi-definite matrices.
\begin{lemma}
For two matrices $M,N\in \mathbb S^d_{+}$, we have  $\langle M, N \rangle \geq 0$.
\label{lemma:positive semi-definite matrices}
\end{lemma}

\begin{proof}
Since $M\in \mathbb S^d_{+}$, it can be represented as $M = Q^\top D Q$, where $D=diag(\lambda_1, \lambda_2, ... \lambda_d)$ is a diagonal matrix with the diagonal entries being the (nonnegative) eigenvalues of $M$, and $Q = (q_1, q_2, ..., q_d)$ is a matrix consisting of the corresponding eigenvectors of $M$.
Then,
\begin{equation}
	\fitdisplay{\begin{aligned}
		\langle M, N \rangle \ &=\langle \sum_{i=1}^d \lambda_i q_i q_i^\top, N \rangle =\sum_{i=1}^d \lambda_i \langle q_i q_i^\top, N\rangle
		\\
		&
		=\sum_{i=1}^d \lambda_i ( q_i^\top N q_i)\geq 0,
	\end{aligned}}
\end{equation}
noting that $\lambda_i\geq 0$ and  $N\in \mathbb{S}^d_+$.
\end{proof}

We now prove Theorem \ref{thm: more random higher variance}. Note that the wealth processes \(x^{u^{\bm\pi}}\) and \(x^{\bm\pi}\) have identical distributions.
It follows from  \eqref{rl_dynamics} that the wealth processes $\{x^{\bm\pi}(t):0\leq t\leq T\}$ and $\{x^{\bm{\hat\pi}}(t):0\leq t\leq T\}$ follow the dynamics:
\[
\dd x^{\bm\pi}(t) = -(\mu-r)^\top \phi_1(x^{\bm\pi}(t) - w) \dd t + \sqrt{ \langle\sigma \sigma^\top,  \phi_1\phi_1^\top(x^{\bm\pi}(t)-w)^2 + C(t) \rangle} \dd W(t),
\]
and
\[
\dd x^{\bm{\hat\pi}}(t) = -(\mu-r)^\top \phi_1(x^{\bm{\hat\pi}}(t) - w) \dd t + \sqrt{ \langle\sigma \sigma^\top,  \phi_1\phi_1^\top(x^{\bm{\hat\pi}}(t)-w)^2 + \hat C(t) \rangle} \dd W(t).
\]
Taking integration and then expectation on both equations and denoting
$g(t) = \E[x^{\bm\pi}(t)]$ and $\hat g(t) = \E [x^{\hat{\bm\pi}}(t)]$, we find that $g$ and $\hat g$ satisfy the same ODE:

%

%
\begin{equation}\label{sameODE}
g ^\prime (t) = -A g(t) + Aw,\;\;g(0)=x_0; \;\;\hat g ^\prime (t) = -A \hat g(t) + Aw,\;\;\hat g(0)=x_0,
\end{equation}
where $A = (\mu-r)^\top \phi_1$.
%
The uniqueness of solution to this  ODE implies  $g\equiv  \hat g$ and, hence, $\E[x^{\bm\pi}(T)] = \E[x^{\hat{\bm\pi}}(T)]$.

Next, applying It\^o's formula to $(x^{\bm\pi}(t)) ^2$, and then integrating and taking expectation, we obtain that $k(t) = \E \left[ (x^{\bm\pi}(t))^2\right]$ satisfies
%
%
\begin{equation}
\label{eq:x_2_moment}
k^\prime(t) = (-2A+B)k(t) + 2w(A-B)g(t) + w^2 B + \langle \sigma \sigma^\top, C(t) \rangle,
\end{equation}
where $B=\langle \sigma \sigma^\top , \phi_1 \phi_1^\top\rangle$.
Similarly, $\hat k(t) = \E \left[ (x^{\hat{\bm\pi}}(t))^2\right]$ satisfies
\begin{equation}\label{eq:x_2_momenthat}
\hat k^\prime(t) = (-2A+B) \hat k(t) + 2w(A-B) \hat g(t) + w^2 B + \langle \sigma \sigma^\top, \hat C(t) \rangle.
\end{equation}
However, Lemma \ref{lemma:positive semi-definite matrices} yields
$\langle \sigma \sigma^\top, C(t) \rangle\geq \langle \sigma \sigma^\top, \hat C(t) \rangle$. Thus it follows from applying the comparison theorem of ODEs to
\eqref{eq:x_2_moment} and \eqref{eq:x_2_momenthat}
that
%
$k(t) \geq \hat k(t)$ $\forall t\in[0,T]$. The desired result that $\operatorname{Var}(x^{\bm\pi}(T)) \geq \operatorname{Var}(x^{\hat{\bm\pi}}(T))$ follows immediately.


\subsection{Proof of Theorem \ref{thm:regret0}}
\label{appendix:proof regret}
We first show that the Sharpe ratio is a function of just  $\phi_1$. For ease of exposition, the wealth process \(x^{\bm u}(t)\) will henceforth be denoted simply as \(x(t)\).
Indeed, under the deterministic policy \eqref{eq:deterministic_policy},  $\E [x(\cdot)]$ satisfies the same ODE \eqref{sameODE}. Solving it we get 
%
$$\E [x(t)] = w + (x_0-w)e^{-At}.$$
Moreover, solving the ODE \eqref{eq:x_2_moment} with $C=0$, we obtain
$$\E[x(t)^2] = e^{(-2A+B)t}(w-1)^2 - 2e^{-At}(w^2-w)+w^2.$$
Hence
\[
\operatorname{Var}(x(t)) = (x_0-w)^2e^{-2At}(e^{Bt} - 1),
\]
leading to
\begin{equation}
\label{eq:sr_def}
\operatorname{SR}(\phi_{1}) = \frac{(\E[x(T)] - x_0)/x_0}{\sqrt{\operatorname{Var}(x(T)/x_0)}} = \frac{e^{A T}-1}{\sqrt{e^{B T}-1}}.
\end{equation}


Next we prove that  $\operatorname{SR}(\phi_1)$ is uniformly bounded in $\phi_1 \in \mathbb{R} ^d$.
To this end, first note that $\operatorname{SR}(\phi_1)$ is a continuous function of $\phi_1$ except at $\phi_1=0$. Denote by $\lambda_{min}$ the smallest eigenvalue of the positive semi-definite matrix $\Sigma$. Then, on one hand, 
\[
\fitdisplay{\begin{aligned}
\limsup\limits_{|\phi_1|\rightarrow0} |\operatorname{SR}(\phi_1)| &\leq \limsup\limits_{|\phi_1|\rightarrow0} \frac{T |(\mu-r)^\top \phi_1 + \frac{1}{2}((\mu-r)^\top \phi_1)^2 + O( |\phi_1|^3)|}{\sqrt{T \phi_1^\top \Sigma \phi_1}}\\
&\leq \limsup\limits_{|\phi_1|\rightarrow0} \frac{|\mu-r| |\phi_1| + \frac{1}{2}|\mu-r|^2 |\phi_1|^2 + O( |\phi_1|^3)}{\sqrt{\lambda_{min} |\phi_1|^2}} \sqrt{T}\\
&= \frac{|\mu-r| \sqrt{T}}{\sqrt{\lambda_{min} }}.
\end{aligned}}
\]
On the other hand, note that $B=\phi_1^\top \Sigma \phi_1\geq \lambda_{min}|\phi_1|^2\to \infty$ as $|\phi_1|\to\infty$. In particular, when $|\phi_1| > \frac{1}{\sqrt{\lambda_{min} T}}$, $e^{BT} - 1 \geq \frac{1}{4}e^{BT}$. Therefore,
\[
\fitdisplay{\begin{aligned}
\limsup\limits_{|\phi_1|\rightarrow \infty} |\operatorname{SR}(\phi_1)| &\leq \limsup\limits_{|\phi_1|\rightarrow \infty} \frac{e^{AT}}{\sqrt{e^{BT}-1}} \leq \limsup\limits_{|\phi_1|\rightarrow \infty}  \frac{e^{|\mu-r||\phi_1| T}}{\sqrt{\frac{1}{4}e^{BT}}} \\
&\leq \limsup\limits_{|\phi_1|\rightarrow \infty}  2 e^{|\mu-r||\phi_1| T - \frac{1}{2} \phi_1^\top \Sigma \phi_1 T} \\
&\leq \limsup\limits_{|\phi_1|\rightarrow \infty}  2 e^{|\mu-r||\phi_1| T - \frac{1}{2} \lambda_{min} |\phi_1|^2 T} =0.   \\
\end{aligned}}
\]
It follows then  $|\operatorname{SR}(\phi_1)| \leq C_1$ $ \forall \phi_1 \in \mathbb{R}^d$ for some constant $C_1>0$.

Now, $\operatorname{SR}$ reaches its maximum at $\phi_1 = \phi_1^*$; hence  $\operatorname{SR}^\prime (\phi_1^*) = 0$. Next we show $\operatorname{SR}^{\prime \prime} (\phi_1^*) \leq 0$. Recall that $k=(\mu-r)^\top \Sigma^{-1}(\mu-r) T$,
\[
\fitdisplay{\begin{aligned}
\operatorname{SR}^{\prime \prime} (\phi_1^*) &= -\frac{1}{2} (e^k-1)^{-\frac{3}{2}} e^k[(e^k-1)\Sigma  T - (\mu-r)(\mu-r)^\top T^2],
\end{aligned}}
\]
where
\[
\fitdisplay{\begin{aligned}
(e^k-1)\Sigma  T - (\mu-r)(\mu-r)^\top T^2 &\geq k\Sigma T - (\mu-r)(\mu-r)^\top T^2\\
&= T^2((\mu-r)^\top \Sigma^{-1}(\mu-r) \Sigma - (\mu-r)(\mu-r)^\top).
\end{aligned}}
\]
Consider the matrix $(\mu-r)^\top \Sigma^{-1}(\mu-r) \Sigma - (\mu-r)(\mu-r)^\top$, by the Cauchy--Schwarz inequality, for any vector $x \in \mathbb{R}^d$,
\[
\fitdisplay{\begin{aligned}
&x^\top ((\mu-r)^\top \Sigma^{-1} (\mu-r) \Sigma - (\mu-r)(\mu-r)^\top) x \\
=& (\mu-r)^\top \Sigma^{-1} (\mu-r) x^\top \Sigma x - x^\top (\mu-r) (\mu-r)^\top x \\
=& (\mu-r)^\top \Sigma^{-1} (\mu-r) (x^\top \Sigma x) - (x^\top (\mu-r))^2 \\
\geq & 0.
\end{aligned}}
\]
Therefore, we have $\operatorname{SR}^{\prime \prime} (\phi_1^*) \leq 0$.

Fix a constant $\delta < |\phi_1^*|$. Then for any $\phi_1$ such that $|\phi_1-\phi_1^*|<\delta$, we have $\operatorname{SR}^{\prime \prime}(\phi_1) \succeq -\bar{C} I$ for some constant $\bar C>0$, because  $\operatorname{SR}^{\prime \prime}$ is continuous in this region.

By Taylor's expansion, for any $\phi_1$ with $|\phi_1-\phi_1^*|<\delta$, we have
\[
\fitdisplay{\begin{aligned}
\operatorname{SR}(\phi_1) - \operatorname{SR}(\phi_1^*) &=  \operatorname{SR}^\prime(\phi_1^*)(\phi_1-\phi_1^*) + \int_0^1 (1-t)(\phi_1-\phi_1^*)^\top \operatorname{SR}^{\prime \prime}(\phi_1^* + t(\phi_1 - \phi_1^*))(\phi_1-\phi_1^*) \dd t\\
&= \int_0^1 (1-t)(\phi_1-\phi_1^*)^\top \operatorname{SR}^{\prime \prime}(\phi_1^* + t(\phi_1 - \phi_1^*))(\phi_1-\phi_1^*) \dd t\\
&\geq -\int_0^1 (1-t)\bar{C} |\phi_1 - \phi_1^*|^2 \dd t= -\frac{1}{2} \bar{C} |\phi_1 - \phi_1^*|^2,
\end{aligned}}
\]
or $\operatorname{SR}(\phi_1^*) - \operatorname{SR}(\phi_1) \leq \frac{1}{2} \bar{C} |\phi_1 - \phi_1^*|^2$.

Recall that  Theorem \ref{thm:convergence_both}-(b) yields that
\[
\fitdisplay{\begin{aligned}
\E [|\phi_{1,n} - \phi_1^*|^2] &\leq C\frac{(\log (n-1))^p \log \log (n-1)}{n-1}\\
&\leq C\frac{(\log n)^p \log \log n}{n-1}\\
&=C\frac{(\log n)^p \log \log n}{n} * \frac{n}{n-1}\\
&\leq \check{C} \frac{(\log n)^p \log \log n}{n},
\end{aligned}}
\]
where \(\check{C}\) is a constant independent of \(n\).

Set $\delta_n ^\prime = (4 \frac{C_1 \check{C} }{\bar{C}} \frac{(\log n)^p \log \log n}{n})^{\frac{1}{4}}$, $n\in \mathbb{N}$, and $n_0 = \inf\{n: \delta_n^\prime < \delta\}$. Further,  define $\delta_n = \delta$ for $n < n_0$, and $\delta_n = \delta_n^\prime$ for $n \geq n_0$.
Then, for  $n\in \mathbb{N}$, we have
\[
\fitdisplay{\begin{aligned}
&\E [\operatorname{SR}(\phi_1^*) - \operatorname{SR}(\phi_{1,n})] \\
&= \int_{|\phi_{1,n}-\phi_1^*| \leq \delta_n} [\operatorname{SR}(\phi_1^*) - \operatorname{SR}(\phi_{1,n})] \dd \mathbb{P} + \int_{|\phi_{1,n}-\phi_1^*| > \delta_n} [\operatorname{SR}(\phi_1^*) - \operatorname{SR}(\phi_{1,n})] \dd \mathbb{P}\\
&\leq \int_{|\phi_{1,n}-\phi_1^*| \leq \delta_n} \frac{1}{2} \bar{C} |\phi_{1,n}-\phi_1^*|^2 \dd \mathbb{P} + \int_{|\phi_{1,n}-\phi_1^*| > \delta_n} 2C_1 \dd \mathbb{P}\\
&\leq \frac{1}{2} \bar{C}\delta_n^2 + 2C_1 \mathbb{P}(|\phi_{1,n}-\phi_1^*| > \delta_n). 
\end{aligned}}
\]
When $n<n_0$, we have $\E [\operatorname{SR}(\phi_1^*) - \operatorname{SR}(\phi_{1,n})] \leq \frac{1}{2} \bar{C}\delta^2 + 2C_1$. When $n>n_0$, we have
\[
\fitdisplay{\begin{aligned}
&\E [\operatorname{SR}(\phi_1^*) - \operatorname{SR}(\phi_{1,n})] \\
&\leq \frac{1}{2} \bar{C}\delta_n^2 + 2C_1 \mathbb{P}(|\phi_{1,n}-\phi_1^*| > \delta_n)\\
&\leq \frac{1}{2} \bar{C}\delta_n^2 + 2C_1 \frac{1}{\delta_n^2} \E [|\phi_{1,n}-\phi_1^*|^2]\\
&\leq \frac{1}{2} \bar{C}\delta_n^2 + 2C_1 \frac{\check{C}}{\delta_n^2} \frac{(\log n)^p \log \log n}{n}\\
&=2 \sqrt{\bar{C} C_1 \check{C} \frac{(\log n)^p \log \log n}{n}}.
\end{aligned}}
\]

Consequently,
\[
\fitdisplay{\begin{aligned}
&\E [\sum_{n=1}^{N} (\operatorname{SR}(\phi_1^*) - \operatorname{SR}(\phi_{1,n}))] \\
&= \sum_{n=1}^{N} \E [\operatorname{SR}(\phi_1^*) - \operatorname{SR}(\phi_{1,n})]\\
&=\sum_{n=1}^{n_0} \E [\operatorname{SR}(\phi_1^*) - \operatorname{SR}(\phi_{1,n})] + \sum_{n=n_0}^{N} \E [\operatorname{SR}(\phi_1^*) - \operatorname{SR}(\phi_{1,n})]\\
&\leq (\frac{1}{2} \bar{C}\delta^2 + 2C_1)n_0 + 2\sum_{n=n_0}^{N} \sqrt{\bar{C}C_1\check{C}\frac{(\log n)^p \log \log n}{n}}\\
&\leq C + C \sqrt{N (\log N)^p \log\log N}.
\end{aligned}}
\]
The proof is complete.

\newpage
\section{Additional Empirical Analysis}
\label{appendix:addition empirical results}

\subsection{Focused wealth trajectory comparison}\label{subsec:focused-wealth}

To provide a clearer visual comparison, we further plot the wealth trajectories of four representative strategies: CTRL,  two model-based  mean--variance strategies (mv and ctmv), and the equally weighted portfolio (ew) which is  a strong and widely used benchmark. For each method, we report the median wealth trajectory across the 100 independent experiments, together with the interquartile range to reflect cross-sectional dispersion.

\begin{figure}[H]
    \centering
    \includegraphics[width=0.9\textwidth]{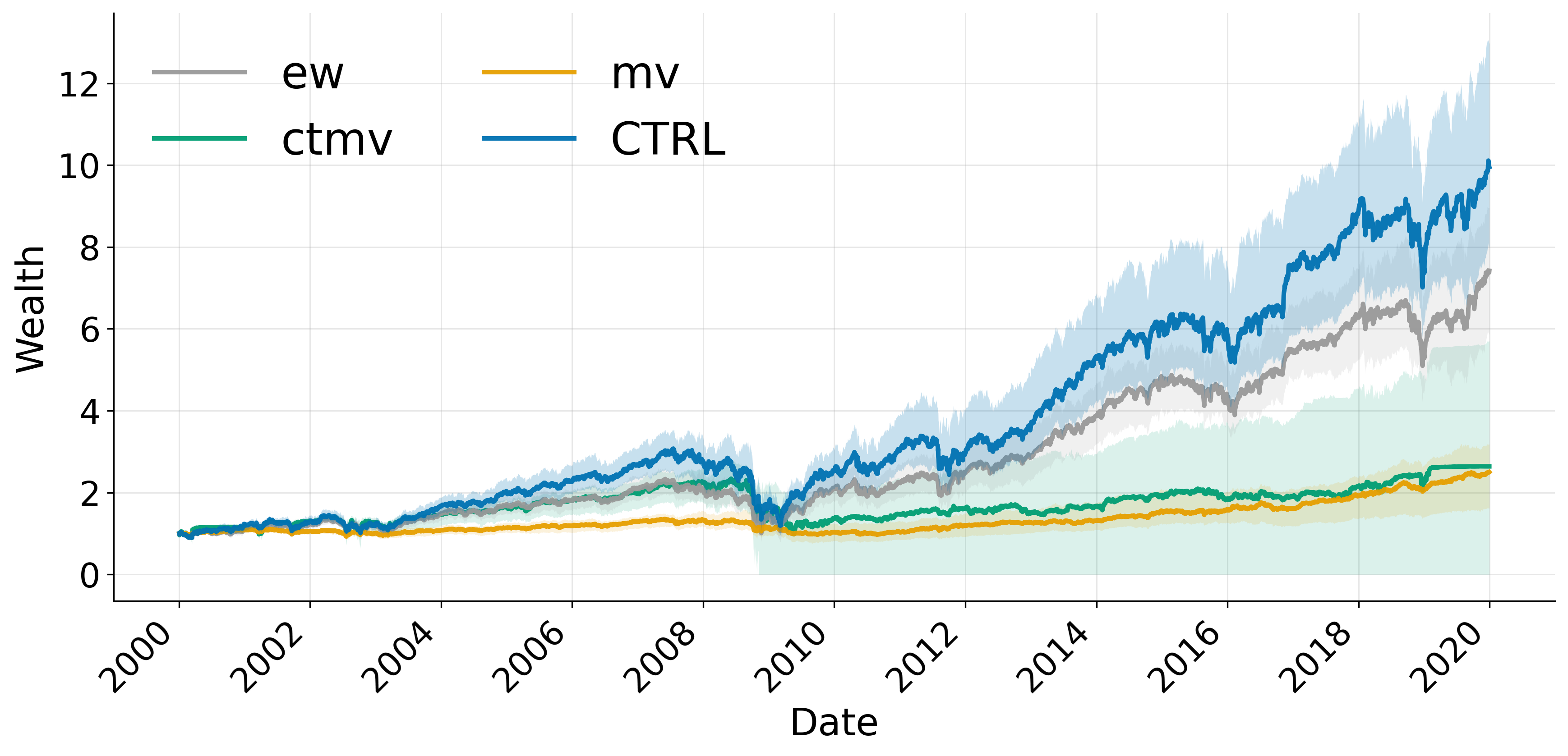}
    \caption{Median wealth trajectories with interquartile ranges for CTRL, mv, ctmv, and ew over the full sample period.}
    \label{fig:selected-wealth}
\end{figure}

Figure~\ref{fig:selected-wealth} shows that the median wealth path of CTRL lies well above those of the other three over nearly the entire period. In particular, the gap between CTRL and the two model-based MV strategies  is persistent, highlighting the advantage of bypassing model parameter estimation. 
Overall, this focused comparison further confirms that the model-free continuous-time RL method delivers materially stronger and more stable long-run growth than its model-based counterparts and other general-purpose RL algorithms.

\newpage
\subsection{Distribution of Sharpe ratios}\label{subsec:box-sharpe}

\begin{figure}[H]
    \centering
    \includegraphics[width=0.95\textwidth]{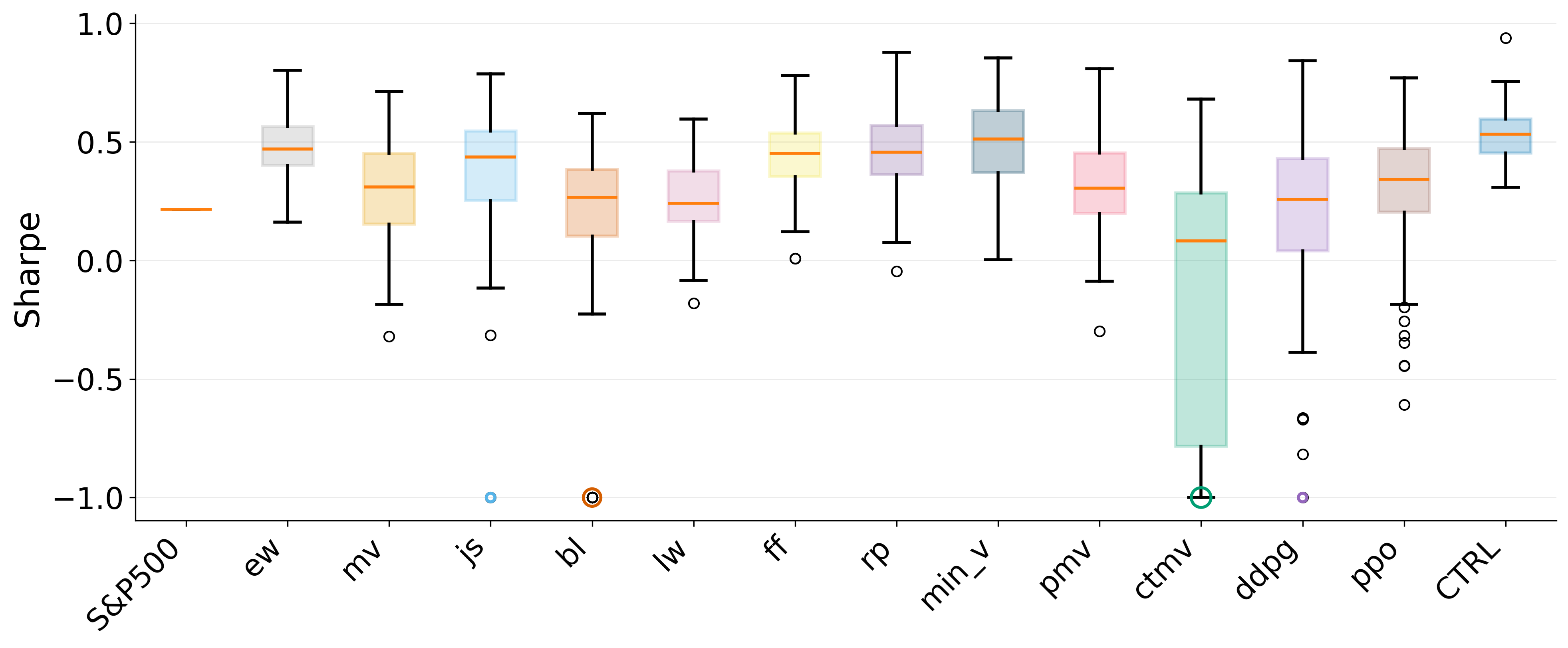}
    \caption{Box plots of Sharpe ratios over the full sample period across 100 experiments. Extremely negative values are truncated at $-1$.}
    \label{fig:box-sharpe}
\end{figure}

To further assess cross-sectional stability, we present in Figure~\ref{fig:box-sharpe} the box plots of Sharpe ratios over the full period across the 100 independent experiments. A box represents the interquartile range, its central line denotes the median, and the whiskers capture the dispersion beyond the quartiles. Moreover, we plot extreme observations as outliers, and truncate extremely negative Sharpe ratio values at $-1$.

Clearly, CTRL has the highest median Sharpe ratio among all the competing methods. In addition, its interquartile range is relatively tight and whiskers short, indicating limited dispersion across experiments and hence strong stability. By contrast, most alternative strategies exhibit wider spreads and pronounced downside outliers. In particular, several model-based mean--variance variants (such as js, bl, and ctmv) and RL-based ddpg display extremely negative Sharpe realizations in some experiments, reflecting substantial instability. CTRL, on the other hand, exhibits only a single mild outlier on the upside, with no extreme downside realizations.

The box plots reinforce the takeaway message of the empirical study: CTRL not only delivers the highest central performance, but also maintains a more concentrated and resilient distribution of risk-adjusted returns across repeated experiments.

\newpage
\subsection{Pairwise test for statistical significance}
\label{appendix:paired test}

We conduct pairwise comparisons using the paired Wilcoxon test, a non-parametric alternative to the paired t-test, to assess the statistical significance of differences in Sharpe ratios among all the strategies during  the full 20-year period.

\begin{table}[H]
\centering
\scriptsize
\resizebox{\textwidth}{!}{%
\begin{tabular}{lrrrrrrrrrrrrrr}
\toprule
 & S\&P500 & ew & mv & js & bl & lw & ff & rp & min\_v & pmv & ctmv & ddpg & ppo & CTRL \\
\midrule
S\&P500 &  & 1 & 1 & 1 & 1 & 0.953 & 1 & 1 & 1 & 0.997 & 0.143 & 0.826 & 0.999 & 1 \\
ew & 2.01e-18 &  & 3.83e-16 & 8.1e-06 & 8.86e-16 & 4.01e-18 & 1.12e-07 & 0.274 & 0.578 & 1.34e-14 & 2.97e-16 & 6.71e-13 & 2.99e-12 & 0.999 \\
mv & 0.000184 & 1 &  & 1 & 0.941 & 9.82e-05 & 1 & 1 & 1 & 0.248 & 0.000963 & 0.0824 & 0.561 & 1 \\
js & 1.98e-12 & 1 & 1.79e-16 &  & 1.1e-06 & 8.77e-15 & 0.974 & 0.999 & 1 & 2.69e-09 & 2.28e-11 & 2.61e-06 & 0.000232 & 1 \\
bl & 0.000381 & 1 & 0.0589 & 1 &  & 0.00166 & 1 & 1 & 1 & 0.0317 & 0.000191 & 0.0463 & 0.373 & 1 \\
lw & 0.047 & 1 & 1 & 1 & 0.998 &  & 1 & 1 & 1 & 0.954 & 0.0527 & 0.462 & 0.958 & 1 \\
ff & 4.37e-17 & 1 & 2.37e-17 & 0.0264 & 9.37e-16 & 2.72e-18 &  & 1 & 1 & 1.65e-11 & 3.86e-14 & 1.74e-08 & 1.49e-06 & 1 \\
rp & 6.87e-18 & 0.726 & 6.01e-14 & 0.00107 & 8.76e-12 & 5.52e-16 & 0.000349 &  & 0.808 & 8.39e-13 & 6.01e-14 & 3.03e-10 & 4.87e-09 & 1 \\
min\_v & 1.99e-17 & 0.422 & 6.28e-18 & 2.78e-12 & 4.39e-13 & 3.36e-17 & 2.53e-06 & 0.192 &  & 1.58e-15 & 1.94e-16 & 7.16e-11 & 2.1e-09 & 0.983 \\
pmv & 0.00282 & 1 & 0.752 & 1 & 0.968 & 0.0463 & 1 & 1 & 1 &  & 0.00177 & 0.118 & 0.732 & 1 \\
ctmv & 0.857 & 1 & 0.999 & 1 & 1 & 0.947 & 1 & 1 & 1 & 0.998 &  & 0.971 & 0.999 & 1 \\
ddpg & 0.174 & 1 & 0.918 & 1 & 0.954 & 0.538 & 1 & 1 & 1 & 0.882 & 0.0293 &  & 0.981 & 1 \\
ppo & 0.000744 & 1 & 0.439 & 1 & 0.627 & 0.0422 & 1 & 1 & 1 & 0.268 & 0.000744 & 0.0191 &  & 1 \\
CTRL & 1.95e-18 & 0.000827 & 1.38e-16 & 3.38e-07 & 2.82e-15 & 4.01e-18 & 1.96e-08 & 2.83e-05 & 0.0168 & 1.34e-15 & 2.17e-17 & 4.4e-14 & 6.01e-14 &  \\
\bottomrule
\end{tabular}
}
\caption{One-sided paired Wilcoxon p-values on per-seed Sharpe. Entry (i,j) tests $Sharpe_i > Sharpe_j$.}
\label{tab:wilcoxon_sharpe_pvals}
\end{table}

Table~\ref{tab:wilcoxon_sharpe_pvals} shows that CTRL significantly outperforms all the competing methods in terms of the Sharpe ratio under the Wilcoxon paired test, with all the p-values below 0.05 and all but one below 0.01.

\subsection{Results under a lower transaction cost (12.5 bps)}
\label{subsec:comp-tc}

We report in this subsection the performance and trading statistics under a lower transaction cost of 12.5 bps in Tables \ref{tab:performance_25bps} and \ref{tab:trade_25bps}. The main conclusions of the paper remain intact: CTRL continues to deliver the highest Sharpe ratio, while maintaining the lowest turnover and a moderate single-asset concentration.

Comparing with the case of 50 bps transaction cost reported in the main paper, we see again that  the outperformance  of CTRL is widened as transaction cost increases, due clearly to its intrinsic stability and efficiency. 

\begin{table}[H]
\centering
\begin{tabular}{lrrrrrrr}
\toprule
 & Return & Volatility & Sharpe & Sortino & Calmar & MDD & RT \\
\midrule
S\&P500 & \makecell[r]{4.077\% \\ (0.000\%)} & \makecell[r]{0.189 \\ (0.000)} & \makecell[r]{0.216 \\ (0.000)} & \makecell[r]{0.341 \\ (0.000)} & \makecell[r]{0.072 \\ (0.000)} & \makecell[r]{0.568 \\ (0.000)} & \makecell[r]{1021 \\ (0)} \\
ew & \makecell[r]{10.247\% \\ (0.185\%)} & \makecell[r]{0.218 \\ (0.002)} & \makecell[r]{0.478 \\ (0.011)} & \makecell[r]{0.782 \\ (0.018)} & \makecell[r]{0.180 \\ (0.005)} & \makecell[r]{0.586 \\ (0.008)} & \makecell[r]{551 \\ (31)} \\
mv & \makecell[r]{3.970\% \\ (0.275\%)} & \makecell[r]{0.148 \\ (0.002)} & \makecell[r]{0.287 \\ (0.020)} & \makecell[r]{0.463 \\ (0.031)} & \makecell[r]{0.106 \\ (0.008)} & \makecell[r]{0.447 \\ (0.013)} & \makecell[r]{1359 \\ (70)} \\
js & \makecell[r]{7.246\% \\ (0.540\%)} & \makecell[r]{0.240 \\ (0.019)} & \makecell[r]{0.386 \\ (0.023)} & \makecell[r]{0.626 \\ (0.037)} & \makecell[r]{0.151 \\ (0.011)} & \makecell[r]{0.579 \\ (0.016)} & \makecell[r]{1126 \\ (65)} \\
bl & \makecell[r]{4.932\% \\ (1.960\%)} & \makecell[r]{0.618 \\ (0.092)} & \makecell[r]{0.189 \\ (0.027)} & \makecell[r]{0.340 \\ (0.041)} & \makecell[r]{0.076 \\ (0.020)} & \makecell[r]{0.851 \\ (0.013)} & \makecell[r]{1250 \\ (79)} \\
lw & \makecell[r]{3.506\% \\ (0.227\%)} & \makecell[r]{0.154 \\ (0.002)} & \makecell[r]{0.242 \\ (0.016)} & \makecell[r]{0.391 \\ (0.026)} & \makecell[r]{0.090 \\ (0.006)} & \makecell[r]{0.462 \\ (0.013)} & \makecell[r]{1487 \\ (78)} \\
ff & \makecell[r]{7.662\% \\ (0.227\%)} & \makecell[r]{0.183 \\ (0.002)} & \makecell[r]{0.430 \\ (0.014)} & \makecell[r]{0.690 \\ (0.024)} & \makecell[r]{0.163 \\ (0.006)} & \makecell[r]{0.503 \\ (0.011)} & \makecell[r]{893 \\ (43)} \\
rp & \makecell[r]{9.278\% \\ (0.258\%)} & \makecell[r]{0.216 \\ (0.004)} & \makecell[r]{0.450 \\ (0.015)} & \makecell[r]{0.729 \\ (0.025)} & \makecell[r]{0.166 \\ (0.006)} & \makecell[r]{0.594 \\ (0.011)} & \makecell[r]{887 \\ (62)} \\
min\_v & \makecell[r]{8.791\% \\ (0.304\%)} & \makecell[r]{0.183 \\ (0.002)} & \makecell[r]{0.494 \\ (0.019)} & \makecell[r]{0.806 \\ (0.031)} & \makecell[r]{0.191 \\ (0.009)} & \makecell[r]{0.499 \\ (0.011)} & \makecell[r]{1001 \\ (45)} \\
pmv & \makecell[r]{1.916\% \\ (0.159\%)} & \makecell[r]{\textbf{0.085} \\ (0.001)} & \makecell[r]{0.231 \\ (0.019)} & \makecell[r]{0.379 \\ (0.031)} & \makecell[r]{0.086 \\ (0.008)} & \makecell[r]{\textbf{0.270} \\ (0.008)} & \makecell[r]{1627 \\ (123)} \\
ctmv & \makecell[r]{-10.849\% \\ (14.325\%)} & \makecell[r]{1.282 \\ (0.206)} & \makecell[r]{0.049 \\ (0.051)} & \makecell[r]{0.259 \\ (0.090)} & \makecell[r]{-0.068 \\ (0.136)} & \makecell[r]{0.834 \\ (0.022)} & \makecell[r]{1922 \\ (112)} \\
ddpg & \makecell[r]{3.865\% \\ (0.777\%)} & \makecell[r]{0.218 \\ (0.012)} & \makecell[r]{0.202 \\ (0.031)} & \makecell[r]{0.330 \\ (0.049)} & \makecell[r]{0.094 \\ (0.012)} & \makecell[r]{0.552 \\ (0.018)} & \makecell[r]{1226 \\ (88)} \\
ppo & \makecell[r]{4.270\% \\ (0.449\%)} & \makecell[r]{0.149 \\ (0.006)} & \makecell[r]{0.259 \\ (0.027)} & \makecell[r]{0.425 \\ (0.044)} & \makecell[r]{0.107 \\ (0.010)} & \makecell[r]{0.443 \\ (0.016)} & \makecell[r]{1173 \\ (110)} \\
CTRL & \makecell[r]{\textbf{12.115\%} \\ (0.219\%)} & \makecell[r]{0.232 \\ (0.003)} & \makecell[r]{\textbf{0.527} \\ (0.010)} & \makecell[r]{\textbf{0.858} \\ (0.017)} & \makecell[r]{\textbf{0.199} \\ (0.005)} & \makecell[r]{0.621 \\ (0.009)} & \makecell[r]{\textbf{523} \\ (30)} \\
\bottomrule
\end{tabular}
\caption{\textbf{Net-of-cost performance over the full sample period (2000--2020) under 12.5 bps proportional transaction costs.} We report annualized return, volatility, Sharpe ratio, Sortino ratio, and Calmar ratio, together with maximum drawdown (MDD) and recovery time (RT) in days, over 100 independent experiments. Each experiment is based on 10 randomly selected stocks (except for S\&P 500 index). Numbers in parentheses are the standard errors across 100 experiments.}
\label{tab:performance_25bps}
\end{table}

\begin{table}[H]
\centering
\begin{tabular}{lrrrrrr}
\toprule
 & BKR & TO & MaxL & MaxS & GE & GEMax \\
\midrule
ew & 0\% & 5.798\% & 10.000\% & 0.000\% & 100.000\% & 100.000\% \\
mv & 0\% & 16.435\% & 113.971\% & 61.028\% & 113.558\% & 260.610\% \\
js & 1\% & 15.344\% & 171.726\% & 106.333\% & 140.192\% & 542.149\% \\
bl & 16\% & 15.903\% & 271.773\% & 150.030\% & 149.242\% & 1000.542\% \\
lw & 0\% & 18.744\% & 116.278\% & 82.500\% & 117.245\% & 389.493\% \\
ff & 0\% & 8.840\% & 116.479\% & 53.619\% & 96.796\% & 274.978\% \\
rp & 0\% & 11.191\% & 100.000\% & 0.000\% & 100.000\% & 100.000\% \\
min\_v & 0\% & 8.805\% & 104.477\% & 31.623\% & 126.789\% & 227.462\% \\
pmv & 0\% & 52.133\% & 75.980\% & 58.536\% & 75.156\% & 231.473\% \\
ctmv & 40\% & 40.668\% & 1877.490\% & 858.394\% & 145.358\% & 7756.333\% \\
ddpg & 1\% & 19.674\% & 77.880\% & 70.666\% & 156.801\% & 532.347\% \\
ppo & 0\% & 20.936\% & 71.240\% & 51.967\% & 94.570\% & 360.815\% \\
CTRL & 0\% & 5.679\% & 18.778\% & 1.903\% & 102.301\% & 124.034\% \\
\bottomrule
\end{tabular}
\caption{\textbf{Trading and exposure characteristics under 12.5 bps proportional transaction costs over the full sample period (2000--2020).}
	BKR denotes the bankruptcy rate across experiments. TO is the average monthly turnover.
	MaxL and MaxS report the maximum single-asset long and short positions, respectively.
	GE is the average gross exposure, and GEMax is the maximum gross exposure observed across all experiments.}
\label{tab:trade_25bps}
\end{table}

\subsection{Different market regimes}\label{subsec:comp-regime}

Here we include the performance and trading tables for both bear (2000--2010, see Tables \ref{tab:performance_bear} and \ref{tab:trade_bear}) and bull (2010--2020, see Tables \ref{tab:performance_bull} and \ref{tab:trade_bull}) subperiods. In the bear period, CTRL achieves the highest average return and the highest risk-adjusted performance including the Sharpe ratio, with a  significant margin over the other methods, and indeed the performance gap is even more pronounced in this adverse regime than in the full period. On the other hand, during the bull market, while many strategies perform well and performance differences narrow, CTRL remains competitive across key metrics.

\begin{table}[H]
\centering
\renewcommand{\arraystretch}{0.94}
\begin{tabular}{lrrrrrrr}
\toprule
 & Return & Volatility & Sharpe & Sortino & Calmar & MDD & RT \\
\midrule
S\&P500 & \makecell[r]{-2.633\% \\ (0.000\%)} & \makecell[r]{0.222 \\ (0.000)} & \makecell[r]{-0.118 \\ (0.000)} & \makecell[r]{-0.190 \\ (0.000)} & \makecell[r]{-0.046 \\ (0.000)} & \makecell[r]{0.568 \\ (0.000)} & \makecell[r]{N/A \\ (N/A)} \\
ew & \makecell[r]{7.467\% \\ (0.286\%)} & \makecell[r]{0.252 \\ (0.003)} & \makecell[r]{0.301 \\ (0.012)} & \makecell[r]{0.497 \\ (0.020)} & \makecell[r]{0.133 \\ (0.006)} & \makecell[r]{0.585 \\ (0.008)} & \makecell[r]{259 \\ (3)} \\
mv & \makecell[r]{-0.166\% \\ (0.358\%)} & \makecell[r]{0.169 \\ (0.004)} & \makecell[r]{0.013 \\ (0.019)} & \makecell[r]{0.027 \\ (0.030)} & \makecell[r]{0.011 \\ (0.008)} & \makecell[r]{0.431 \\ (0.012)} & \makecell[r]{1488 \\ (17)} \\
js & \makecell[r]{0.242\% \\ (0.636\%)} & \makecell[r]{0.282 \\ (0.026)} & \makecell[r]{0.064 \\ (0.020)} & \makecell[r]{0.105 \\ (0.031)} & \makecell[r]{0.027 \\ (0.010)} & \makecell[r]{0.570 \\ (0.016)} & \makecell[r]{476 \\ (5)} \\
bl & \makecell[r]{-3.817\% \\ (1.830\%)} & \makecell[r]{0.731 \\ (0.123)} & \makecell[r]{-0.030 \\ (0.020)} & \makecell[r]{-0.021 \\ (0.032)} & \makecell[r]{-0.030 \\ (0.018)} & \makecell[r]{0.850 \\ (0.014)} & \makecell[r]{\textbf{223} \\ (2)} \\
lw & \makecell[r]{0.411\% \\ (0.338\%)} & \makecell[r]{0.178 \\ (0.003)} & \makecell[r]{0.041 \\ (0.019)} & \makecell[r]{0.071 \\ (0.030)} & \makecell[r]{0.024 \\ (0.008)} & \makecell[r]{0.440 \\ (0.012)} & \makecell[r]{1790 \\ (43)} \\
ff & \makecell[r]{2.348\% \\ (0.343\%)} & \makecell[r]{0.212 \\ (0.004)} & \makecell[r]{0.124 \\ (0.016)} & \makecell[r]{0.203 \\ (0.026)} & \makecell[r]{0.057 \\ (0.007)} & \makecell[r]{0.499 \\ (0.011)} & \makecell[r]{264 \\ (2)} \\
rp & \makecell[r]{5.345\% \\ (0.462\%)} & \makecell[r]{0.261 \\ (0.006)} & \makecell[r]{0.226 \\ (0.017)} & \makecell[r]{0.371 \\ (0.028)} & \makecell[r]{0.101 \\ (0.008)} & \makecell[r]{0.592 \\ (0.011)} & \makecell[r]{1198 \\ (30)} \\
min\_v & \makecell[r]{3.495\% \\ (0.376\%)} & \makecell[r]{0.214 \\ (0.003)} & \makecell[r]{0.175 \\ (0.018)} & \makecell[r]{0.287 \\ (0.030)} & \makecell[r]{0.084 \\ (0.009)} & \makecell[r]{0.496 \\ (0.011)} & \makecell[r]{1254 \\ (23)} \\
pmv & \makecell[r]{0.743\% \\ (0.201\%)} & \makecell[r]{\textbf{0.089} \\ (0.001)} & \makecell[r]{0.093 \\ (0.023)} & \makecell[r]{0.155 \\ (0.037)} & \makecell[r]{0.056 \\ (0.013)} & \makecell[r]{\textbf{0.232} \\ (0.007)} & \makecell[r]{1213 \\ (46)} \\
ctmv & \makecell[r]{-7.662\% \\ (13.857\%)} & \makecell[r]{1.310 \\ (0.172)} & \makecell[r]{0.027 \\ (0.048)} & \makecell[r]{0.223 \\ (0.089)} & \makecell[r]{-0.033 \\ (0.132)} & \makecell[r]{0.819 \\ (0.024)} & \makecell[r]{1178 \\ (49)} \\
ddpg & \makecell[r]{0.794\% \\ (0.558\%)} & \makecell[r]{0.169 \\ (0.011)} & \makecell[r]{0.053 \\ (0.025)} & \makecell[r]{0.093 \\ (0.040)} & \makecell[r]{0.033 \\ (0.011)} & \makecell[r]{0.442 \\ (0.017)} & \makecell[r]{799 \\ (57)} \\
ppo & \makecell[r]{1.732\% \\ (0.301\%)} & \makecell[r]{0.146 \\ (0.005)} & \makecell[r]{0.111 \\ (0.019)} & \makecell[r]{0.182 \\ (0.031)} & \makecell[r]{0.050 \\ (0.007)} & \makecell[r]{0.402 \\ (0.015)} & \makecell[r]{856 \\ (57)} \\
CTRL & \makecell[r]{\textbf{9.788\%} \\ (0.297\%)} & \makecell[r]{0.269 \\ (0.004)} & \makecell[r]{\textbf{0.367} \\ (0.011)} & \makecell[r]{\textbf{0.600} \\ (0.018)} & \makecell[r]{\textbf{0.161} \\ (0.005)} & \makecell[r]{0.621 \\ (0.009)} & \makecell[r]{264 \\ (2)} \\
\bottomrule
\end{tabular}
\caption{\textbf{Comparison of out-of-sample performance of different allocation methods in the bear period (2000-2010)} We report annualized return, volatility, Sharpe ratio, Sortino ratio, and Calmar ratio, together with maximum drawdown (MDD) and recovery time (RT) in days, over 100 independent experiments. Each experiment is based on 10 randomly selected stocks (except for S\&P 500 index). For each cell, the upper number is the average (over the 100 experiments) while the lower one with parentheses is the standard error. }
\label{tab:performance_bear}
\end{table}

\begin{table}[H]
\centering
\begin{tabular}{lrrrrrr}
\toprule
 & BKR & TO & MaxL & MaxS & GE & GEMax \\
\midrule
ew & 0\% & 6.559\% & 10.000\% & 0.000\% & 100.000\% & 100.000\% \\
mv & 0\% & 13.854\% & 81.364\% & 61.028\% & 94.358\% & 260.610\% \\
js & 1\% & 13.370\% & 136.498\% & 106.333\% & 114.192\% & 461.708\% \\
bl & 16\% & 18.404\% & 271.773\% & 150.030\% & 130.610\% & 1000.542\% \\
lw & 0\% & 16.854\% & 100.599\% & 82.500\% & 103.348\% & 389.493\% \\
ff & 0\% & 10.359\% & 93.438\% & 53.619\% & 94.511\% & 274.978\% \\
rp & 0\% & 17.600\% & 100.000\% & 0.000\% & 100.000\% & 100.000\% \\
min\_v & 0\% & 8.233\% & 71.733\% & 28.276\% & 112.341\% & 187.259\% \\
pmv & 0\% & 44.051\% & 59.308\% & 49.418\% & 62.703\% & 216.246\% \\
ctmv & 39\% & 43.790\% & 1532.915\% & 865.850\% & 143.487\% & 6332.816\% \\
ddpg & 1\% & 13.557\% & 54.976\% & 52.558\% & 86.592\% & 321.775\% \\
ppo & 0\% & 15.316\% & 56.158\% & 47.462\% & 69.909\% & 259.993\% \\
CTRL & 0\% & 6.482\% & 18.778\% & 1.903\% & 102.096\% & 123.150\% \\
\bottomrule
\end{tabular}
\caption{\textbf{Trading and exposure statistics over the bear period 2000--2010.}
	BKR is the bankruptcy rate across experiments,  TO is the average monthly turnover,
	MaxL and MaxS are the maximum single-asset long and short positions, respectively,
	GE is the average gross exposure, and GEMax is the maximum gross exposure observed across all experiments.}
\label{tab:trade_bear}
\end{table}

\begin{table}[H]
\centering
\renewcommand{\arraystretch}{0.94}
\begin{tabular}{lrrrrrrr}
\toprule
 & Return & Volatility & Sharpe & Sortino & Calmar & MDD & RT \\
\midrule
S\&P500 & \makecell[r]{11.070\% \\ (0.000\%)} & \makecell[r]{0.148 \\ (0.000)} & \makecell[r]{0.750 \\ (0.000)} & \makecell[r]{1.173 \\ (0.000)} & \makecell[r]{0.560 \\ (0.000)} & \makecell[r]{0.198 \\ (0.000)} & \makecell[r]{\textbf{81} \\ (0)} \\
ew & \makecell[r]{13.335\% \\ (0.248\%)} & \makecell[r]{0.177 \\ (0.002)} & \makecell[r]{0.772 \\ (0.020)} & \makecell[r]{1.249 \\ (0.032)} & \makecell[r]{0.526 \\ (0.018)} & \makecell[r]{0.270 \\ (0.006)} & \makecell[r]{195 \\ (11)} \\
mv & \makecell[r]{8.873\% \\ (0.375\%)} & \makecell[r]{0.123 \\ (0.001)} & \makecell[r]{0.736 \\ (0.031)} & \makecell[r]{1.224 \\ (0.052)} & \makecell[r]{0.515 \\ (0.031)} & \makecell[r]{0.207 \\ (0.008)} & \makecell[r]{303 \\ (25)} \\
js & \makecell[r]{\textbf{15.379\%} \\ (0.602\%)} & \makecell[r]{0.177 \\ (0.006)} & \makecell[r]{0.941 \\ (0.031)} & \makecell[r]{1.564 \\ (0.053)} & \makecell[r]{0.761 \\ (0.034)} & \makecell[r]{0.240 \\ (0.011)} & \makecell[r]{177 \\ (17)} \\
bl & \makecell[r]{15.348\% \\ (2.260\%)} & \makecell[r]{0.419 \\ (0.049)} & \makecell[r]{0.647 \\ (0.047)} & \makecell[r]{1.074 \\ (0.071)} & \makecell[r]{0.477 \\ (0.038)} & \makecell[r]{0.494 \\ (0.025)} & \makecell[r]{242 \\ (29)} \\
lw & \makecell[r]{7.359\% \\ (0.330\%)} & \makecell[r]{0.123 \\ (0.002)} & \makecell[r]{0.608 \\ (0.027)} & \makecell[r]{1.004 \\ (0.045)} & \makecell[r]{0.382 \\ (0.023)} & \makecell[r]{0.225 \\ (0.008)} & \makecell[r]{430 \\ (37)} \\
ff & \makecell[r]{13.602\% \\ (0.276\%)} & \makecell[r]{0.148 \\ (0.002)} & \makecell[r]{0.929 \\ (0.020)} & \makecell[r]{1.507 \\ (0.033)} & \makecell[r]{0.683 \\ (0.019)} & \makecell[r]{0.207 \\ (0.004)} & \makecell[r]{112 \\ (8)} \\
rp & \makecell[r]{13.813\% \\ (0.223\%)} & \makecell[r]{0.155 \\ (0.002)} & \makecell[r]{0.908 \\ (0.021)} & \makecell[r]{1.467 \\ (0.034)} & \makecell[r]{0.649 \\ (0.020)} & \makecell[r]{0.227 \\ (0.005)} & \makecell[r]{155 \\ (10)} \\
min\_v & \makecell[r]{14.708\% \\ (0.367\%)} & \makecell[r]{0.145 \\ (0.002)} & \makecell[r]{\textbf{1.034} \\ (0.028)} & \makecell[r]{\textbf{1.712} \\ (0.048)} & \makecell[r]{\textbf{0.830} \\ (0.034)} & \makecell[r]{0.198 \\ (0.006)} & \makecell[r]{154 \\ (12)} \\
pmv & \makecell[r]{4.757\% \\ (0.244\%)} & \makecell[r]{\textbf{0.079} \\ (0.001)} & \makecell[r]{0.615 \\ (0.031)} & \makecell[r]{1.018 \\ (0.052)} & \makecell[r]{0.394 \\ (0.028)} & \makecell[r]{\textbf{0.152} \\ (0.006)} & \makecell[r]{403 \\ (27)} \\
ctmv & \makecell[r]{-5.834\% \\ (13.875\%)} & \makecell[r]{1.063 \\ (0.178)} & \makecell[r]{0.303 \\ (0.065)} & \makecell[r]{0.670 \\ (0.106)} & \makecell[r]{0.148 \\ (0.137)} & \makecell[r]{0.607 \\ (0.039)} & \makecell[r]{625 \\ (40)} \\
ddpg & \makecell[r]{7.898\% \\ (1.193\%)} & \makecell[r]{0.255 \\ (0.013)} & \makecell[r]{0.348 \\ (0.044)} & \makecell[r]{0.575 \\ (0.072)} & \makecell[r]{0.233 \\ (0.026)} & \makecell[r]{0.480 \\ (0.019)} & \makecell[r]{741 \\ (43)} \\
ppo & \makecell[r]{8.333\% \\ (0.779\%)} & \makecell[r]{0.163 \\ (0.007)} & \makecell[r]{0.484 \\ (0.042)} & \makecell[r]{0.799 \\ (0.069)} & \makecell[r]{0.321 \\ (0.027)} & \makecell[r]{0.311 \\ (0.014)} & \makecell[r]{753 \\ (64)} \\
CTRL & \makecell[r]{14.539\% \\ (0.308\%)} & \makecell[r]{0.189 \\ (0.003)} & \makecell[r]{0.783 \\ (0.020)} & \makecell[r]{1.276 \\ (0.032)} & \makecell[r]{0.526 \\ (0.017)} & \makecell[r]{0.291 \\ (0.006)} & \makecell[r]{188 \\ (11)} \\
\bottomrule
\end{tabular}
\caption{\textbf{Comparison of out-of-sample performance of different allocation methods in the bull period (2010-2020).} We report annualized return, volatility, Sharpe ratio, Sortino ratio, and Calmar ratio, together with maximum drawdown (MDD) and recovery time (RT) in days, over 100 independent experiments. Each experiment is based on 10 randomly selected stocks (except for S\&P 500 index). For each cell, the upper number is the average (over the 100 experiments) while the lower one with parentheses is the standard error. }
\label{tab:performance_bull}
\end{table}

\begin{table}[H]
\centering
\begin{tabular}{lrrrrrr}
\toprule
 & BKR & TO & MaxL & MaxS & GE & GEMax \\
\midrule
ew & 0\% & 5.036\% & 10.000\% & 0.000\% & 100.000\% & 100.000\% \\
mv & 0\% & 19.019\% & 113.971\% & 58.252\% & 132.757\% & 260.003\% \\
js & 1\% & 17.320\% & 171.726\% & 82.796\% & 166.192\% & 542.149\% \\
bl & 16\% & 13.396\% & 168.732\% & 85.625\% & 167.874\% & 574.338\% \\
lw & 0\% & 20.636\% & 116.278\% & 62.322\% & 131.142\% & 295.426\% \\
ff & 0\% & 7.323\% & 116.479\% & 38.422\% & 99.081\% & 225.771\% \\
rp & 0\% & 4.793\% & 100.000\% & 0.000\% & 100.000\% & 100.000\% \\
min\_v & 0\% & 9.379\% & 104.477\% & 31.623\% & 141.237\% & 227.462\% \\
pmv & 0\% & 60.260\% & 75.980\% & 58.536\% & 87.609\% & 231.473\% \\
ctmv & 39\% & 36.135\% & 548.376\% & 328.037\% & 141.127\% & 2712.849\% \\
ddpg & 1\% & 25.795\% & 77.880\% & 70.666\% & 227.010\% & 532.347\% \\
ppo & 0\% & 27.899\% & 69.267\% & 51.967\% & 122.017\% & 296.168\% \\
CTRL & 0\% & 4.876\% & 18.586\% & 1.754\% & 102.505\% & 124.034\% \\
\bottomrule
\end{tabular}
\caption{\textbf{Trading and exposure statistics over the bull period 2010--2020.}
	BKR is the bankruptcy rate across experiments,  TO is the average monthly turnover,
	MaxL and MaxS are the maximum single-asset long and short positions, respectively,
	GE is the average gross exposure, and GEMax is the maximum gross exposure observed across all experiments.}
\label{tab:trade_bull}
\end{table}

\subsection{Sensitivity analysis on hyperparameters}
\label{appendix:sensitivity}

To evaluate the robustness of the CTRL algorithm with respect to the choice of hyperparameters,  we conduct a sensitivity analysis by varying three groups of hyperparameters: (i) the learning rates, (ii) the temperature parameter, and (iii) the (non-trainable) hyperparameters $\phi_3$ and $\theta_3$, which are the only hyperparameters appearing in the policy and value functions, respectively. The learning rates and the temperature parameter are each scaled by multiplicative factors of $1.5$, $2$, $2/3$, and $0.5$ relative to their baseline values used in the previously reported experiments.

We use the following notations in our tables of results:
\begin{itemize}
    \item \textbf{CTRL-baseline}: The exact CTRL Algorithm~\ref{alg:offline} in the main paper.
    \item \textbf{CTRL-lrX}: A variant where the learning rates are scaled by a factor of $X$ (e.g., \texttt{lr5} uses 5 times the baseline learning rate). 
    \item \textbf{CTRL-gammaX}: A variant where the temperature $\gamma$ is scaled by $X$, with larger values representing more exploration.
    \item \textbf{CTRL-approxX}: A variant where the only two non-trainable hyperparameters in the policy and value function ($\phi_3$ and $\theta_3$) are scaled by $X$.
\end{itemize}

The following Tables \ref{tab:sensitivity_gross} to \ref{tab:sensitivity_trade_tc100bps} report the resulting performance under proportional transaction costs of 0, 12.5, and 50 bps.

\begin{table}[H]
		\centering
\begin{tabular}{llllllll}
\toprule
 & Return & Volatility & Sharpe & Sortino & Calmar & MDD & RT \\
\midrule
CTRL-baseline & \makecell[r]{12.211\%\\(0.219\%)} & \makecell[r]{0.232\\(0.003)} & \makecell[r]{0.531\\(0.010)} & \makecell[r]{0.865\\(0.017)} & \makecell[r]{0.200\\(0.005)} & \makecell[r]{0.621\\(0.009)} & \makecell[r]{521\\(30)} \\
CTRL-lr3/2 & \makecell[r]{12.202\%\\(0.220\%)} & \makecell[r]{0.232\\(0.003)} & \makecell[r]{0.530\\(0.010)} & \makecell[r]{0.864\\(0.017)} & \makecell[r]{0.200\\(0.005)} & \makecell[r]{0.621\\(0.009)} & \makecell[r]{523\\(30)} \\
CTRL-lr2/1 & \makecell[r]{12.192\%\\(0.221\%)} & \makecell[r]{0.232\\(0.003)} & \makecell[r]{0.530\\(0.010)} & \makecell[r]{0.863\\(0.017)} & \makecell[r]{0.200\\(0.005)} & \makecell[r]{0.621\\(0.009)} & \makecell[r]{529\\(30)} \\
CTRL-lr2/3 & \makecell[r]{12.217\%\\(0.219\%)} & \makecell[r]{0.233\\(0.003)} & \makecell[r]{0.531\\(0.010)} & \makecell[r]{0.865\\(0.017)} & \makecell[r]{0.201\\(0.005)} & \makecell[r]{0.620\\(0.009)} & \makecell[r]{519\\(30)} \\
CTRL-lr1/2 & \makecell[r]{12.220\%\\(0.219\%)} & \makecell[r]{0.233\\(0.003)} & \makecell[r]{0.531\\(0.010)} & \makecell[r]{0.866\\(0.017)} & \makecell[r]{0.201\\(0.005)} & \makecell[r]{0.620\\(0.009)} & \makecell[r]{513\\(30)} \\
CTRL-gamma3/2 & \makecell[r]{12.211\%\\(0.219\%)} & \makecell[r]{0.232\\(0.003)} & \makecell[r]{0.531\\(0.010)} & \makecell[r]{0.865\\(0.017)} & \makecell[r]{0.200\\(0.005)} & \makecell[r]{0.621\\(0.009)} & \makecell[r]{521\\(30)} \\
CTRL-gamma2/1 & \makecell[r]{12.210\%\\(0.219\%)} & \makecell[r]{0.232\\(0.003)} & \makecell[r]{0.531\\(0.010)} & \makecell[r]{0.865\\(0.017)} & \makecell[r]{0.200\\(0.005)} & \makecell[r]{0.621\\(0.009)} & \makecell[r]{521\\(30)} \\
CTRL-gamma2/3 & \makecell[r]{12.212\%\\(0.219\%)} & \makecell[r]{0.232\\(0.003)} & \makecell[r]{0.531\\(0.010)} & \makecell[r]{0.865\\(0.017)} & \makecell[r]{0.200\\(0.005)} & \makecell[r]{0.621\\(0.009)} & \makecell[r]{521\\(30)} \\
CTRL-gamma1/2 & \makecell[r]{12.212\%\\(0.219\%)} & \makecell[r]{0.232\\(0.003)} & \makecell[r]{0.531\\(0.010)} & \makecell[r]{0.865\\(0.017)} & \makecell[r]{0.200\\(0.005)} & \makecell[r]{0.621\\(0.009)} & \makecell[r]{521\\(30)} \\
CTRL-approx3/2 & \makecell[r]{12.165\%\\(0.220\%)} & \makecell[r]{0.232\\(0.003)} & \makecell[r]{0.530\\(0.010)} & \makecell[r]{0.863\\(0.017)} & \makecell[r]{0.200\\(0.005)} & \makecell[r]{0.621\\(0.009)} & \makecell[r]{533\\(30)} \\
CTRL-approx2/1 & \makecell[r]{12.097\%\\(0.223\%)} & \makecell[r]{0.231\\(0.003)} & \makecell[r]{0.528\\(0.010)} & \makecell[r]{0.858\\(0.017)} & \makecell[r]{0.198\\(0.005)} & \makecell[r]{0.622\\(0.009)} & \makecell[r]{569\\(30)} \\
CTRL-approx2/3 & \makecell[r]{12.229\%\\(0.219\%)} & \makecell[r]{0.233\\(0.003)} & \makecell[r]{0.531\\(0.010)} & \makecell[r]{0.865\\(0.017)} & \makecell[r]{0.201\\(0.005)} & \makecell[r]{0.621\\(0.009)} & \makecell[r]{519\\(30)} \\
CTRL-approx1/2 & \makecell[r]{12.235\%\\(0.219\%)} & \makecell[r]{0.233\\(0.003)} & \makecell[r]{0.531\\(0.010)} & \makecell[r]{0.866\\(0.017)} & \makecell[r]{0.201\\(0.005)} & \makecell[r]{0.621\\(0.009)} & \makecell[r]{513\\(30)} \\
\bottomrule
\end{tabular}
\caption{Performance summary: No transaction cost.}
\label{tab:sensitivity_gross}
\end{table}

\begin{table}[H]
		\centering
\begin{tabular}{llllllll}
\toprule
 & Return & Volatility & Sharpe & Sortino & Calmar & MDD & RT \\
\midrule
CTRL-baseline & \makecell[r]{12.115\%\\(0.219\%)} & \makecell[r]{0.232\\(0.003)} & \makecell[r]{0.527\\(0.010)} & \makecell[r]{0.858\\(0.017)} & \makecell[r]{0.199\\(0.005)} & \makecell[r]{0.621\\(0.009)} & \makecell[r]{523\\(30)} \\
CTRL-lr3/2 & \makecell[r]{12.106\%\\(0.220\%)} & \makecell[r]{0.232\\(0.003)} & \makecell[r]{0.526\\(0.010)} & \makecell[r]{0.857\\(0.017)} & \makecell[r]{0.198\\(0.005)} & \makecell[r]{0.622\\(0.009)} & \makecell[r]{528\\(30)} \\
CTRL-lr2/1 & \makecell[r]{12.097\%\\(0.220\%)} & \makecell[r]{0.232\\(0.003)} & \makecell[r]{0.526\\(0.010)} & \makecell[r]{0.856\\(0.017)} & \makecell[r]{0.198\\(0.005)} & \makecell[r]{0.622\\(0.009)} & \makecell[r]{537\\(30)} \\
CTRL-lr2/3 & \makecell[r]{12.121\%\\(0.218\%)} & \makecell[r]{0.233\\(0.003)} & \makecell[r]{0.527\\(0.010)} & \makecell[r]{0.859\\(0.017)} & \makecell[r]{0.199\\(0.005)} & \makecell[r]{0.621\\(0.009)} & \makecell[r]{522\\(30)} \\
CTRL-lr1/2 & \makecell[r]{12.124\%\\(0.218\%)} & \makecell[r]{0.233\\(0.003)} & \makecell[r]{0.527\\(0.010)} & \makecell[r]{0.859\\(0.017)} & \makecell[r]{0.199\\(0.005)} & \makecell[r]{0.621\\(0.009)} & \makecell[r]{521\\(30)} \\
CTRL-gamma3/2 & \makecell[r]{12.115\%\\(0.219\%)} & \makecell[r]{0.232\\(0.003)} & \makecell[r]{0.527\\(0.010)} & \makecell[r]{0.858\\(0.017)} & \makecell[r]{0.199\\(0.005)} & \makecell[r]{0.621\\(0.009)} & \makecell[r]{523\\(30)} \\
CTRL-gamma2/1 & \makecell[r]{12.115\%\\(0.219\%)} & \makecell[r]{0.232\\(0.003)} & \makecell[r]{0.527\\(0.010)} & \makecell[r]{0.858\\(0.017)} & \makecell[r]{0.199\\(0.005)} & \makecell[r]{0.621\\(0.009)} & \makecell[r]{523\\(30)} \\
CTRL-gamma2/3 & \makecell[r]{12.116\%\\(0.219\%)} & \makecell[r]{0.232\\(0.003)} & \makecell[r]{0.527\\(0.010)} & \makecell[r]{0.858\\(0.017)} & \makecell[r]{0.199\\(0.005)} & \makecell[r]{0.621\\(0.009)} & \makecell[r]{523\\(30)} \\
CTRL-gamma1/2 & \makecell[r]{12.116\%\\(0.219\%)} & \makecell[r]{0.232\\(0.003)} & \makecell[r]{0.527\\(0.010)} & \makecell[r]{0.858\\(0.017)} & \makecell[r]{0.199\\(0.005)} & \makecell[r]{0.621\\(0.009)} & \makecell[r]{523\\(30)} \\
CTRL-approx3/2 & \makecell[r]{12.070\%\\(0.219\%)} & \makecell[r]{0.232\\(0.003)} & \makecell[r]{0.526\\(0.010)} & \makecell[r]{0.856\\(0.017)} & \makecell[r]{0.198\\(0.005)} & \makecell[r]{0.621\\(0.009)} & \makecell[r]{536\\(30)} \\
CTRL-approx2/1 & \makecell[r]{12.003\%\\(0.222\%)} & \makecell[r]{0.231\\(0.003)} & \makecell[r]{0.524\\(0.010)} & \makecell[r]{0.852\\(0.017)} & \makecell[r]{0.196\\(0.005)} & \makecell[r]{0.623\\(0.009)} & \makecell[r]{575\\(31)} \\
CTRL-approx2/3 & \makecell[r]{12.133\%\\(0.219\%)} & \makecell[r]{0.233\\(0.003)} & \makecell[r]{0.527\\(0.010)} & \makecell[r]{0.859\\(0.017)} & \makecell[r]{0.199\\(0.005)} & \makecell[r]{0.621\\(0.009)} & \makecell[r]{521\\(30)} \\
CTRL-approx1/2 & \makecell[r]{12.139\%\\(0.219\%)} & \makecell[r]{0.233\\(0.003)} & \makecell[r]{0.527\\(0.010)} & \makecell[r]{0.859\\(0.017)} & \makecell[r]{0.199\\(0.005)} & \makecell[r]{0.621\\(0.009)} & \makecell[r]{519\\(30)} \\
\bottomrule
\end{tabular}
\caption{Performance summary: Transaction cost 12.5 bps.}
\label{tab:sensitivity_tc25bps}
\end{table}

\begin{table}[H]
		\centering
\begin{tabular}{llllllll}
\toprule
 & Return & Volatility & Sharpe & Sortino & Calmar & MDD & RT \\
\midrule
CTRL-baseline & \makecell[r]{11.828\%\\(0.218\%)} & \makecell[r]{0.233\\(0.003)} & \makecell[r]{0.514\\(0.010)} & \makecell[r]{0.837\\(0.017)} & \makecell[r]{0.193\\(0.005)} & \makecell[r]{0.623\\(0.009)} & \makecell[r]{538\\(31)} \\
CTRL-lr3/2 & \makecell[r]{11.820\%\\(0.219\%)} & \makecell[r]{0.232\\(0.003)} & \makecell[r]{0.514\\(0.010)} & \makecell[r]{0.837\\(0.017)} & \makecell[r]{0.193\\(0.005)} & \makecell[r]{0.624\\(0.009)} & \makecell[r]{548\\(31)} \\
CTRL-lr2/1 & \makecell[r]{11.812\%\\(0.219\%)} & \makecell[r]{0.232\\(0.003)} & \makecell[r]{0.514\\(0.010)} & \makecell[r]{0.836\\(0.017)} & \makecell[r]{0.193\\(0.005)} & \makecell[r]{0.624\\(0.009)} & \makecell[r]{556\\(31)} \\
CTRL-lr2/3 & \makecell[r]{11.834\%\\(0.217\%)} & \makecell[r]{0.233\\(0.003)} & \makecell[r]{0.515\\(0.010)} & \makecell[r]{0.838\\(0.017)} & \makecell[r]{0.194\\(0.005)} & \makecell[r]{0.623\\(0.009)} & \makecell[r]{537\\(31)} \\
CTRL-lr1/2 & \makecell[r]{11.836\%\\(0.217\%)} & \makecell[r]{0.233\\(0.003)} & \makecell[r]{0.515\\(0.010)} & \makecell[r]{0.838\\(0.017)} & \makecell[r]{0.194\\(0.005)} & \makecell[r]{0.623\\(0.009)} & \makecell[r]{536\\(31)} \\
CTRL-gamma3/2 & \makecell[r]{11.828\%\\(0.218\%)} & \makecell[r]{0.232\\(0.003)} & \makecell[r]{0.514\\(0.010)} & \makecell[r]{0.837\\(0.017)} & \makecell[r]{0.193\\(0.005)} & \makecell[r]{0.623\\(0.009)} & \makecell[r]{538\\(31)} \\
CTRL-gamma2/1 & \makecell[r]{11.828\%\\(0.218\%)} & \makecell[r]{0.232\\(0.003)} & \makecell[r]{0.514\\(0.010)} & \makecell[r]{0.837\\(0.017)} & \makecell[r]{0.193\\(0.005)} & \makecell[r]{0.623\\(0.009)} & \makecell[r]{538\\(31)} \\
CTRL-gamma2/3 & \makecell[r]{11.829\%\\(0.218\%)} & \makecell[r]{0.233\\(0.003)} & \makecell[r]{0.514\\(0.010)} & \makecell[r]{0.837\\(0.017)} & \makecell[r]{0.193\\(0.005)} & \makecell[r]{0.623\\(0.009)} & \makecell[r]{538\\(31)} \\
CTRL-gamma1/2 & \makecell[r]{11.829\%\\(0.218\%)} & \makecell[r]{0.233\\(0.003)} & \makecell[r]{0.514\\(0.010)} & \makecell[r]{0.837\\(0.017)} & \makecell[r]{0.193\\(0.005)} & \makecell[r]{0.623\\(0.009)} & \makecell[r]{538\\(31)} \\
CTRL-approx3/2 & \makecell[r]{11.786\%\\(0.218\%)} & \makecell[r]{0.232\\(0.003)} & \makecell[r]{0.514\\(0.010)} & \makecell[r]{0.836\\(0.017)} & \makecell[r]{0.193\\(0.005)} & \makecell[r]{0.623\\(0.009)} & \makecell[r]{556\\(31)} \\
CTRL-approx2/1 & \makecell[r]{11.721\%\\(0.221\%)} & \makecell[r]{0.231\\(0.003)} & \makecell[r]{0.512\\(0.010)} & \makecell[r]{0.831\\(0.017)} & \makecell[r]{0.191\\(0.004)} & \makecell[r]{0.625\\(0.008)} & \makecell[r]{593\\(31)} \\
CTRL-approx2/3 & \makecell[r]{11.845\%\\(0.218\%)} & \makecell[r]{0.233\\(0.003)} & \makecell[r]{0.515\\(0.010)} & \makecell[r]{0.838\\(0.017)} & \makecell[r]{0.194\\(0.005)} & \makecell[r]{0.623\\(0.009)} & \makecell[r]{537\\(31)} \\
CTRL-approx1/2 & \makecell[r]{11.851\%\\(0.218\%)} & \makecell[r]{0.233\\(0.003)} & \makecell[r]{0.515\\(0.010)} & \makecell[r]{0.838\\(0.017)} & \makecell[r]{0.194\\(0.005)} & \makecell[r]{0.623\\(0.009)} & \makecell[r]{541\\(31)} \\
\bottomrule
\end{tabular}
\caption{Performance summary: Transaction cost 50 bps.}
\label{tab:sensitivity_tc100bps}
\end{table}


\begin{table}[H]
		\centering
\begin{tabular}{lllllll}
\toprule
 & BKR & TO & MaxL & MaxS & GE & GEMax \\
\midrule
CTRL-baseline & \makecell[r]{0\%} & \makecell[r]{5.679\%} & \makecell[r]{18.778\%} & \makecell[r]{1.903\%} & \makecell[r]{102.301\%} & \makecell[r]{124.034\%} \\
CTRL-lr3/2 & \makecell[r]{0\%} & \makecell[r]{5.658\%} & \makecell[r]{18.903\%} & \makecell[r]{1.930\%} & \makecell[r]{102.479\%} & \makecell[r]{125.026\%} \\
CTRL-lr2/1 & \makecell[r]{0\%} & \makecell[r]{5.637\%} & \makecell[r]{18.901\%} & \makecell[r]{1.957\%} & \makecell[r]{102.657\%} & \makecell[r]{126.016\%} \\
CTRL-lr2/3 & \makecell[r]{0\%} & \makecell[r]{5.693\%} & \makecell[r]{18.653\%} & \makecell[r]{1.885\%} & \makecell[r]{102.179\%} & \makecell[r]{123.366\%} \\
CTRL-lr1/2 & \makecell[r]{0\%} & \makecell[r]{5.700\%} & \makecell[r]{18.590\%} & \makecell[r]{1.876\%} & \makecell[r]{102.118\%} & \makecell[r]{122.977\%} \\
CTRL-gamma3/2 & \makecell[r]{0\%} & \makecell[r]{5.679\%} & \makecell[r]{18.780\%} & \makecell[r]{1.903\%} & \makecell[r]{102.297\%} & \makecell[r]{124.026\%} \\
CTRL-gamma2/1 & \makecell[r]{0\%} & \makecell[r]{5.679\%} & \makecell[r]{18.782\%} & \makecell[r]{1.904\%} & \makecell[r]{102.293\%} & \makecell[r]{124.019\%} \\
CTRL-gamma2/3 & \makecell[r]{0\%} & \makecell[r]{5.679\%} & \makecell[r]{18.777\%} & \makecell[r]{1.903\%} & \makecell[r]{102.303\%} & \makecell[r]{124.039\%} \\
CTRL-gamma1/2 & \makecell[r]{0\%} & \makecell[r]{5.679\%} & \makecell[r]{18.776\%} & \makecell[r]{1.903\%} & \makecell[r]{102.304\%} & \makecell[r]{124.041\%} \\
CTRL-approx3/2 & \makecell[r]{0\%} & \makecell[r]{5.630\%} & \makecell[r]{18.915\%} & \makecell[r]{2.005\%} & \makecell[r]{102.400\%} & \makecell[r]{126.072\%} \\
CTRL-approx2/1 & \makecell[r]{0\%} & \makecell[r]{5.590\%} & \makecell[r]{18.919\%} & \makecell[r]{4.065\%} & \makecell[r]{102.977\%} & \makecell[r]{132.107\%} \\
CTRL-approx2/3 & \makecell[r]{0\%} & \makecell[r]{5.698\%} & \makecell[r]{18.586\%} & \makecell[r]{1.880\%} & \makecell[r]{102.277\%} & \makecell[r]{123.440\%} \\
CTRL-approx1/2 & \makecell[r]{0\%} & \makecell[r]{5.705\%} & \makecell[r]{18.534\%} & \makecell[r]{1.876\%} & \makecell[r]{102.267\%} & \makecell[r]{123.214\%} \\
\bottomrule
\end{tabular}
\caption{Trading statistics: No transaction cost. }
\label{tab:sensitivity_trade_gross}
\end{table}

\begin{table}[H]
		\centering
\begin{tabular}{lllllll}
\toprule
 & BKR & TO & MaxL & MaxS & GE & GEMax \\
\midrule
CTRL-baseline & \makecell[r]{0\%} & \makecell[r]{5.679\%} & \makecell[r]{18.778\%} & \makecell[r]{1.903\%} & \makecell[r]{102.301\%} & \makecell[r]{124.034\%} \\
CTRL-lr3/2 & \makecell[r]{0\%} & \makecell[r]{5.658\%} & \makecell[r]{18.903\%} & \makecell[r]{1.930\%} & \makecell[r]{102.479\%} & \makecell[r]{125.026\%} \\
CTRL-lr2/1 & \makecell[r]{0\%} & \makecell[r]{5.638\%} & \makecell[r]{18.901\%} & \makecell[r]{1.957\%} & \makecell[r]{102.657\%} & \makecell[r]{126.016\%} \\
CTRL-lr2/3 & \makecell[r]{0\%} & \makecell[r]{5.693\%} & \makecell[r]{18.653\%} & \makecell[r]{1.885\%} & \makecell[r]{102.179\%} & \makecell[r]{123.366\%} \\
CTRL-lr1/2 & \makecell[r]{0\%} & \makecell[r]{5.701\%} & \makecell[r]{18.590\%} & \makecell[r]{1.876\%} & \makecell[r]{102.118\%} & \makecell[r]{122.977\%} \\
CTRL-gamma3/2 & \makecell[r]{0\%} & \makecell[r]{5.679\%} & \makecell[r]{18.780\%} & \makecell[r]{1.903\%} & \makecell[r]{102.297\%} & \makecell[r]{124.026\%} \\
CTRL-gamma2/1 & \makecell[r]{0\%} & \makecell[r]{5.679\%} & \makecell[r]{18.782\%} & \makecell[r]{1.904\%} & \makecell[r]{102.293\%} & \makecell[r]{124.019\%} \\
CTRL-gamma2/3 & \makecell[r]{0\%} & \makecell[r]{5.679\%} & \makecell[r]{18.777\%} & \makecell[r]{1.903\%} & \makecell[r]{102.303\%} & \makecell[r]{124.039\%} \\
CTRL-gamma1/2 & \makecell[r]{0\%} & \makecell[r]{5.680\%} & \makecell[r]{18.776\%} & \makecell[r]{1.903\%} & \makecell[r]{102.304\%} & \makecell[r]{124.041\%} \\
CTRL-approx3/2 & \makecell[r]{0\%} & \makecell[r]{5.630\%} & \makecell[r]{18.915\%} & \makecell[r]{2.005\%} & \makecell[r]{102.400\%} & \makecell[r]{126.072\%} \\
CTRL-approx2/1 & \makecell[r]{0\%} & \makecell[r]{5.590\%} & \makecell[r]{18.919\%} & \makecell[r]{4.065\%} & \makecell[r]{102.977\%} & \makecell[r]{132.107\%} \\
CTRL-approx2/3 & \makecell[r]{0\%} & \makecell[r]{5.698\%} & \makecell[r]{18.586\%} & \makecell[r]{1.880\%} & \makecell[r]{102.277\%} & \makecell[r]{123.440\%} \\
CTRL-approx1/2 & \makecell[r]{0\%} & \makecell[r]{5.705\%} & \makecell[r]{18.534\%} & \makecell[r]{1.876\%} & \makecell[r]{102.267\%} & \makecell[r]{123.214\%} \\
\bottomrule
\end{tabular}
\caption{Trading statistics: Transaction cost 12.5 bps.}
\label{tab:sensitivity_trade_tc25bps}
\end{table}

\begin{table}[H]
	\centering
\begin{tabular}{lllllll}
\toprule
 & BKR & TO & MaxL & MaxS & GE & GEMax \\
\midrule
CTRL-baseline & \makecell[r]{0\%} & \makecell[r]{5.680\%} & \makecell[r]{18.778\%} & \makecell[r]{1.903\%} & \makecell[r]{102.301\%} & \makecell[r]{124.034\%} \\
CTRL-lr3/2 & \makecell[r]{0\%} & \makecell[r]{5.659\%} & \makecell[r]{18.903\%} & \makecell[r]{1.930\%} & \makecell[r]{102.479\%} & \makecell[r]{125.026\%} \\
CTRL-lr2/1 & \makecell[r]{0\%} & \makecell[r]{5.638\%} & \makecell[r]{18.901\%} & \makecell[r]{1.957\%} & \makecell[r]{102.657\%} & \makecell[r]{126.016\%} \\
CTRL-lr2/3 & \makecell[r]{0\%} & \makecell[r]{5.694\%} & \makecell[r]{18.653\%} & \makecell[r]{1.885\%} & \makecell[r]{102.179\%} & \makecell[r]{123.366\%} \\
CTRL-lr1/2 & \makecell[r]{0\%} & \makecell[r]{5.701\%} & \makecell[r]{18.590\%} & \makecell[r]{1.876\%} & \makecell[r]{102.118\%} & \makecell[r]{122.977\%} \\
CTRL-gamma3/2 & \makecell[r]{0\%} & \makecell[r]{5.680\%} & \makecell[r]{18.780\%} & \makecell[r]{1.903\%} & \makecell[r]{102.297\%} & \makecell[r]{124.026\%} \\
CTRL-gamma2/1 & \makecell[r]{0\%} & \makecell[r]{5.680\%} & \makecell[r]{18.782\%} & \makecell[r]{1.904\%} & \makecell[r]{102.293\%} & \makecell[r]{124.019\%} \\
CTRL-gamma2/3 & \makecell[r]{0\%} & \makecell[r]{5.680\%} & \makecell[r]{18.777\%} & \makecell[r]{1.903\%} & \makecell[r]{102.303\%} & \makecell[r]{124.039\%} \\
CTRL-gamma1/2 & \makecell[r]{0\%} & \makecell[r]{5.680\%} & \makecell[r]{18.776\%} & \makecell[r]{1.903\%} & \makecell[r]{102.304\%} & \makecell[r]{124.041\%} \\
CTRL-approx3/2 & \makecell[r]{0\%} & \makecell[r]{5.631\%} & \makecell[r]{18.915\%} & \makecell[r]{2.005\%} & \makecell[r]{102.400\%} & \makecell[r]{126.072\%} \\
CTRL-approx2/1 & \makecell[r]{0\%} & \makecell[r]{5.591\%} & \makecell[r]{18.919\%} & \makecell[r]{4.065\%} & \makecell[r]{102.977\%} & \makecell[r]{132.107\%} \\
CTRL-approx2/3 & \makecell[r]{0\%} & \makecell[r]{5.699\%} & \makecell[r]{18.586\%} & \makecell[r]{1.880\%} & \makecell[r]{102.277\%} & \makecell[r]{123.440\%} \\
CTRL-approx1/2 & \makecell[r]{0\%} & \makecell[r]{5.706\%} & \makecell[r]{18.534\%} & \makecell[r]{1.876\%} & \makecell[r]{102.267\%} & \makecell[r]{123.214\%} \\
\bottomrule
\end{tabular}
\caption{Trading statistics: Transaction cost 50 bps.}
\label{tab:sensitivity_trade_tc100bps}
\end{table}

Evidently, across all the three transaction cost settings,  CTRL is robust to variations in learning rates, the temperature parameter, and the non-trainable hyperparameters in  policy and value functions. Both performance metrics and trading statistics are very stable with only minor fluctuations under parameter perturbations. This sensitivity analysis shows that the empirical performance of CTRL is not driven by fine tuning but by its underlying distinctive approach. 







\end{document}